\documentclass [letter, 12pt]{article}
\usepackage{a4wide, amsmath, amsfonts, amssymb, geometry}
\usepackage{bm}
\usepackage{fancyhdr, dsfont}
\usepackage[latin1]{inputenc}
\usepackage[pdftex]{graphicx} 
\usepackage{graphicx}
\usepackage{float}
\usepackage{slashed}
\usepackage{rotating}
\usepackage{color}
\definecolor{dgreen}{rgb}{0,0.70,0.30}
\definecolor{gold}{rgb}{0.85,.66,0}
\definecolor{purple}{rgb}{1.0,0.3,0.6}

\usepackage{tikz}
\usetikzlibrary{calc} \usetikzlibrary{patterns} \usetikzlibrary{decorations.pathreplacing} \usetikzlibrary{decorations.markings} \usetikzlibrary{decorations.pathmorphing} \usetikzlibrary{positioning}

\restylefloat{figure}
\usepackage{epstopdf}

\newcommand{\nwc}{\newcommand}
\nwc{\ba}  {\begin{array}}
\nwc{\ea}  {\end{array}}
\nwc{\bdm} {\begin{displaymath}}
\nwc{\edm} {\end{displaymath}}

\nwc{\bea} {\begin{equation}\ba{lcl}}
\nwc{\eea} {\ea\end{equation}}

\nwc{\bda} {\bdm\ba{lcl}}
\nwc{\eda} {\ea\edm}

\nwc{\bc}  {\begin{center}}
\nwc{\ec}  {\end{center}}

\nwc{\ds}  {\displaystyle}
\nwc{\bmat}{\left(\ba}
\nwc{\emat}{\ea\right)}
\nwc{\nn}  {\nonumber}
\nwc{\nnn} {\nonumber \vspace{.2cm} \\ }
\nwc{\ra}  {\rightarrow}
\nwc{\lra} {\longrightarrow}

\nwc{\p} {\partial}

\def\beq{\begin{equation}}
\def\eeq{\end{equation}}

\newcommand{\vecb}{\left(\begin{array}{c}}
\newcommand{\vece}{\end{array}\right)}
\newcommand{\ccb}{\left(\begin{array}{cc}}
\newcommand{\cce}{\end{array}\right)}
\newcommand{\cccb}{\left(\begin{array}{ccc}}
\newcommand{\ccce}{\end{array}\right)}
\newcommand{\ccccb}{\left(\begin{array}{cccc}}
\newcommand{\cccce}{\end{array}\right)}
\newcommand{\cccccb}{\left(\begin{array}{ccccc}}
\newcommand{\ccccce}{\end{array}\right)}

\newcommand{\pa}{\partial}

\newcommand{\al}{\alpha}
\newcommand{\be}{\beta}
\newcommand{\ga}{\gamma}
\newcommand{\de}{\delta}

\newcommand{\vep}{\varepsilon}

\newcommand{\si}{\sigma}
\newcommand{\la}{\lambda}

\newcommand{\om}{\omega}

\newcommand{\Si}{\Sigma}
\newcommand{\La}{\Lambda}
\newcommand{\Om}{\Omega}

\newcommand{\te}{\textrm}
\newcommand{\eq}{ \ \ = \ \ }
\newcommand{\co}{\ , \ \ \ \ \ \ }
\newcommand{\dd}{\mathrm{d}}
\newcommand{\ee}{\mathrm{e}}

\newcommand{\ap}{\alpha'}

\newcommand{\dal}{\dot{\alpha}}
\newcommand{\dbe}{\dot{\beta}}
\newcommand{\dga}{\dot{\gamma}}

\newcommand{\sib}{\bar{\sigma}}
\newcommand{\gab}{\bar{\gamma}}

\newcommand{\tspinb}{\Theta^{\vec{a}}_{\vec{b}} \left[ \begin{smallmatrix}}
\newcommand{\tspine}{\end{smallmatrix} \right]}

\allowdisplaybreaks

\begin{document}


\title{\textbf{Massive Supermultiplets in Four-Dimensional} \\
 \textbf{Superstring Theory}\\[0.5cm]}
\author{Wan-Zhe Feng$^{\te{a,b}}$, Dieter L\"ust$^{\te{b,c}}$,  Oliver
Schlotterer$^{\te{b,d}}$}
\date{\today}
\smallskip
\maketitle
\centerline{\it $^{\rm a}$  \hskip -1.5mm Department of Physics, Northeastern University,
Boston, MA 02115, USA}
\centerline{\it $^{\rm b}$ \hskip -1.5mm Max--Planck--Institut f\"ur Physik
Werner--Heisenberg--Institut,
\it 80805 M\"unchen, Germany}
\centerline{\it $^{\rm c}$  \hskip -1.5mm Arnold--Sommerfeld--Center for Theoretical
Physics,
}
\centerline{\it Ludwig--Maximilians--Universit\"at, 80333 M\"unchen, Germany}
\centerline{\it $^{\rm d}$  \hskip -1.5mm Max--Planck--Institut f\"ur Gravitationsphysik Albert--Einstein--Institut, 14476 Potsdam, Germany}

\medskip\bigskip\vskip2cm {\abstract \noindent We extend the discussion of \cite{Feng:2010yx} on massive Regge excitations on the first mass level of four-dimensional superstring theory. For the lightest massive modes of the open string sector, universal supermultiplets common to all four-dimensional compactifications with ${\cal N}=1, 2$ and ${\cal N}=4$ spacetime supersymmetry are constructed respectively -- both their vertex operators and their supersymmetry variations. Massive spinor helicity methods shed light on the interplay between individual polarization states.
} 

\begin{flushright}
{\small LMU-ASC 10/12} \\
{\small MPP-2012-11} \\
{\small AEI-2012-018}
\end{flushright}

\thispagestyle{empty}

\newpage
\setcounter{tocdepth}{2}
\tableofcontents

\numberwithin{equation}{section}

\newpage

\section{Introduction}

String theory as seen from the point particle perspective contains an infinite number of massive, higher spin states.
As known already from the early days of string theory, these massive states lie on the so-called Regge trajectories that display the linear
relation between the (mass)$^2$ and the spin $J$ of all states. The existence of the infinitely many higher spin states is essential for the
ultra-violet behavior of string scattering amplitudes. It unitarizes all string amplitudes among the massless modes
in the UV via the exchange of the infinite tower of massive states, as it was first shown in the famous Veneziano amplitude that describes
the scattering of four massless open string states. In addition the consistency of quantum gravity in string theory completely relies
on the massive higher spin states. Since the size of a string grows with its excitation energy, larger and larger
states are produced at higher and higher energies. Hence, the UV properties of string scattering amplitudes are non-Wilsonian,
which is also manifest in the UV-IR mixing in string theory. As it was argued in \cite{Amati:1988tn} this might lead to a reformulation of the Heisenberg uncertainty
principle in string theory with the result that the string scale appears at the shortest possible length scale, which can be dissolved in string scattering experiments.

\medskip
The existence of massive higher spin states in string theory is not only crucial  for the consistency of the theory, but is possibly  also interesting from
the phenomenological point of view.
Since the masses of the higher spin states are all multiples of the string scale $M_s=\sqrt{\alpha'^{-1}}$, D-brane compactifications
with a low string scale $M_s$ in the TeV region and with large extra dimensions offer the exciting possibility that the lightest Regge excitations
of massless open strings can be directly produced and detected at the LHC. As it was shown in  \cite{Lust:2008qc, Lust:2009pz}, four- and five-point  string scattering amplitudes among standard model gauge bosons (gluons, $W,Z$-bosons, photons) and  at most two external massless fermions (quarks or leptons) are completely
independent from any geometrical details of the underlying  D-brane model in four dimensions. Hence, the production of the first
heavy colored string states (e.g. excited gluons with $J=0,2$)
from gluon fusion  and their subsequent decay into two or three hadronic jets leads to completely model independent cross sections
and decay rates at the LHC.
Based on these calculations the recent LHC searches for  non-standard dijet events due to heavy new resonances can now exclude massive string states
with masses below about 4 TeV.

\medskip
This paper is not so much concerned about the phenomenological implications of massive higher spins states, but we rather like
to exploit some of the basic supersymmetry properties of higher spin states in four dimensions, originating
from supersymmetric type II compactifications.
In  \cite{Feng:2010yx}   we already computed string scattering amplitudes not only with massless external string states, but also
three- and four-point amplitudes with one massive excited open string state as external field. For this purpose
we constructed in \cite{Feng:2010yx}  the corresponding covariant vertex operators for the lowest massive open states
in four dimensions,\footnote{Additional and previous work
on vertex operators for massive higher spin excitations includes
\cite{Koh:1987hm,Berkovits:2002qx,Sagnotti:2010at,Bianchi:2010es,Park:2011if,Feng:2011qc}.}
focusing in particular on those universal open Regge states,
which are present in any D-brane compactification to four dimensions (excited gluons and gluinos).
Working within the world-sheet NS-R formalism, physical, massive states belong to the cohomology of the BRST operator. In addition,
besides world-sheet conformal invariance, supersymmetry plays a key role for the consistency of string theory, both on the world-sheet as well
as in target space. In ten spacetime dimensions, the type IIB(A) superstring exhibits extended (non-)chiral ${\cal N}=2$ spacetime
supersymmetry with in total
32 supersymmetry charges. It follows that all  massless as well as all massive closed string states are organized in supermultiplets of the ten-dimensional
${\cal N}=2$ supersymmetry algebra. This leads to a very subtle interplay between massive string excitations with different higher spins
that belong to common supersymmetry multiplets. In fact, the covariant world-sheet vertex operators of the higher spin states
must transform into each other when acting on them with the supersymmetry charge operators. Hence, spacetime supersymmetry must be
reflected in the structure of the world-sheet BRST cohomology on each mass level of the higher spin excitations.

\medskip
Going from ten to lower dimensions, parts or all of spacetime supersymmetry can be preserved during the compactification process.
As it is known already for several years \cite{FMS,BD1,BD2,BD3},
there exists a deep relation between the number of spacetime supersymmetries, preserved by
the compactification, and the number of world-sheet supersymmetries of the corresponding internal superconformal field theory.
Specifically, for type II compactifications on six-dimensional Calabi-Yau spaces, which correspond to $\hat c=6$ SCFT's with $(2,2)$ world
sheet supersymmetry, one obtains in the closed string sector four-dimensional ${\cal N}=2$ effective supergravity theories with 8 preserved supercharges in the bulk.
Second, type II compactification on $K3\times T^2$ with four-dimensional ${\cal N}=4$ spacetime
supersymmetry (16 bulk supercharges) can be described by the direct product of two SCFT's with central charges $\hat c=4$ and $\hat c=2$,
where the $\hat c=4$ part possesses $(4,4)$ supersymmetry on the world-sheet.
Finally, compactifications on a six-dimensional torus leads to effective type II supergravity theories with maximal ${\cal N}=8$ supersymmetry
(32 bulk supercharges).

\medskip
However, when also including D-branes and open strings, the number of spacetime supersymmetries is reduced by half compared to the closed string bulk sector, we just
discussed above. First, the effective, four-dimensional Yang-Mills theories of type IIB, Calabi-Yau orientifolds with D3/D7-branes
or with D5/D9-branes (or  type IIA Calabi-Yau orientifolds with intersecting D6-branes) possess just ${\cal N}=1$ supersymmetry. Next
the IIB $K3\times T^2$ orientifolds with D5/D9-branes lead to ${\cal N}=2$ supersymmetric Yang-Mills
theories in four dimensions.\footnote{These theories originate upon compactification on $T^2$ from $D=6$, IIB theories on $K3$ with (1,1)
spacetime  supersymmetry.}  And finally, toroidal compactifications of type II superstrings lead to Yang-Mills open string sectors with
${\cal N}=4$ supersymmetry in $D=4$.

\medskip
It is the aim of this paper to extend the work of \cite{Feng:2010yx} in order to systematically construct the covariant vertex operators of the lowest massive open string
supermultiplets for all three cases of ${\cal N}=4,2,1$ spacetime supersymmetry on the corresponding D-branes.
We will focus in particular on those massive supermultiplets and their SUSY transformations in the universal sector, which are always present in any four-dimensional orientifold models:
\begin{itemize}
\item  For  ${\cal N}=4$ super Yang-Mills, there is a single massive, spin two supermultiplet with 128 bosonic as well as 128 fermionic degrees of freedom.
\item The supermultiplets of the universal ${\cal N}=1$ sector contains one spin two supermultiplet and two spin 1/2 representations with in
total 12 + 12 bosonic and fermion degrees of freedom.
\item Finally, for ${\cal N}=2$ super Yang-Mills we are dealing with 40 + 40 massive open string
states, being organized in one spin two plus two spin one massive supermultiplets.
\end{itemize}
In this way we extend  the analysis of \cite{BD3} about the relation between world-sheet and spacetime supersymmetries
and their closed string (massless) supermultiplet structure to the case of the massive, open string supermultiplets.
At the same time we are giving here a massive version of the SUSY multiplet analysis in \cite{MHV}, where it was shown that SUSY Ward identities among scattering amplitudes are valid to all orders in $\ap$, and where the spinor helicity methods were
 applied to make efficient use of these Ward identities.

\medskip
The paper is organized as follows. As a warm-up case, in section \ref{sec:10dim} we first construct the covariant NS and R vertex operators of the ten-dimensional type I open string states at the first mass level. They comprise in total 128 + 128 bosonic as well as fermionic states. We verify that these states form a massive representation of the ten-dimensional (type I) ${\cal N}=1$ SUSY algebra.
Next, in section \ref{sec:SCFTuniv} we consider the SCFT's of string vacua in four dimensions, and discuss the relation between the extended world-sheet
superconformal algebras and the spacetime ${\cal N}=4,1,2$ SUSY algebras and the covariant vertex operators for the corresponding
supercharge operators. Sections \ref{sec:n4}, \ref{sec:n1} and \ref{sec:n2} are devoted to construct the massive open string supermultiplets, their vertex operators and their supersymmetry transformations
for the three cases of ${\cal N}=4$,  ${\cal N}=1$ and ${\cal N}=2$ supersymmetry in four dimensions respectively.
Finally, in section \ref{sec:hel} we study in more detail in helicity structure of the various on-shell supermultiplets.

\section{The first mass level in $D=10$}
\label{sec:10dim}

The lightest Regge excitations of open superstring theory in ten-dimensional Minkowski spacetime were firstly constructed in 1987 \cite{Koh:1987hm}. Let us briefly review the general method to construct heavy string excitations as well as the explicit results of \cite{Koh:1987hm} and then offer a covariant approach to the excited Ramond sector states.

\subsection{The general method}

Physical states belong to the cohomology of the BRST operator $Q_{\te{BRST}} $. In the world-sheet variables of the RNS formalism, it splits into three pieces of different superghost charge:
\begin{align}
Q_{\te{BRST}} \ \ &= \ \   Q_0 \ + \ Q_1 \ + \ Q_2 \label{BRST1} \ , \\
Q_0 \ \ &= \ \ \oint \frac{ \dd z }{2\pi i} \; \bigl( \, c \, (T + T_{\be,\ga}) \ + \ b \, c \, \pa c \, \bigr) \label{BRST2} \ , \\
Q_1 \ \ &= \ \ - \, \oint \frac{ \dd z }{2\pi i} \; \ga \, G \eq
-  \, \oint \frac{ \dd z }{2\pi i} \;  \ee^{\phi} \, \eta \, G \label{BRST3} \ , \\
Q_2 \ \ &= \ \ - \, \frac{1}{4} \, \oint \frac{ \dd z }{2\pi i} \;  b \, \ga^2 \eq - \, \frac{1}{4} \, \oint \frac{ \dd z }{2\pi i} \; b \, \ee^{2\phi} \, \eta \, \pa \eta \ . \label{BRST4}
\end{align}
We denote the $c=15$ stress tensor and supercurrent of the matter fields\footnote{Our normalization conventions for the world-sheet matter fields are fixed by
\beq
i \pa X^m (z) \, i \pa X^n(w) \ \ \sim \ \ \frac{ 2\ap \,\eta^{mn} }{(z-w)^2}  \ + \ \ldots \co \psi^m(z) \, \psi^n(w) \ \ \sim \ \ \frac{ \eta^{mn} }{z-w} \ + \ \ldots \ .
\eeq} $i\pa X^m,\psi^n$ by $T$ and $G$, respectively, whereas $T_{\be,\ga}$ captures the $\be,\ga$ superghost system of $c=11$. The latter is partially bosonized in terms of exponentials $\ee^{q\phi}$ (with $\phi$ denoting a free chiral boson) and completed by a pair of $h=1,0$ fermions $\eta,\xi$. The Grassmann odd ghost system $(b,c)$ is well-known from the bosonic string.

\medskip
States of uniform superghost charge are BRST closed only if they are annihilated by $Q_0, Q_1$ and $Q_2$ separately. Closure under $Q_0$ forces vertex operators to be a Virasoro primary of unit weight, while $Q_2$ does not contribute in the ghost pictures considered in this paper. Hence, given a vertex operator ansatz of suitable conformal weight, only the $Q_1$ constraint involving the supercurrent
\beq
G(z) \eq \frac{1}{2\sqrt{2\ap}} \; i \pa X_m(z) \, \psi^m(z)
\label{scurrent}
\eeq
has to be evaluated separately.

\subsection{The NS sector}

The lowest mass $m^2 = -k^2 = 1/\ap$ for Regge excitations assigns conformal weight $h=-1$ to the plane wave $\ee^{ik\cdot X}$ which introduces spacetime momentum into vertex operators. In the NS sector of canonical superghost charge $-1$, it can combine with the $h=\frac{1}{2}$ field $\ee^{-\phi}$ and a $h=\frac{3}{2}$ combination of $i\pa X^m,\psi^n$ oscillators to form a Virasoro primary of unit conformal weight in total. (Hence, neglecting the plane wave $\ee^{ik\cdot X}$ contribution, the massive states at first mass level always correspond to vertex operators with conformal dimension $h=2$.)

\medskip
The most general $h=1$ ansatz for the first massive NS sector states involves three\footnote{The addition of $\xi_m \psi^m \pa \phi \ee^{-\phi}$ is neglected because it can be absorbed into a total derivative.} $h=\frac{3}{2}$ operators $i\pa X^m \psi^n , \, \psi^m  \psi^n \psi^p$ and $\pa \psi^m$ along with polarization wavefunctions $B_{mn}, \, E_{mnp}, \, H_m$:
\beq
V^{(-1)}(B,E,H,k,z) \eq \Bigl( \, B_{mn} \, i\pa X^m \, \psi^n \ + \ E_{mnp } \, \psi^m \, \psi^n \, \psi^p \ + \ H_m \, \pa \psi^m \, \Bigr) \, \ee^{-\phi} \, \ee^{ik\cdot X} \ .
\eeq
The BRST constraints arising from $Q_1$ admit two physical solutions,\footnote{Throughout this paper, we are setting the vertex operator normalization factor $g_{\te{A}}= \sqrt{2\ap} g_{\te{YM}}$ from \cite{Lust:2008qc, Lust:2009pz} to unity.} namely a (traceless and symmetric) spin two tensor $B_{mn}$ and a three-form $E_{mnp}$:
\begin{align}
V^{(-1)}(B,k,z) &\eq  \frac{1}{\sqrt{2\ap}} \; B_{mn} \, i\pa X^m \, \psi^n \, \ee^{-\phi} \, \ee^{ik\cdot X}  \co k^m \, B_{mn}  \ = \  B_{m}{}^{m} \ = \ B_{[mn]} \ = \ 0 \ , \label{4,3}
 \\
V^{(-1)}(E,k,z) &\eq \frac{1}{6} \; E_{mnp } \, \psi^m \, \psi^n \, \psi^p \, \ee^{-\phi} \, \ee^{ik\cdot X}  \co k^m \, E_{mnp} \eq 0 \ . \label{4,4}
\end{align}
Both polarizations are transverse and therefore naturally fall into representations of the stabilizer group $SO(9)$ of massive momenta. The number of degrees of freedom is $\frac{9\cdot 10}{2} -1 = 44$ for $B_{mn}$ and $\frac{9\cdot 8 \cdot 7}{1\cdot 2 \cdot 3} = 84$ for $E_{mnp}$, i.e. we have $44 + 84 =128$ bosonic states in total.

\medskip
Some of the solutions to the BRST constraint turn out to be $Q_{\te{BRST}}$ exact:
\begin{align}
\bigl[ \, Q_{\te{BRST}} \; , \; \ee^{-2\phi} \,  \Si_{[mn]} \, \psi^m \, \psi^n \, \pa \xi \, \ee^{ik \cdot X} \, \bigr] \ \ &\sim \ \ \Bigl( \, 2 \, \Si_{[mn]} \, i \pa X^m \, \psi^n \ + \ \Si_{[ mn} \, k_{p]} \, \psi^m\, \psi^n \, \psi^p \, \Bigr) \, \ee^{-\phi} \, \ee^{ik\cdot X} \ , \notag \\
\bigl[ \, Q_{\te{BRST}} \; , \; \ee^{-2\phi} \,  \pi_m \, i \pa X^m \, \pa \xi \, \ee^{ik \cdot X} \, \bigr] \ \ &\sim \ \ \Bigl( \, \pi_m \, \pa \psi^m \ + \ \pi_m \, k_n \, i \pa X^m \, \psi^n \, \Bigr) \, \ee^{-\phi} \, \ee^{ik\cdot X} \label{4,15aa} \ , \\
\bigl[ \, Q_{\te{BRST}} \; , \; \pa \ee^{-2\phi} \, \pa \xi  \, \ee^{ik \cdot X} \, \bigr] \ \ &\sim \ \ \Bigl( \, \Big[\frac{\eta_{mn}}{2\ap} \, + \, 2\, k_m \, k_n\Big] \, i\pa X^m \, \psi^n \ + \ 3 \, k_m \, \pa \psi^m \, \Bigr) \, \ee^{-\phi} \, \ee^{ik\cdot X} \ . \notag
\end{align}
These spurious states parametrized by a two-form $\Si_{[mn]}$, a vector $\pi_m$ and a scalar of $SO(9)$ (i.e. subject to $k^m \Si_{mn} = k^m \pi_m =0$) decouple from physical states.

\subsection{Excited spin fields}
\label{sec:excitspin}

In the R sector, the canonical superghost vacuum is created by the $h=\frac{3}{8}$ field $\ee^{-\phi/2}$. Masses $m^2 = 1/\ap$ allow for an $h=\frac{13}{8}$ operator to complete fermionic vertex operators for the first mass level. The matter sector of the R ground states corresponds to $h=\frac{5}{8}$ spin fields $S_\al$ transforming as left-handed spinors of the Lorentz group \cite{Cohn:1986bn, Kostelecky:1986xg}. The right-handed chirality is forbidden by GSO projection. The role of $S_\al$ to open or close branch cuts for the $\psi^m$ is reflected in the OPE
\beq
\psi^m(z) \, S_\al (w) \  \ \sim \ \ \frac{ \ga^m_{\al \dbe} }{\sqrt{2} \, (z-w)^{1/2}} \; S^{\dbe}(w) \ + \ \ldots \ .
\label{psi-S-OPE}
\eeq
The nontrivial three-point interactions between $\psi^m$ and $S_\al$ render their covariant correlation functions inaccessible to the Wick theorem, one has to use techniques of \cite{Haertl:2009yf, Hartl:2010ks, Thesis} instead to compute higher order correlators. Only by breaking $SO(1,9)$ to its $SU(5)$ subgroup, one can relate the $\psi^m$ and $S_\al$ to a free field system of chiral bosons $H_{1,2,\ldots,5}$:
\beq
i\pa H_k(z) \, i \pa H_l(w) \ \ \sim \ \ \frac{ \de_{kl}}{(z-w)^2} \ + \ i\pa H_k(w) \, i \pa H_l(w) \ + \ \ldots
\label{1,4}
\eeq
This technique is known as bosonization\footnote{
We should admit that our discussion neglects Jordan-Wigner cocycle factors \cite{Kostelecky:1986xg}. These are additional algebraic objects accompanying the exponentials to ensure that $\ee^{\pm iH_k}$ and $\ee^{\pm iH_l}$ associated with different bosons $k \neq l$ anticommute. We drop cocycle factors to simplify the notation, it suffices to remember that they are implicitly present and that the bosonized representation of $\psi^\mu$ still obeys Fermi statistics. The instance where they contribute a phase is commented on above (\ref{N2spin2mulVac1}).} \cite{Kostelecky:1986xg}:
\beq
\psi^m \ \ \leftrightarrow \ \ \ee^{\pm i H_m} \co S_\al  \ \ \leftrightarrow \ \ \ee^{\pm i H_1/2} \, \ee^{\pm i H_2/2} \, \ee^{\pm i H_3/2} \, \ee^{\pm i H_4/2} \, \ee^{\pm i H_5/2} \ .
\eeq
It is clear from this bosonized representation that the subleading term $\sim (z-w)^{1/2}$ of the OPE (\ref{psi-S-OPE}) involves $\ee^{\pm 3iH_k/2}$ primary operators, in addition to the derivatives $\pa \ee^{\pm iH_k/2}$. The covariant description of these new excited primary fields requires an irreducible vector spinor
\beq
S_m^{\dbe} \ \ \leftrightarrow \ \ \ee^{\pm i 3 H_1/2} \, \ee^{\pm i H_2/2} \, \ee^{\pm i H_3/2} \, \ee^{\pm i H_4/2} \, \ee^{\pm i H_5/2} \co \ga^m_{\al \dbe} \, S_m^{\dbe} \eq 0
\label{exspin10}
\eeq
of weight $h=\frac{13}{8}$, where the gamma tracelessness condition subtracts the descendant components $\pa S_{\al} \leftrightarrow \pa(\ee^{\pm i H_1/2} \ee^{\pm i H_2/2} \ee^{\pm i H_3/2}  \ee^{\pm i H_4/2} \ee^{\pm i H_5/2})$. The introduction of $S_m^{\dbe}$ and $\pa S_\al$ is the covariant way to disentangle the primary field- and descendant components within the operator $\psi_m \psi^n S_\al \ga_n^{\al \dbe}$ used in \cite{Koh:1987hm}. The completion of the OPE (\ref{psi-S-OPE}) to the subleading level reads
\beq
\psi^m(z) \, S_\al(w) \ \ \sim \ \ \frac{ \ga^m_{\al \dbe} \, S^{\dbe}(w) }{\sqrt{2} \, (z-w)^{1/2}}  \ + \ (z-w)^{1/2} \, \left[ \, S^m_{\al}(w) \ + \ \frac{ 2  }{\sqrt{2} \, 5 } \; \ga^m_{\al \dbe}  \, \pa S^{\dbe}(w) \, \right]  \ + \ \ldots
\eeq
in $D=10$. A more exhaustive list of OPEs involving $\psi^m, S_\al$ and $S_m^{\dbe}$ (and their counterparts of opposite $SO(1,9)$ chirality) can be found in appendix \ref{appD=10cft}. A covariant treatment of generic higher spin primary fields will be given in \cite{progress}.

\subsection{The operator content of the R sector}

After the GSO projection, the most general vertex operator for spacetime fermions at the first mass level involves the $h=\frac{13}{8}$ operators $i \pa X^m S_\al, \, S_m^{\dbe}$ and $\pa S_\al$ and therefore two vector spinor wavefunctions $v_m^\al, \bar \rho^m_{\dbe}$ as well as spinor wavefunction $u^\al$:
\beq
V^{(-1/2)}(v,\bar \rho,u,k,z) \eq \Big( \, v_m^\al \, i \pa X^m \, S_\al \ + \ \bar \rho^m_{\dbe} \, S_m^{\dbe} \ + \ u^\al \, \pa S_\al \, \Big) \, \ee^{-\phi/2} \, \ee^{ik\cdot X} \ .
\label{4,8}
\eeq
Since $\bar \rho$ is contracted with the excited spin field $S_m^{\dbe}$, we can regard it as $\ga$ traceless, i.e. $\bar \rho^m_{\dbe} \gab_m^{\dbe \al}=0$. The independent $Q_1$ BRST constraints for (\ref{4,8}) can be summarized as
\begin{align}
0 &\eq 2\ap \, v_m^\al \not \! k_{\al \dbe} \ + \ \sqrt{2} \, \bar \rho_{m,\dbe} \ + \ \tfrac{1}{2} \, u^\al \, \ga_{m\al \dbe} \ ,\\
0 &\eq 2 \sqrt{2} \, k_\mu \, \bar \rho^\mu_{\dbe} \ - \ \tfrac{3}{2} \, u^\al \not \! k_{\al \dbe} \ .
\label{brs10}
\end{align}
Disentangling the $SO(1,9)$ irreducibles of the former allows to express $u^\al$ and $\bar \rho^m_{\dbe}$ in terms of $v^\al_m$,
\begin{align}
\bar \rho^m_{\dbe} \ \ &= \ \ - \, \sqrt{2} \, \ap \, \left( \, v^{m\al} \not \! k_{\al \dbe} \ + \ \frac{1}{10} \; v_p^\al \, (\not \! k \, \gab^p \, \ga^m)_{\al \dbe} \, \right) \label{4,10} \ , \\
u^\al \ \ &= \ \ \frac{2\ap}{5} \; v_m^{\be} \, ( \not \! k \, \ga^m)_\be{}^\al \ ,
\label{4,11}
\end{align}
whereas (\ref{brs10}) yields an extra constraint on the only independent polarization $v_m^\al$:
\beq
v_m^\al \, \ga^m_{\al \dbe} \eq 2\ap \, k^m \, v_m^{\al} \not \! k_{\al \dbe} \ .
\label{4,12}
\eeq
As recognized in \cite{Koh:1987hm}, there is a physical solution $v_m^{\al} \equiv \chi_m^{\al}$ of spin 3/2
\begin{align}
V^{(-1/2)}(\chi,k) &\eq \frac{1}{\sqrt{2} \ap^{1/4}} \; \Big( \, \chi_m^\al \, i \pa X^m \, S_\al \ - \ \sqrt{2} \, \ap \, \chi^{m\al} \not \! k_{\al \dbe} \, S_m^{\dbe} \, \Big) \, \ee^{-\phi/2} \, \ee^{ik\cdot X}
\label{4,13} \\
0 &\eq k^m \, \chi_m^\al \eq \chi_m^\al  \, \ga^m_{\al \dbe} \notag
\end{align}
and one spurious state associated with the gamma trace choice $v_m^{\al}=k_m \Theta^\al + \frac{1}{4} \Theta^\beta (\not \! k \ga_m)_\be{}^\al$
\begin{align}
\bigl[ \, Q_{\te{BRST}} \; , \; \ee^{-3\phi/2} \, \pa \xi \, \Theta^\al \not \! k_{\al \dbe} \, S^{\dbe} \, \ee^{ik \cdot X} \, \bigr] \ \ \sim \ \ \Big( \, \big[ \, k_m \, \Theta^\al \ + \ \tfrac{1}{4} \; &\Theta^\be \, (\not \! k \, \ga_m)_\be {}^\al \, \big] \, i \pa X^m \, S_\al \  \notag \\ -  \ \frac{1}{\sqrt{2}} \, \big[ \, \ap \, k^m \, \Theta^\al \not \! k_{\al \dbe} \ + \ \tfrac{1}{10} \, \Theta^\al \, \ga^m_{\al \dbe} \, \big] \, S_m^{\dbe} \ & + \ \frac{6}{5} \ \Theta^\al \, \pa S_\al \, \Big) \, \ee^{-\phi/2} \, \ee^{ik\cdot X}
\label{4,15}
\end{align}
which allows to gauge away the $u^\al$ wavefunction.

\subsection{Ten-dimensional SUSY transformations}

The SUSY charge in open superstring theory is given by the massless gaugino vertex at zero momentum \cite{FMS}:
\beq
{\cal Q}^{(-1/2)}_\al \eq \frac{1}{\ap^{1/4}} \oint \frac{ \dd z }{2\pi i} \; S_\al \, \ee^{-\phi/2} \ .
\eeq
It transforms R sector states in their canonical $-1/2$ superghost picture into canonical NS vertex operators $\sim \ee^{-\phi}$. The contour integral is evaluated by performing OPEs between the $S_\al$ and $\ee^{-\phi/2}$ fields from the supercharge at point $z$ and the vertex operator $V^{(-1/2)}(w)$ of the fermion in question. Appendix \ref{appD=10cft} gathers the required OPEs for the $D=10$ case.

\medskip
The inverse transformation from the NS sector to the R sector requires the $+1/2$ picture representative of the SUSY generator
\beq
{\cal Q}^{(+1/2)}_\al \eq \frac{1}{2\ap^{3/4}} \oint \frac{ \dd z }{2\pi i} \; i \pa X_m \, \ga^m_{\al \dbe} \, S^{\dbe} \, \ee^{+\phi/2}  \ .
\eeq
The latter allows to write down the ghost neutral ${\cal N}=1$ SUSY algebra in ten dimensions,
\beq
\big\{ \, {\cal Q}_\al^{(+1/2)} \; , \;  {\cal Q}^{(-1/2)}_{\beta} \, \big\} \eq (\ga^m \, C)_{\alpha \beta} \, P_m \co P_m \eq \frac{1}{2\ap} \oint \frac{ \dd z }{2\pi i} \; i \pa X_m \ .
\eeq
Let us list the SUSY variations of the physical $D=10$ vertex operators. The NS sector states (\ref{4,3}) and (\ref{4,4}) have already been discussed in \cite{Koh:1987hm}
\begin{align}
\bigl[ \, \eta^\al \, {\cal Q}_\al^{(+1/2)} \; , \; V^{(-1)}(B,k) \, \bigr] \ \ &= \ \ V^{(-1/2)} \left( \, \chi_m^\al \ = \ \tfrac{1}{\sqrt{2}} \, B_{mn} \, (\eta \not \! k \, \ga^n)^\al \, \right) \label{D=10,1} \ ,\\
\bigl[ \, \eta^\al \, {\cal Q}_\al^{(+1/2)} \; , \; V^{(-1)}(E,k) \, \bigr] \ \ &= \ \ V^{(-1/2)} \Big( \, \chi_m^\al \ = \ \tfrac{1}{12 \, \sqrt{\ap} } \, \big[ E_{mnp} \, (\eta \, \ga^{np})^\al \, - \, \tfrac{1}{3} \, E_{npq} \, (\eta \, \ga_m \, \ga^{npq})^\al \notag\\
& \ \ \ \ \ \ \ \ \ \  \ \ \ \ \ \ \ \ \ \ \ \ \ \ \ \  - \, \tfrac{\ap}{3} \, k_m \, E_{npq} \, (\eta \not \! k \, \ga^{npq})^\al \, \big] \, \Big) \ . \label{D=10,2}
\end{align}
In addition, we use the covariant OPEs from appendix \ref{appD=10cft} to compute the SUSY variation of the massive gravitino (\ref{4,13}):
\begin{align}
\bigl[ \, \eta^\al \, {\cal Q}_\al^{(-1/2)} \; , \;& V^{(-1/2)}(\chi,k) \, \bigr] \ \ = \ \ V^{(-1)} \left( \, B_{mn} \, = \, \tfrac{\ap}{\sqrt{2}} \, (\eta \not \! k \, \chi_{(m}) \, k_{n)} \, + \, \tfrac{1}{\sqrt{2}} \, (\eta \, \ga_{(m} \, \chi_{n)})  \, \right) \notag \\
& \ \ \  + \ V^{(-1)} \left( \, E_{mnp} \, = \, 3 \ap^{1/2} \, (\eta \, \ga_{[m} \, \chi_{n}) \, k_{p]} \, - \, \tfrac{3}{2 } \, \ap^{1/2} \, (\eta \, \ga_{[np} \not \! k \, \chi_{m]}) \, \right) \ . \label{D=10,3}
\end{align}

\section{CFTs of supersymmetric string vacua in four dimensions}
\label{sec:SCFTuniv}

In this section we will first review some basic facts about extended supersymmetry algebras in four spacetime dimensions and about the
general relation between extended spacetime supersymmetries and world-sheet supersymmetries.
In part, we are following the work in references \cite{BD1,BD2,BD3}. Our conventions for indices w.r.t. Lorentz symmetry $SO(1,3)$ and R-symmetries $SO(6)$ or $SU(2)$ are gathered in appendix \ref{appCONV}.

\subsection{The $D=4$ spacetime supersymmetry algebra}
\label{sec:D=4scft}

The ${\cal N}$ supercharges ${\cal Q}^I_a$ as well as the complex conjugate operators $\bar{\cal Q}^{\dot a}_{\bar I}$
satisfy the ${\cal N}$-extended supersymmetry algebra $(I,\bar I=1\dots, {\cal N})$
\begin{eqnarray}
\lbrace{\cal Q}_a^I \, , \, \bar{\cal Q}^{\dot b}_{\bar J}\rbrace&=& \, C^I_{\bar J} \, (\si^\mu \, \vep)_{a}{}^{\dot b} \, P_\mu\, ,\nonumber\\
\lbrace{\cal Q}^I_a \, , \, {\cal Q}^{J}_b \rbrace&=&\vep_{ab} \, {\cal Z}^{IJ}\, .
\end{eqnarray}
$P^\mu$ is the momentum operator and the ${\cal Z}^{IJ}$ are central charges, which are antisymmetric in $I,J$ and can therefore appear in the ${\cal N}\geq 2$ supersymmetry algebra only.

\medskip
Next let us discuss the representations of the extended supersymmetry algebras, namely how the supercharges in general act on massless and on massive states. Let us first recall the case of massless states. Here we can choose a frame where the momenta are $k^\mu=(E,0,0,E)$, the supercharges are
\begin{align}
{\cal Q}^I_1 \ \ \equiv \ \ {\cal Q}^I\co \bar{\cal Q}^{\dot 1}_{\bar I}\ \ \equiv \ \ \bar{\cal Q}_I \co \te{whereas} \ \ \ 0 \eq  {\cal Q}^2_1 \eq \bar{\cal Q}^{\dot 2}_{\bar I} \ .
\end{align}
In terms of ${\cal Q}^I$ and $\bar {\cal Q}_I$ the supersymmetry algebra takes the form
\begin{eqnarray}
\big\{  {\cal Q}^I \, , \,  \bar{ {\cal Q} }_{J} \big\}&=&\delta^I_J\, ,\nonumber\\
\big\{ {\cal Q}^I \, , \, {\cal Q}^{J}\big\}&=&\big\{ \bar{ {\cal Q} }_I \, , \,  \bar{ {\cal Q} }_{J} \big\} \ =\ 0\, ,
\end{eqnarray}
where we have rescaled the supersymmetry charges by $\sqrt E$.
The $2{\cal N}$ supercharges  ${\cal Q}^{I}$ and $\bar {\cal Q}_{I}$ build an $SO(2{\cal N})$ Clifford algebra
\begin{eqnarray}
\Gamma_{2I-1}&=&{\cal Q}^{I}+\bar {\cal Q}_{ \bar I}\, ,\quad \Gamma_{2I} \ = \ i({\cal Q}^{I}-\bar {\cal Q}_{\bar I})\, ,\nonumber\\
\lbrace \Gamma_i,\Gamma_j\rbrace&=&2\delta_{ij}\, ,\quad i,j=1,\dots ,2{\cal N}
\label{Clifford}
\end{eqnarray}
whose representations have dimension $2^{{\cal N}}$. The generators for $SO(2{\cal N})$ rotations are
\begin{equation}
\Lambda_{ij} \ \ = \ \ {1\over 4 i} \; \lbrack\Gamma_i \, , \, \Gamma_j\rbrack \ .
\end{equation}
This group contains a $SU({\cal N})\times U(1)$ subgroup specified by the following generators
\begin{eqnarray}
\Lambda^I_J&=& {1\over 2} \; \lbrack {\cal Q}^I \, ,\, \bar{\cal Q}_{\bar J}\rbrack \ - \ {1\over 2{\cal N}} \;\delta^I_J \, \lbrack{\cal Q}^K \, , \, \bar {\cal Q}_{\bar K}\rbrack \,\,\,{\rm for}\,\,\,
SU({\cal N})\ , \nonumber\\
\Lambda&=&{1\over 4} \; \lbrack{\cal Q}^I \, , \, \bar {\cal Q}_I\rbrack \,\,\,{\rm for}\,\,\,
U(1)\ .
\label{lambdas}
\end{eqnarray}
For massless states, this  $SU({\cal N})$ commutes with the $SO(2)$ helicity group. Hence this group classifies massless states.
The eigenvalue of the supercharge under the $U(1)$, which is called intrinsic helicity,
is the same as under spacetime helicity. Therefore one can define a new generator $\Lambda'$ through a shift by the $z$ component $j^3$ of the spin, called superhelicity,
\begin{equation}
\Lambda' \eq j^3 \ - \ \Lambda\, ,
\end{equation}
which commutes with ${\cal Q}^I$.

\medskip
Next let us consider massive states rotated into their rest frame $k^\mu= (m,0,0,0)$. Now also the second helicity components of the supercharge spinors become active, i.e. give rise to nonzero supersymmetry transformations on massive
states. We will denote them as follows:
\begin{equation}
{\cal Q}^I_2 \ \ \equiv \ \  \tilde {\cal Q}^I\, ,\quad \bar{\cal Q}^{\dot 2}_{\bar I}\ \ \equiv \ \ \bar{ \tilde{ {\cal Q}}}_{\bar I} \, .
\end{equation}
The supersymmetry algebra between the ${\cal \tilde Q}$ looks like
\begin{equation}
\big\{   \tilde {\cal Q}^I \, , \,  \bar{ \tilde{ {\cal Q}}}_{ \bar J}\big\} \eq m \, C^I_{\bar J}\, ,\quad
\big\{ \tilde{ \cal Q}^I \, , \, \tilde {\cal  Q}^{J} \big\} \eq \big\{ \bar{ \tilde{ {\cal Q}}}_I \, , \, \bar{ \tilde {{\cal Q}}}_{J}\big\} \eq 0 \ .
\end{equation}
Now the $({\cal Q}^I , {\bar Q}_{\bar I})$ and $({\cal \tilde Q}^I, \bar { \tilde{ {\cal Q}}}_{\bar I})$ build an $SO(4{\cal N})$ Clifford algebra on the states without central charges. Consequently, the dimension of massive representations is a multiple of $2^{2{\cal N}}$. The maximal subalgebra
that commutes with the $SO(3)$ little group of the massive states is $USp(2{\cal N})$. Therefore massive states without central charges build
representations of $USp(2{\cal N})$. As for the massless states one can consider an
$SU({\cal N})\times U(1)$ subgroup with generators $\Lambda_{tot}=\Lambda+\tilde \Lambda$ where the $\tilde \Lambda$ are defined from the
$\cal {\tilde Q}$ as in (\ref{lambdas}). In section \ref{sec:hel} we will introduce an organization scheme for massive SUSY representations based on spinor helicity methods which keeps track of the spin quantum numbers along a reference axis of choice.

\medskip
However, in the presence of central charges ${\cal Z}^{IJ}$, the operators ${\cal Q}^I$ and ${\cal\tilde Q}^I$ generate a smaller
$SO(2{\cal N})$ Clifford algebra, whose maximal subalgebra is
$SO(3)\times Sp({\cal N})$. Therefore states with central charges only build representations of $Sp({\cal N})$.

\subsection{CFT realization of extended $D=4$ SUSY}

As it is well known, there is a  beautiful  relation between the ${\cal N}$-extended spacetime supersymmetry algebras and the
$n$-extended internal superconformal algebras with corresponding Kac-Moody symmetry $g$.
We will assume in the following that we are dealing with holomorphic spacetime supercharges that all originate from the right-moving
sector of the compactified string theory, as it is always the case for heterotic string compactifications.
As we will discuss, for purely holomorphic supercharges, the massive BPS states with non-vanishing central charges are of
perturbative nature. However
in type II compactifications, the supercharges can originate from the left-moving as well as the right-moving sector of the string theory.
In this case, some of the massive BPS states with central charges are non-perturbative, as they are given in terms of wrapped type II D-branes.
These non-perturbative states will not be discussed in this paper.

\medskip
In SCFT, the holomorphic supercharges ${\cal Q}^I$ and $\bar{\cal Q}_{\bar I}$ can be always realized by the world-sheet fields of the uncompactified four-dimensional
Minkowski spacetime together with those of the internal Kac-Moody symmetries. This fact allows for a completely
model-independent realization of the spacetime supersymmetry algebra without any reference to "geometrical" details of the internal SCFT.
To be more specific,
compactifications to four-dimensional Minkowski spacetime which allow for a CFT description, still have $SO(1,3)$ vectors $i\pa X^\mu$ and $\psi^\mu$ in their world-sheet theory, the first four components of the ten-dimensional ancestors $i \pa X^m$ and $\psi^m$. Similarly, the ten-dimensional $SO(1,9)$ spin field $S_\al$ factorizes into separate $h=\frac{1}{4}$ and $h=\frac{3}{8}$ primaries $S_a$ and $\Si$, the former being a Weyl spinor of $SO(1,3)$ and the latter falling into representations of the R-symmetry. In fact, both $SO(1,3)$ chiralities can occur, i.e.
\beq
S_\al \ \ \equiv \ \ S_a \, \Si^I \ \oplus \ S^{\dot b} \, \bar \Si_I \ .
\eeq
The number of $(\Si^I,\bar \Si_I)$ species coincides with the number of spacetime supersymmetries, we will discuss the ${\cal N}=4,1,2$ cases below. In each case, the (left- and right-handed) supercharges in their canonical ghost picture are given by
\beq
{\cal Q}^{(-1/2)I}_a \eq \frac{1}{\ap^{1/4}} \oint \frac{ \dd z }{2\pi i} \; S_a \, \Si^I \, \ee^{-\phi/2} \co \bar {\cal Q}^{(-1/2), \dot b}_{\bar J} \eq \frac{1}{\ap^{1/4}} \oint \frac{ \dd z }{2\pi i} \; S^{\dot b} \, \bar \Si_{\bar J} \, \ee^{-\phi/2} \ .
\label{3,0}
\eeq
Independent on the fate of the internal spin fields $\Si^I,\bar \Si_I$, the interactions of the $h=\frac{1}{4}$ spacetime spin fields $S_a,S^{\dot b}$ with the NS fermions is governed by
\beq \psi^\mu(z) \, S_a(w) \ \ \sim \ \ \frac{ \si^\mu_{a \dot b} \, S^{\dot b}(w) }{\sqrt{2} \, (z-w)^{1/2}}  \ + \ (z-w)^{1/2} \, \left[ \, S^\mu_{a}(w) \ + \ \frac{ 1  }{\sqrt{2}} \; \si^\mu_{a \dot b} \, \pa S^{\dot b}(w) \, \right]  \ + \ \ldots \ .
\label{moreOPEs}
 \eeq
In lines with the discussion of subsection \ref{sec:excitspin}, one can bosonize the left- and right-handed spin fields as $\ee^{\pm i(H_1+H_2)/2}$ and $\ee^{\pm i (H_1- H_2)/2}$, respectively. In order to reconcile bosonization techniques with $SO(1,3)$ symmetry, we align $\ee^{\pm 3iH_j/2} $ components showing up in the subleading term of the OPE (\ref{moreOPEs}) into covariant excited spin fields $S_\mu^{\dot b}, S^\mu_a$ of weight $h=\frac{5}{4}$:
\beq
S^\mu_a \ \ \leftrightarrow \ \ \ee^{\pm 3iH_1/2} \, \ee^{\pm i H_2/2} \co
S_\mu^{\dot b} \ \ \leftrightarrow \ \ \ee^{\pm 3iH_1/2} \, \ee^{\mp i H_2/2} \co
 \sib_\mu^{\dot b a} \,  S^\mu_a \eq \si^\mu_{a \dot b} \, S_\mu^{\dot b} \eq 0 \ .
 \label{exspin4}
\eeq
A large list of OPEs between $(\psi^\mu,S_a,S^{\dot b} , S_\mu^{\dot b}, S^\mu_a)$ including subleading singularities can be found in appendix \ref{appD=4cft}.

\subsection{CFT operators in ${\cal N}=4$ compactifications}
\label{sec:N=4cft}

The internal SCFT in maximally supersymmetric ${\cal N}=4$ compactifications to $D=4$ dimensions can be understood in terms of free fields $i\pa Z^m,\Psi^m$ with $m=4,5,\ldots,9$ which represent the internal components of the ten-dimensional $i \pa X^{m=0,1,\ldots,9},\psi^{m=0,1,\ldots,9}$ and transform as vectors of the internal rotation group $SO(6)$. The corresponding $h=\frac{3}{8}$ spin fields $\Si^I$ and $\bar \Si_{\bar J}$, responsible for branch cuts of $\Psi^m$, transform as spinors of the $SO(6) \equiv SU(4)$ with left-handed (right-handed) index $I$ ($\bar J$). They enter the dimensional reduction $SO(1,9) \rightarrow SO(1,3) \times SO(6)$ of the $D=10$ SUSY charges
\beq
{\cal Q}^{(-1/2)I}_a \eq \frac{1}{\ap^{1/4}} \oint \frac{ \dd z }{2\pi i} \; S_a \, \Si^I \, \ee^{-\phi/2} \co \bar {\cal Q}^{(-1/2), \dot b}_{\bar J} \eq \frac{1}{\ap^{1/4}} \oint \frac{ \dd z }{2\pi i} \; S^{\dot b} \, \bar \Si_{\bar J} \, \ee^{-\phi/2}
\label{3,1}
\eeq
where the internal $SO(6) \equiv SU(4)$ is interpreted as the R-symmetry group. The ten-dimensional bosonization prescription can be straightforwardly applied to $\Psi^m,\Si^I,\bar \Si_{\bar J}$ (e.g. $\Si^I \leftrightarrow \ee^{\pm i(H_3+H_4+H_5)/2}$), and excited spin fields $\Si^I_m$ and $\bar \Si^m_{\bar J}$ of weight $h=\frac{11}{8}$ are constructed in close analogy to their ten- and four-dimensional counterparts (\ref{exspin10}) and (\ref{exspin4}):
\beq
\Si_m^I \ \ \leftrightarrow \ \ \ee^{\pm 3iH_3/2} \, \ee^{\pm i H_4/2} \, \ee^{\pm i H_5/2} \co
 \ga^m_{\bar J I} \, \Si_m^I \eq \gab_m^{I \bar J} \, \bar \Si_{\bar J}^m \eq 0 \ .
 \label{exspin6}
\eeq
The internal supercurrent is built from the $m=4,5,\ldots ,9$ components of its ten-dimensional ancestor (\ref{scurrent})
\beq
G_{\te{int}} \eq \frac{1}{2 \sqrt{2\ap}} \; i\pa Z_m \, \Psi^m
\label{3,0}
\eeq
and gives rise to internal central charge\footnote{The underlying OPEs are
\beq
i\pa Z_m(z) \, i \pa Z_n(w) \ \ \sim \ \ \frac{ 2\ap \, \de^{(6)}_{mn}}{(z-w)^{2}}  \ + \ \ldots \co
\Psi_m(z)\, \Psi_n(w) \ \ \sim \ \  \frac{ \de^{(6)}_{mn} }{z-w} \ + \ \ldots \ .
\eeq} $c=9$. OPEs among the $\Psi^k, \Si^I,\bar \Si_{\bar J}$ and $\Si^I_m, \bar \Si^m_{\bar J}$ are gathered in appendix \ref{appN=4SUSY}. Identities between six-dimensional gamma and charge conjugation matrices can for instance be found in the appendix of \cite{Hartl:2010ks}. The following figure \ref{N=4} aims to give an overview of the conformal fields in the spacetime and ${\cal N}=4$ internal CFTs\footnote{The fermionic bilinear states $\psi^\nu \psi^\la$ and $\Psi^n \Psi^p$ at weight $h=1$ by themselves should be eliminated by the GSO projection, but trilinear combinations $\Psi^m \psi^\nu \psi^\la$ and $\psi^\mu \Psi^n \Psi^p$ which mix between spacetime components and internal fields would survive after the GSO projection. That is why we include the bilinears into the bookkeeping.}

\begin{figure}[h]
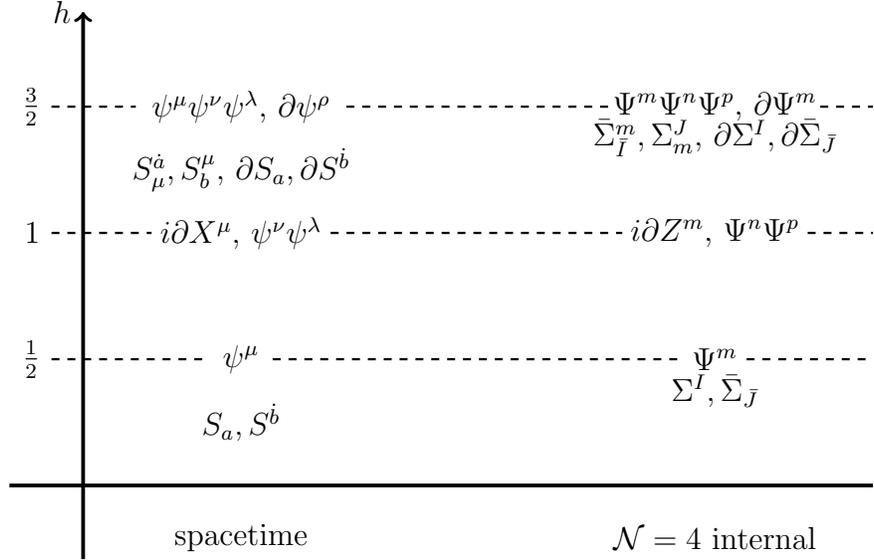

\centerline{
\tikzpicture [scale=1.4,line width=0.30mm]
\draw[line width=0.50mm,->] (0,-0.7) -- (0,4.5) node[left]{$h$};
\draw[line width=0.50mm] (-0.7,0) -- (7.5,0);
%
%
\draw[dashed] (-0.3,1.2) node[left]{$\tfrac{1}{2}$}  -- (1.2,1.2);
\draw[dashed] (1.8,1.2) -- (5.7,1.2);
\draw[dashed] (6.3,1.2) -- (7.5,1.2);
\draw[dashed] (-0.3,2.4) node[left]{$1$}  -- (0.65,2.4);
\draw[dashed] (2.35,2.4)  --(5.12,2.4);
\draw[dashed] (6.88,2.4) --(7.5,2.4);
\draw[dashed] (-0.3,3.6) node[left]{$\tfrac{3}{2}$}  -- (0.5,3.6);
\draw[dashed] (2.5,3.6) -- (5,3.6);
\draw[dashed] (7,3.6) --(7.5,3.6);
\draw (1.5,-0.5) node{spacetime};
\draw (1.5,0.6) node{$S_a ,S^{\dot b}$};
\draw (1.5,1.2) node{$\psi^\mu$};
\draw (1.5,3) node{$S_\mu^{\dot a} ,S^\mu_{ b}, \, \pa S_a , \pa S^{\dot b}$};
\draw (1.5,2.4) node{$i\pa X^\mu, \, \psi^\nu \psi^\la$};
\draw (1.5,3.6) node{$\psi^\mu\psi^\nu \psi^\la, \, \pa \psi^\rho$};
\draw (6,-0.5) node{${\cal N}=4$ internal};
\draw (6,0.9) node{$\Si^I ,\bar \Si_{\bar J}$};
\draw (6,1.2) node{$\Psi^m$};
\draw (6,3.3) node{$\bar \Si^m_{\bar I} , \Si_m^J, \, \pa \Si^I , \pa \bar \Si_{\bar J}$};
\draw (6,2.4) node{$i\pa Z^m, \, \Psi^n \Psi^p$};
\draw (6,3.6) node{$\Psi^m\Psi^n \Psi^p, \, \pa \Psi^m$};
\endtikzpicture
}
\caption{Conformal fields in the spacetime CFT and the internal CFT of ${\cal N}=4$ supersymmetric compactifications}
\label{N=4}
\end{figure}

\medskip
The higher ghost picture version of the SUSY generators (\ref{3,1}) is given by
\begin{align}
{\cal Q}^{(+1/2),I}_a &\eq \frac{1}{2\ap^{3/4}} \oint \frac{ \dd z }{2\pi i} \; \left[ \,  i\pa X_\mu \, \si^\mu_{a \dot b} \, S^{\dot b} \, \Si^I \, \ + \ S_a \, i \pa Z^m \, \ga_m^{I \bar J} \, \bar \Si_{\bar J} \, \right] \, \ee^{+\phi/2} \ ,
\label{3,2a} \\
\bar {\cal Q}^{(+1/2), \dot b}_{\bar J} &\eq \frac{1}{2\ap^{3/4}} \oint \frac{ \dd z }{2\pi i} \; \left[ \,  i\pa X^\mu \, \si_\mu^{\dot b a} \, S_{ a} \, \bar \Si_{\bar J} \, \ + \ S^{\dot b} \, i \pa Z_m \, \bar \ga^m_{\bar J I} \, \Si^{I} \, \right] \, \ee^{+\phi/2} \ ,
\label{3,2b}
\end{align}
their anticommutator with the $(-1/2)$ picture analogues (\ref{3,1}) yields the following ghost-neutral SUSY algebra with nontrivial central charges ${\cal Z}^{IJ}$ and $\bar {\cal Z}_{\bar I \bar J}$:
\begin{align}
\big\{ \, {\cal Q}_a^{(+1/2),I} \; , \; \bar {\cal Q}_{\bar J}^{(-1/2),\dot b} \, \big\} &\eq  C^I{}_{\bar J} \, (\si^\mu \, \vep)_a{}^{\dot b} \, P_\mu \co P_\mu \eq \frac{1}{2\ap} \oint \frac{ \dd z }{2\pi i} \; i \pa X_\mu \ ,
\label{moreQ+2} \\
\big\{ \, {\cal Q}_a^{(+1/2),I} \; , \; {\cal Q}_b^{(-1/2),J} \, \big\} &\eq \vep_{ab} \, {\cal Z}^{IJ} \co {\cal Z}^{IJ} \eq \frac{1}{2\ap} \oint \frac{ \dd z }{2\pi i} \; i \pa Z^m \, (\ga_m \, C)^{IJ} \ ,
\label{moreQ+3} \\
\big\{ \, \bar {\cal Q}^{(+1/2),\dot a}_{\bar I} \; , \; \bar {\cal Q}^{(-1/2), \dot b }_{\bar J} \, \big\} &\eq \vep^{\dot a \dot b} \, \bar {\cal Z}_{\bar I \bar J} \co \bar {\cal Z}_{\bar I \bar J} \eq \frac{1}{2\ap} \oint \frac{ \dd z }{2\pi i} \; i \pa Z_m \, (\gab^m \, C)_{\bar I \bar J} \ .
\label{moreQ+4}
\end{align}
The central charges arise due to poles in the operator product expansion of ${\cal Q}_a^{(+1/2),I}$ and ${\cal Q}_b^{(-1/2),J}$ caused by internal free fermions and bosons $\Psi^m$ and $\partial Z_m$. The latter
appear in the internal supercurrent $G_{\te{int}} \sim i \pa Z_m \Psi^m$ and generate an internal Kac-Moody algebra
\begin{equation}
g \eq SO(6)\times \lbrack U(1)\rbrack^6
\end{equation}
with dimension one currents
\begin{equation}
j_{SO(6)}^{mn}(z) \eq \Psi^m \, \Psi^n(z)\co j^m_{U(1)^6}(z) \eq i\partial Z^m(z)\, .
\end{equation}
The fields $Z_m(z)$ can be viewed as the coordinates of a (holomorphic) torus compactification on a six-dimensional torus $T^6$.
Their world-sheet superpartners $\Psi^m$ generate a $U(1)^6$ spacetime gauge symmetry, and the six spacetime gauge bosons are
the six graviphotons, which arise in any compactification on
a (holomorphic) six-torus. States that carry non-vanishing internal momenta $p^m$ on the (holomorphic) six-torus always
have the following field as part of their vertex operator:
\begin{equation}
|p^m\rangle \ \ \sim  \ \ \ee^{ip^mZ_m(z)}\, .
\end{equation}
Switching to the more convenient bispinor basis,
the six central charge operators (in the zero ghost picture) of the ${\cal N}=4$ supersymmetry algebra are nothing else than the free bosons $Z^m$:
\begin{equation}
{\cal Z}^{IJ}(z) \eq \frac{1}{2\ap} \; (\ga_m\,  C)^{IJ} \, i\partial Z^m(z) \ .
\end{equation}
It follows that the internal momentum states  $|p^m\rangle$ are precisely those states that carry non-vanishing ${\cal N}=4$ central charges.
They break the internal world-sheet $SO(6)$ symmetry to $SO(5)$. At the same time, states with non-vanishing momenta $p^m$ build
representations of
 the spacetime automorphism group for massive states with central charges, which is $Sp(4)\cong SO(5)$.
 On the other hand, states with vanishing internal momenta, $|p^m=0\rangle$, build internal $SO(6)$ representations,
 respectively at the same time representations of the group $USp(8)$, which classifies massive states without central charges. The subsequent discussions only take into account the states at zero internal momentum ($p^m=0$).

\subsection{CFT operators in ${\cal N}=1$ compactifications}
\label{sec:N=1cft}

In this subsection, we summarize universal aspects of internal $c=9$ SCFTs describing $D=4$ superstring compactifications which preserve ${\cal N}=1$ SUSY in spacetime \cite{BD1,BD2,BD3}. The existence of one supercharge species
\beq
{\cal Q}^{(-1/2)}_a \eq \frac{1}{\ap^{1/4}} \oint \frac{ \dd z }{2\pi i} \; S_a \, \Si^+ \, \ee^{-\phi/2} \co \bar {\cal Q}^{(-1/2) \dot b} \eq \frac{1}{\ap^{1/4}} \oint \frac{ \dd z }{2\pi i} \; S^{\dot b} \, \Si^- \, \ee^{-\phi/2}
\label{1,1}
\eeq
with $h=\frac{3}{8}$ spin fields $\Si^{\pm}$ implies that the world-sheet supersymmetry is enhanced to ${\cal N}=2$. This can be traced back to the existence of a $U(1)$ Kac-Moody current ${\cal J}$ of $h=1$ which emerges from the mutual OPEs of spin fields with opposite charge:
\beq
\Si^{\pm} (z) \, \Si^{\mp}(w) \ \ \sim \ \ \frac{1}{(z-w)^{3/4}} \ \pm \ \frac{\sqrt{3}}{2} \; (z-w)^{1/4} \, {\cal J}(w) \ + \ \ldots
\label{1,3a} \eeq
The internal supercurrents $G_{\te{int}}^{\pm}$ can be split into two components of opposite $U(1)$ charge,
\beq
G_{\te{int}} \eq \frac{1}{\sqrt{2}} \; \bigl( \, G_{\te{int}}^+ \ + \  G_{\te{int}}^- \, \bigr) \ ,
\label{1,0}
\eeq
subject to the superconformal ${\cal N}=2$ algebra\footnote{In contrast to \cite{BD1,BD2,BD3}, we normalize ${\cal J}$ such that it has canonical two-point functions $\langle {\cal J}(z) {\cal J}(w) \rangle = 1 \cdot (z-w)^{-2}$. This simplifies (subleading) OPE coefficients and normalization factors in vertex operators.}
\begin{align}
{\cal J}(z) \, {\cal J}(w) \ \ &\sim \ \ \frac{1}{(z-w)^2} \ + \ {\cal J}(w) \, {\cal J}(w) \ + \ \ldots
\label{1,2a} \\
{\cal J}(z) \, G^{\pm}_{\te{int}}(w) \ \ &\sim \ \ \pm \, \frac{ G^{\pm}_{\te{int}}(w)}{\sqrt{3} \, (z-w)} \ + \ {\cal J}(w) \, G^{\pm}_{\te{int}}(w) \ + \ \ldots \label{1,2b} \\
G^{\pm}_{\te{int}}(z) \, G^{\pm}_{\te{int}}(w) \ \ &\sim \ \ G^{\pm}_{\te{int}}(w) \, G^{\pm}_{\te{int}}(w) \ + \ \ldots \label{1,2c} \\
G^{\pm}_{\te{int}}(z) \, G^{\mp}_{\te{int}}(w) \ \ &\sim \ \ \frac{3/2}{(z-w)^3} \ \pm \ \frac{ \sqrt{3} \, {\cal J}(w) }{2 \, (z-w)^2} + \ \frac{ 2 \, T_{\te{int}}(w) \ \pm \ \sqrt{3} \, \pa {\cal J}(w) }{4 \, (z-w)} \ + \ \ldots \label{1,2d}
\end{align}
with internal $c=9$ energy momentum tensor $T_{\te{int}}$. The OPE of alike spin fields gives rise to new $h=\frac{3}{2}$ Virasoro primary operators
\beq
\Si^{\pm} (z) \, \Si^{\pm}(w) \ \ \sim \ \ (z-w)^{3/4} \, {\cal O}^{\pm}(w) \ + \ \ldots
\label{1,3b}
\eeq
with twice the $U(1)$ charge of the spin fields, and iterated OPEs with $\Si^{\pm}$ create an infinite tower of further conformal primaries with higher weights and charges.

\begin{figure}[h]
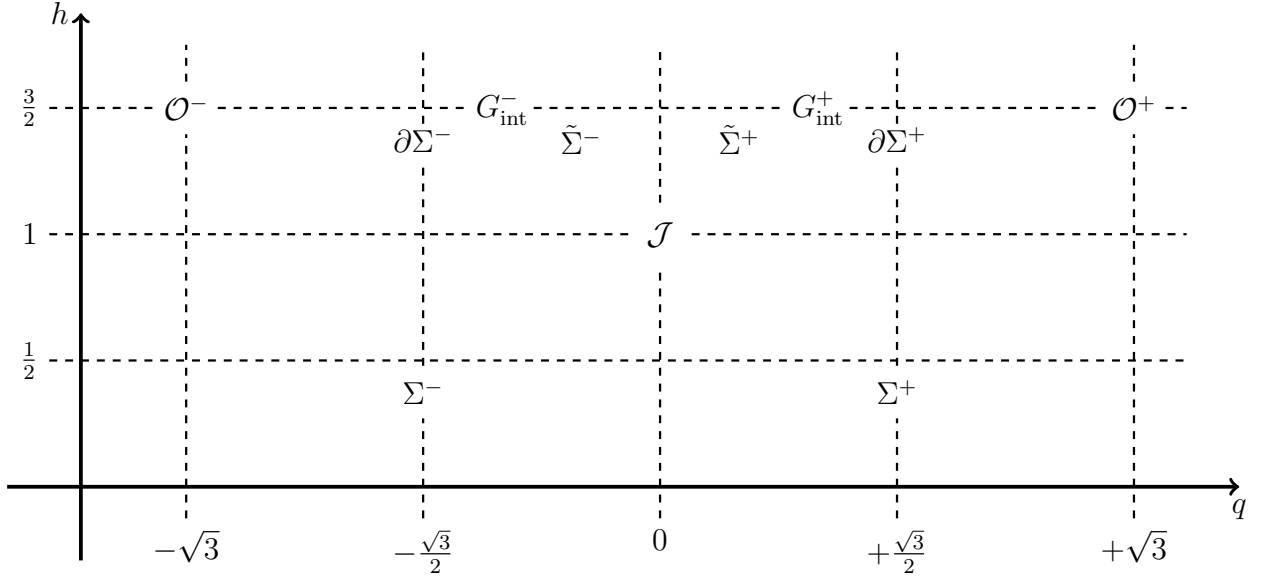

\centerline{
\tikzpicture [scale=1.4,line width=0.30mm]
\draw[line width=0.50mm,->] (0,-0.7) -- (0,4.5) node[left]{$h$};
\draw[line width=0.50mm,->] (-0.7,0) -- (11,0) node[below]{$q$};
\draw[dashed] (5.5,-0.3)node[below]{$0$} -- (5.5,2.1) ;
\draw[dashed] (5.5,2.7) -- (5.5,4.2) ;
\draw[dashed] (3.25,-0.3)node[below]{$-\tfrac{\sqrt{3}}{2}$} -- (3.25,0.65) ;
\draw[dashed] (3.25,1.15) -- (3.25,3.05) ;
\draw[dashed] (3.25,3.45) -- (3.25,4.2) ;
\draw[dashed] (7.75,-0.3)node[below]{$+\tfrac{\sqrt{3}}{2}$} -- (7.75,0.65) ;
\draw[dashed] (7.75,1.15) -- (7.75,3.05) ;
\draw[dashed] (7.75,3.45) -- (7.75,4.2) ;
\draw[dashed] (1,-0.3)node[below]{$-\sqrt 3$} -- (1,3.35) ;
\draw[dashed] (1,3.85)  -- (1,4.2) ;
\draw[dashed] (10,-0.3)node[below]{$+ \sqrt{3} $} -- (10,3.35) ;
\draw[dashed] (10,3.85) -- (10,4.2) ;
\draw (1,3.6) node{${\cal O}^-$};
\draw (10,3.6) node{${\cal O}^+$};
\draw (4,3.6) node{$G_{\te{int}}^-$};
\draw (7,3.6) node{$G_{\te{int}}^+$};
\draw (3.25,0.9) node{$\Si^-$};
\draw (7.75,0.9) node{$\Si^+$};
\draw (3.25,3.3) node{$\pa \Si^-$};
\draw (7.75,3.3) node{$\pa \Si^+$};
\draw (4.75,3.3) node{$\tilde \Si^-$};
\draw (6.25,3.3) node{$\tilde \Si^+$};
\draw (5.5,2.4) node{${\cal J}$};
\draw[dashed] (-0.3,1.2) node[left]{$\tfrac{1}{2}$}  --  (10.5,1.2);
\draw[dashed] (-0.3,2.4) node[left]{$1$}  -- (5.2,2.4);
\draw[dashed] (5.8,2.4) --(10.5,2.4);
\draw[dashed] (-0.3,3.6) node[left]{$\tfrac{3}{2}$}  -- (0.7,3.6);
\draw[dashed] (1.3,3.6) -- (3.7,3.6);
\draw[dashed] (4.3,3.6) --(6.7,3.6);
\draw[dashed] (7.3,3.6) --(9.7,3.6);
\draw[dashed] (10.3,3.6) --(10.5,3.6);
\endtikzpicture
}
\caption{Conformal fields in the ${\cal N}=1$ internal CFT, together with their weight $h$ and $U(1)$ charge $q$.}
\label{N=1}
\end{figure}

\medskip
A large sector of the internal CFT can be captured by bosonization. Let $H(z)$ denote a canonically normalized free \& chiral boson, then we have the following representation for some for the aforementioned operators:
\beq
{\cal J} \ \ \equiv \ \ i \pa H \co \Si^{\pm} \ \ \equiv \ \ \ee^{\pm i\sqrt{3}H/2} \co {\cal O}^{\pm} \ \ \equiv \ \ \ee^{\pm i \sqrt{3} H} \ .
\label{1,5}
\eeq
However, the internal supercurrent (or energy momentum tensor) cannot be fully bosonized. Instead, we can represent $G_{\te{int}}^{\pm}$ as
\beq
G^{\pm} _{\te{int}} \eq \sqrt{ \frac{3}{2}} \; \ee^{\pm \frac{i}{\sqrt{3}} H} \, g^{\pm} \ ,
\label{1,7}
\eeq
where the $h=\frac{4}{3}$ operators $g^{\pm}$ are local with respect to $H$ and satisfy
\begin{align}
g^{\pm}(z) \, g^{\mp}(w) \ \ &\sim \ \ \frac{1}{(z-w)^{8/3} } \ + \ \frac{0}{(z-w)^{5/3}} \ + \ \ldots
\label{1,8a} \\
g^{\pm}(z) \, g^{\pm}(w) \ \ &\sim \ \ \frac{ g^{\pm}(w) \, g^{\pm}(w) }{(z-w)^{1/3}} \ + \ \ldots
\label{1,8b} \ .
\end{align}
On these grounds, we can understand the OPE of the supercurrent with internal spin fields,
\begin{align}
G^\pm_{\te{int}}(z) \, \Si^{\mp}(w) \ \ &\sim \ \ \sqrt{ \frac{3}{2} } \; \frac{ \tilde \Si^{\mp}(w)}{(z-w)^{1/2}} \ + \ \ldots \label{1,9a}
\\
G^\pm_{\te{int}}(z) \, \Si^{\pm}(w) \ \ &\sim \ \ (z-w)^{1/2} \, g^{\pm} \, \ee^{\pm \frac{5i}{2\sqrt{3}}}(w) \ + \ \ldots \label{1,9b}
\end{align}
which introduces excited spin fields $\tilde \Si^{\pm}$ of $h=\frac{11}{8}$ in case of opposite $U(1)$ charges $G^\pm_{\te{int}} \leftrightarrow \Si^{\mp}$,
\beq
\tilde \Si^{\pm} \ \ \equiv \ \ g^{\mp} \, \ee^{\pm \frac{i}{2\sqrt{3}} H} \ .
\label{1,10}
\eeq
Figure \ref{N=1} gives an overview of the universal Virasoro primaries in the internal $c=9$ SCFT. More detailed OPEs including subleading singularities can be found in appendix \ref{appN=1SUSY}.


\medskip
From these OPEs, we obtain the following $+1/2$ ghost picture version for the SUSY charge
\begin{align}
{\cal Q}^{(+1/2)}_a &\eq  \oint \frac{ \dd z }{2\pi i} \; \left[ \, \frac{\sqrt{3}}{\ap^{1/4}} \;  S_a \, \tilde \Si^+ \ + \ \frac{1}{2\ap^{3/4}} \; i\pa X_\mu \, \si^\mu_{a \dot b} \, S^{\dot b} \, \Si^+ \, \right] \, \ee^{+\phi/2} \ ,
\label{1,14a} \\
\bar {\cal Q}^{(+1/2) \dot b} &\eq  \oint \frac{ \dd z }{2\pi i} \; \left[ \, \frac{\sqrt{3}}{\ap^{1/4}} \;  S^{\dot b} \, \tilde \Si^- \ + \ \frac{1}{2\ap^{3/4}} \; i\pa X^\mu \, \sib_\mu^{\dot b a} \, S_{a} \, \Si^- \, \right] \, \ee^{+\phi/2} \ ,
\label{1,14b}
\end{align}
which yield the ${\cal N}=1$ SUSY algebra
\beq
\big\{ \, {\cal Q}_a^{(+1/2)} \; , \; \bar {\cal Q}^{(-1/2),\dot b} \, \big\} \eq (\si^\mu \, \vep)_a{}^{\dot b} \, P_\mu \co P_\mu \eq \frac{1}{2\ap} \oint \frac{ \dd z }{2\pi i} \; i \pa X_\mu \ \label{N=1SUSYalg}.
\eeq

\subsection{CFT operators in ${\cal N}=2$ compactifications}
\label{sec:N=2cft}

In superstring compactifications which preserve ${\cal N}=2$ spacetime SUSY, it can be shown along the lines of \cite{BD2,BD3} that the internal CFT splits into two decoupled sectors with central charges $c=6$ and $c=3$, respectively. Starting point are the two supercharges
\beq
{\cal Q}^{(-1/2),i}_a \eq \frac{1}{\ap^{1/4}} \oint \frac{ \dd z }{2\pi i} \; S_a \, \Si^i \, \ee^{-\phi/2} \co \bar {\cal Q}^{(-1/2), \dot b i} \eq \frac{1}{\ap^{1/4}} \oint \frac{ \dd z }{2\pi i} \; S^{\dot b} \, \bar \Si^i \, \ee^{-\phi/2} \ ,
\label{2,1}
\eeq
containing two species of spin fields $\Si^{i=1,2}$ and $\bar \Si^{i=1,2}$. The latter turn out to factorize into decoupled primaries $\la^i$ and $\ee^{\pm iH/2}$ from the $c=6$ and $c=3$ sector, respectively:
\beq
\Si^i \eq \la^i \, \ee^{+iH/2} \co \bar \Si^i \eq \la^i \, \ee^{- iH/2} \ .
\label{2,2}
\eeq
The $c=3$ part can be represented in terms of a single free chiral boson $H$ subject to (\ref{1,4}). Its contribution $\frac{1}{2} (i\pa H )^2$ to the $c=3$ energy momentum tensor assigns conformal weight $h(\ee^{\pm iH/2}) = 1/8$ (or more generally, $h(\ee^{iqH}) = q^2/2$). Moreover, OPEs of the partial spin fields $\ee^{\pm iH/2}$ introduce $h=\frac{1}{2}$ fermions $\ee^{\pm i H}$ and excited spin fields $\ee^{\pm 3iH/2}$ of weight $h=\frac{9}{8}$.

\medskip
On the other hand, the $\la^i$ fields from the $c=6$ sector have weight $h(\la^i)=1/4$ and form an $SU(2)$ doublet. Their operator algebra\footnote{The contraction rules for the antisymmetric $\vep^{ij}, \vep_{ij}$ tensors introduce signs in some of the OPEs:
\beq
\la_i(z) \,  \la ^j(w) \ \sim  \ \frac{+ \, \de_i^j }{(z-w)^{1/2}} \ , \ \ \ \la_i (z) \, \la_j (w) \ \sim \ \frac{ - \, \vep_{ij} }{(z-w)^{1/2}}  \ , \ \ \ \la^i(z) \, \la_j(w) \ \sim \ \frac{-  \, \de^i_j }{(z-w)^{1/2}} \ . \eeq} gives rise to an $SU(2)$ triplet of $h=1$ currents ${\cal J}^{A=1,2,3}$:
\beq
\la^i(z) \, \la^j(w) \ \ \sim \ \ \frac{ \vep^{ij} }{(z-w)^{1/2}} \ + \ \frac{1}{\sqrt{2}} \; (z-w)^{1/2} \, (\tau_A \, \vep)^{ij} \, {\cal J}^{A}(w) \ +\ \ldots
\label{2,3}
\eeq
The $\tau_A$ denote the standard (traceless) $SU(2)$ Pauli matrices $\left\{ \left( \begin{smallmatrix} 0 &1 \\ 1 &0 \end{smallmatrix} \right), \ \left( \begin{smallmatrix} 0 &-i \\ i &0 \end{smallmatrix} \right), \ \left( \begin{smallmatrix} 1 &0 \\ 0 &-1 \end{smallmatrix} \right) \right\}$ subject to the multiplication rule $\tau_A \tau_B = \de_{AB} + i \vep_{ABC} \tau^C$.

\medskip
The currents obey the $SU(2)$ current algebra at level $k=1$, we use normalization conventions
\beq
{\cal J}^A(z) \, {\cal J}^B(w) \ \ \sim \ \ \frac{\de^{AB}}{(z-w)^2} \ + \ \frac{ i \, \sqrt{2} \, \vep^{ABC} \, {\cal J}_C(w) }{z-w}  \ + \ \ldots
\label{2,4}
\eeq
in which their interaction with the spin fields is governed by
\begin{align}
{\cal J}^A(z) \, \la^i(w) & \ \ \sim \ \ \frac{ (\tau^A)^i{}_j \, \la^j(w) }{\sqrt{2} \, (z-w)} \ + \ \sqrt{2} \, (\tau^A)^i{}_j \, \pa \la^j(w) \ + \ \ldots
\label{2,5a} \\
\la^i(z) \, {\cal J}^A(w) & \ \ \sim \ \ \frac{ (\tau^A)^i{}_j \, \la^j(w) }{\sqrt{2} \, (z-w)} \ - \ \frac{1}{\sqrt{2}} \; (\tau^A)^i{}_j \, \pa \la^j(w) \ + \ \ldots \ .
\label{2,5b}
\end{align}
Note that also the $\la^i$ and ${\cal J}^A$ fit into a bosonization scheme according to
\beq
{\cal J}^{A=3} \ \ \equiv \ \ i \pa H_3 \co {\cal J}^{A=1} \ \pm \ i {\cal J}^{A=2} \ \ \equiv \ \ \sqrt{2} \, \ee^{\pm i\sqrt{2} H_3} \co \la^{i=1,2} \eq \ee^{\pm i H_3 / \sqrt{2}}
\label{2,6}
\eeq
with $H_3$ being nonsingular with respect to the $c=3$ boson $H$. This fixes the choice of the $SU(2)$ Cartan subalgebra.

\medskip
The world-sheet supercurrents associated with the two decoupled CFTs,
\beq
G_{\te{int}} \ \ \equiv \ \ G_{c=3} \ + \ G_{c=6} \ ,
\label{2,7}
\eeq
can be split according to their charges under the $h=1$ currents. In the $c=3$ sector, we find a free field representation in terms of internal $h=1$ coordinates\footnote{As usual, the OPEs between $i\pa Z^{\pm}$ are normalized as
\beq
i \pa Z^{\pm}(z) \, i \pa Z^{\mp}(w) \ \ \sim \ \ \frac{2\ap}{(z-w)^2} \ + \ \ldots \co i \pa Z^{\pm}(z) \, i \pa Z^{\pm}(w) \ \ \sim \ \  i \pa Z^{\pm}(w) \, i \pa Z^{\pm}(w) \ + \ \ldots \ .
\label{2,9}
\eeq} $i \pa Z^{\pm}$,
\beq
G_{c=3} \eq \frac{1}{2\, \sqrt{2\ap}} \; \big( \, i\pa Z^+ \, \ee^{-iH} \ + \ i \pa Z^- \, \ee^{iH} \, \big) \ ,
\label{2,8}
\eeq
The fermions $\Psi^{\pm}(z)=e^{\pm iH(z)}$ together with the free bosons $Z^\pm$ generate
an internal Kac-Moody algebra
\begin{equation}
g \eq SO(2)\times \lbrack U(1)\rbrack^2
\end{equation}
with dimension one currents
\begin{equation}
j_{SO(2)}(z) \eq \Psi^+ \, \Psi^-(z) \eq i \pa H(z)\co j^\pm_{U(1)}(z) \eq i\partial Z^\pm(z)\ .
\end{equation}
As for the ${\cal N}=4$ case, the fields $Z_\pm(z)$ can be viewed as the coordinates of a (holomorphic) torus compactification on a two-dimensional torus $T^2$.

\begin{figure}[h]
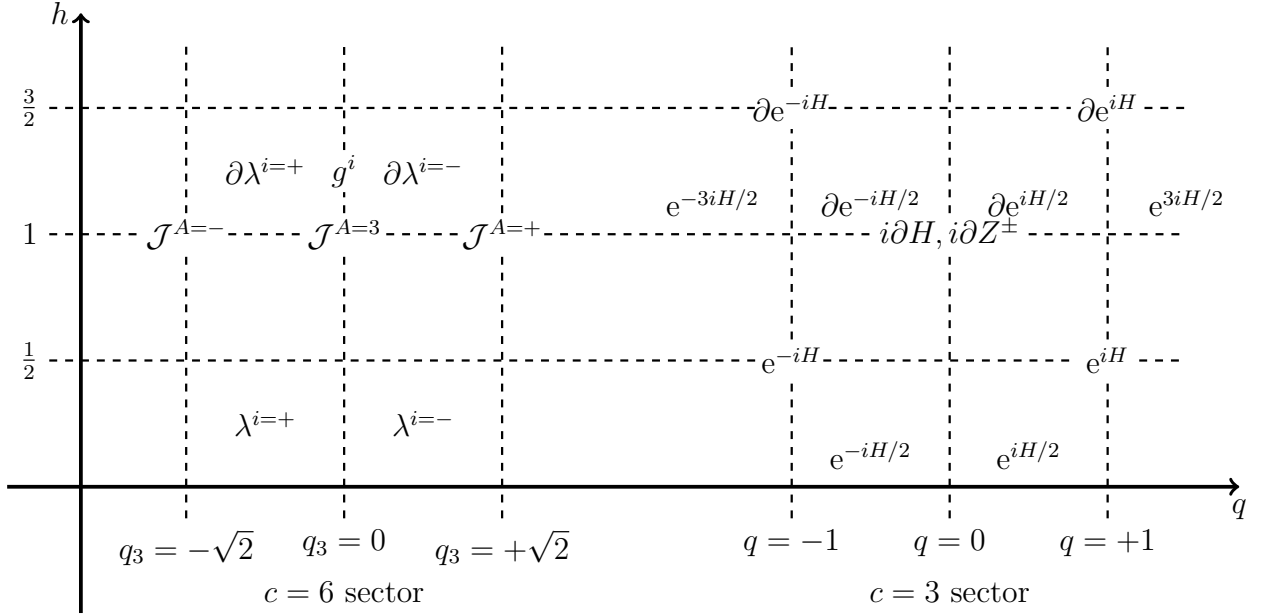

\centerline{
\tikzpicture [scale=1.4,line width=0.30mm]
\draw[line width=0.50mm,->] (0,-1.2) -- (0,4.5) node[left]{$h$};
\draw[line width=0.50mm,->] (-0.7,0) -- (11,0) node[below]{$q$};
\draw (2.5,-1) node{$c=6$ sector} ;
\draw[dashed] (2.5,-0.3)node[below]{$q_3=0$} -- (2.5,2.2) ;
\draw[dashed] (2.5,2.6) -- (2.5,2.8) ;
\draw[dashed] (2.5,3.2) -- (2.5,4.2) ;
\draw[dashed] (1,-0.3)node[below]{$q_3=-\sqrt{2}$} -- (1,2.2) ;
\draw[dashed] (1,2.6) -- (1,4.2) ;
\draw[dashed] (4,-0.3)node[below]{$q_3=+\sqrt{2}$} -- (4,2.2) ;
\draw[dashed] (4,2.6) -- (4,4.2) ;
\draw (1.75,0.6) node{$\la^{i=+}$};
\draw (3.25,0.6) node{$\la^{i=-}$};
\draw (1.75,3) node{$\pa \la^{i=+}$};
\draw (3.25,3) node{$\pa \la^{i=-}$};
\draw (2.5,3) node{$g^i$};
\draw (2.5,2.4) node{${\cal J}^{A=3}$};
\draw (1,2.4) node{${\cal J}^{A=-}$};
\draw (4,2.4) node{${\cal J}^{A=+}$};
\begin{scope}[xshift=0.5cm]
\draw (7.75,-1) node{$c=3$ sector} ;
\draw[dashed] (7.75,-0.3)node[below]{$q=0$} -- (7.75,2.2) ;
\draw[dashed] (7.75,2.6) -- (7.75,4.2) ;
\draw[dashed] (9.25,-0.3)node[below]{$q=+1$} -- (9.25,1) ;
\draw[dashed] (9.25,1.4) -- (9.25,3.4) ;
\draw[dashed] (9.25,3.8) -- (9.25,4.2) ;
\draw[dashed] (6.25,-0.3)node[below]{$q=-1$} -- (6.25,1) ;
\draw[dashed] (6.25,1.4) -- (6.25,3.4) ;
\draw[dashed] (6.25,3.8) -- (6.25,4.2) ;
\draw (7.75,2.4) node{$ i\pa H, i \pa Z^{\pm}$};
\draw (8.5,0.3) node{$\ee^{iH/2}$};
\draw (7,0.3) node{$\ee^{-iH/2}$};
\draw (8.5,2.7) node{$\pa \ee^{iH/2}$};
\draw (7,2.7) node{$\pa \ee^{-iH/2}$};
\draw (10,2.7) node{$\ee^{3iH/2}$};
\draw (5.5,2.7) node{$\ee^{-3iH/2}$};
\draw (9.25,1.2) node{$\ee^{iH}$};
\draw (6.25,1.2) node{$\ee^{-iH}$};
\draw (9.25,3.6) node{$\pa \ee^{iH}$};
\draw (6.25,3.6) node{$\pa \ee^{-iH}$};
\end{scope}
\draw[dashed] (-0.3,1.2) node[left]{$\tfrac{1}{2}$}  --  (6.45,1.2);
\draw[dashed] (7.05,1.2)  --  (9.45,1.2);
\draw[dashed] (10.05,1.2)  --  (10.5,1.2);
\draw[dashed] (-0.3,2.4) node[left]{$1$}  -- (0.6,2.4);
\draw[dashed] (1.4,2.4) --(2.1,2.4);
\draw[dashed] (2.9,2.4) --(3.6,2.4);
\draw[dashed] (4.4,2.4) --(7.5,2.4);
\draw[dashed] (9,2.4) --(10.5,2.4);
\draw[dashed] (-0.3,3.6) node[left]{$\tfrac{3}{2}$}  -- (6.4,3.6);
\draw[dashed] (7.1,3.6) -- (9.4,3.6);
\draw[dashed] (10.1,3.6) --(10.5,3.6);
\endtikzpicture
}
\caption{Universal operator content of the internal CFT associated with ${\cal N}=2$ spacetime SUSY, including weight $h$ and charges $q_3,q$ under $i \pa H_3$ and $i\pa H$, respectively.}
\label{N=2}
\end{figure}

\medskip
Also the supercurrent of the $c=6$ sector cannot be fully built from the bosonization prescription (\ref{2,6}), it additionally requires the introduction of an $SU(2)$ doublet of $h=5/4$ fields $g_i$:
\beq
G_{c=6} \eq \frac{1}{\sqrt{2}} \; \big( \, \ee^{iH_3/\sqrt{2}} \, g_1 \ + \ \ee^{-iH_3/\sqrt{2}} \, g_2 \, \big) \eq  \frac{1}{\sqrt{2}} \;  \la^i \, g_i \ . \label{2,10}
\eeq
The $g_i$ decouple from the $\la^i$ and ${\cal J}^A$, and their OPE\footnote{$\vep$ contractions yield signs opposite to the $\la^i\la_j$ case:
\beq
g^i(z) \, g_j(w) \ \sim \ \frac{ + \, \de^i_j }{(z-w)^{5/2}} \ , \ \ \ g^i(z) \,  g^j(w) \  \sim \ \frac{ - \, \vep^{ij}}{(z-w)^{5/2}} \ , \ \ \ g_i(z) \,  g^j(w) \  \sim \ \frac{ - \,  \de_i^j }{(z-w)^{5/2}} \ . \eeq}
\beq
g_i(z) \, g_j(w) \ \ \sim \ \ \frac{ \vep_{ij} }{(z-w)^{5/2}} \ + \ \frac{0}{(z-w)^{3/2}} \ + \ \ldots
\label{2,11}
\eeq
makes sure that the supercurrents satisfy the required ${\cal N}=4$ superconformal algebra at $c=6$. A summary of operators in the internal SCFTs common to ${\cal N}=2$ compactifications are presented in figure \ref{N=2}.

\medskip
The internal supercurrent yields the following higher ghost picture SUSY charges:
\begin{align}
{\cal Q}_a^{(+1/2),i} &\eq  \frac{1}{\sqrt{2} \ap^{3/4}} \oint \frac{ \dd z }{2\pi i} \; \bigg[ \, \frac{1}{\sqrt{2}} \; i \pa X_\mu \, \si^\mu_{a \dot b} \, S^{\dot b} \, \la^i \, \ee^{iH/2} \ + \ i \pa Z^+ \, S_a \, \la^i \, \ee^{-iH/2} \notag \\
& \hskip6cm - \ 2\sqrt{\ap} \;  g^i \, S_a \, \ee^{iH/2} \, \bigg] \, \ee^{\phi/2} \ ,
\label{Q+1} \\
\bar {\cal Q}^{(+1/2),\dot b i} &\eq  \frac{1}{\sqrt{2} \ap^{3/4}} \oint \frac{ \dd z }{2\pi i} \; \bigg[ \, \frac{1}{\sqrt{2}} \; i \pa X^\mu \, \sib_\mu^{\dot b a} \, S_{a} \, \la^i \, \ee^{-iH/2} \ + \  i \pa Z^- \, S^{\dot b} \, \la^i \, \ee^{iH/2} \notag \\
& \hskip6cm - \ 2\sqrt{\ap} \;  g^i \, S^{\dot b} \, \ee^{-iH/2} \, \bigg] \, \ee^{\phi/2} \ .
\label{Q+2}
\end{align}
The anticommutator of equal chirality generators gives rise to  a complex central charge operator, which can be written in terms
of the free bosons $Z^\pm$:
\begin{align}
\big\{ \, {\cal Q}_a^{(+1/2),i} \; , \; {\cal Q}_b^{(-1/2),j} \, \big\} &\eq \vep_{ab} \, {\cal Z}^{ij} \co {\cal Z}^{ij} \eq \frac{\vep^{ij}}{\sqrt{2} \,\ap} \oint \frac{ \dd z }{2\pi i} \; i \pa Z ^+ \ ,
\label{Q+3} \\
\big\{ \, \bar {\cal Q}^{(+1/2),\dot ai} \; , \; \bar {\cal Q}^{(-1/2), \dot b j} \, \big\} &\eq \vep^{\dot a \dot b} \, \bar {\cal Z}^{ij} \co \bar {\cal Z}^{ij} \eq \frac{\vep^{ij}}{\sqrt{2} \,\ap} \oint \frac{ \dd z }{2\pi i} \; i \pa Z^- \ .
\label{Q+4}
\end{align}
It again follows that the internal momentum states  $|p^\pm\rangle$ of the two-torus are precisely those states that carry non-vanishing  ${\cal N}=2$
central charges.
They completely break the internal world-sheet $SO(2)$ symmetry.  On the other hand, states with vanishing internal momenta, $|p^\pm=0\rangle$, build internal $SO(2)$ representations,
resp. representations of the group $USp(4)$, which classifies the ${\cal N}=2$ massive states without central charges.

\subsection{Summary of CFT operators}

\begin{figure}[h]
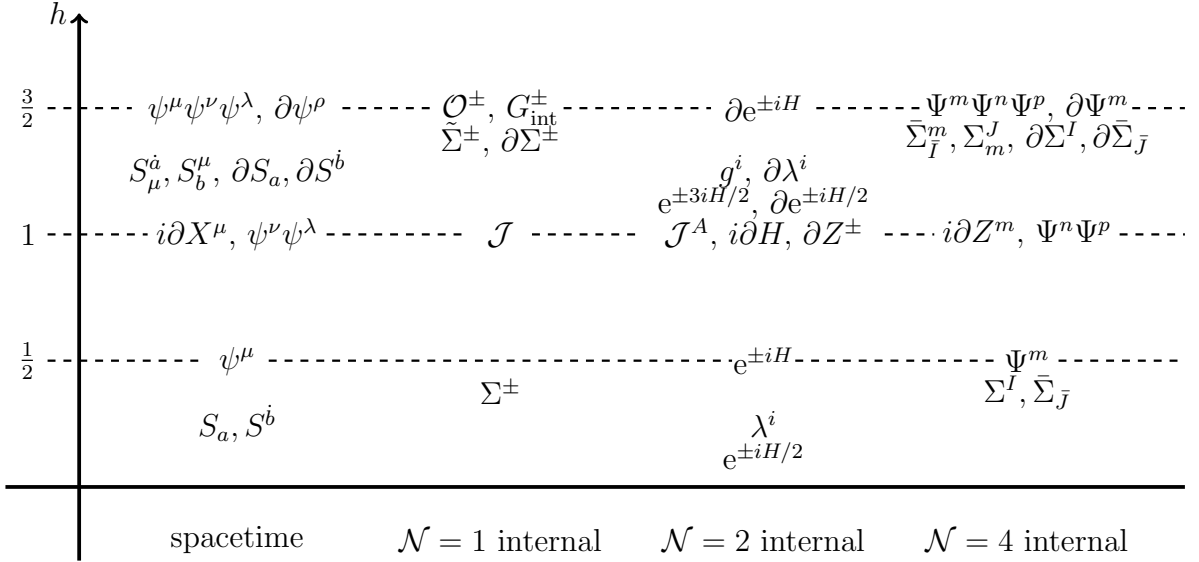

\centerline{
\tikzpicture [scale=1.4,line width=0.30mm]
\draw[line width=0.50mm,->] (0,-0.7) -- (0,4.5) node[left]{$h$};
\draw[line width=0.50mm] (-0.7,0) -- (10.5,0);
%
%
\draw[dashed] (-0.3,1.2) node[left]{$\tfrac{1}{2}$}  -- (1.2,1.2);
\draw[dashed] (1.8,1.2) -- (6.2,1.2);
\draw[dashed] (6.8,1.2) -- (8.7,1.2);
\draw[dashed] (9.3,1.2) -- (10.5,1.2);
\draw[dashed] (-0.3,2.4) node[left]{$1$}  -- (0.65,2.4);
\draw[dashed] (2.35,2.4) -- (3.7,2.4);
\draw[dashed] (4.3,2.4) --(5.35,2.4);
\draw[dashed] (7.65,2.4) --(8.12,2.4);
\draw[dashed] (9.88,2.4) --(10.5,2.4);
\draw[dashed] (-0.3,3.6) node[left]{$\tfrac{3}{2}$}  -- (0.5,3.6);
\draw[dashed] (2.5,3.6) -- (3.4,3.6);
\draw[dashed] (4.6,3.6) --(6,3.6);
\draw[dashed] (7,3.6) --(8,3.6);
\draw[dashed] (10,3.6) --(10.5,3.6);
\draw (1.5,-0.5) node{spacetime};
\draw (1.5,0.6) node{$S_a ,S^{\dot b}$};
\draw (1.5,1.2) node{$\psi^\mu$};
\draw (1.5,3) node{$S_\mu^{\dot a} ,S^\mu_{ b}, \, \pa S_a , \pa S^{\dot b}$};
\draw (1.5,2.4) node{$i\pa X^\mu, \, \psi^\nu \psi^\la$};
\draw (1.5,3.6) node{$\psi^\mu\psi^\nu \psi^\la, \, \pa \psi^\rho$};
\draw (4,0.9) node{$\Si^{\pm}$};
\draw (4,2.4) node{${\cal J}$};
\draw (4,3.6) node{${\cal O}^{\pm}, \, G_{\te{int}}^{\pm}$};
\draw (4,3.3) node{$\tilde \Si^{\pm}, \, \pa \Si^{\pm}$};
\draw (4,-0.5) node{${\cal N}=1$ internal};
\draw (6.5,-0.5) node{${\cal N}=2$ internal};
\draw (9,-0.5) node{${\cal N}=4$ internal};
\draw (6.5,0.6) node{$\la^i$};
\draw (6.5,2.4) node{${\cal J}^A, \, i \pa H, \, \pa Z^{\pm}$};
\draw (6.5,1.2) node{$\ee^{\pm iH}$};
\draw (6.5,0.3) node{$\ee^{\pm iH/2}$};
\draw (6.5,2.7) node{$\ee^{\pm 3iH/2}, \, \pa \ee^{\pm i H/2}$};
\draw (6.5,3) node{$g^i, \, \pa \la^i$};
\draw (6.5,3.6) node{$\pa \ee^{\pm iH}$};
\draw (9,0.9) node{$\Si^I ,\bar \Si_{\bar J}$};
\draw (9,1.2) node{$\Psi^m$};
\draw (9,3.3) node{$\bar \Si^m_{\bar I} , \Si_m^J, \, \pa \Si^I , \pa \bar \Si_{\bar J}$};
\draw (9,2.4) node{$i\pa Z^m, \, \Psi^n \Psi^p$};
\draw (9,3.6) node{$\Psi^m\Psi^n \Psi^p, \, \pa \Psi^m$};
\endtikzpicture
}
\caption{Conformal fields together with their weight in various decoupling CFT sectors}
\label{appetizer}
\end{figure}

To conclude this section on the internal SCFTs associated with $D=4$ compactifications of different supercharges, figure \ref{appetizer} summarizes the field content of the different sectors. This is a good reference to build the most general ansatz for physical vertex operators.

\section{Massive supermultiplets for ${\cal N}=4$ SUSY}
\label{sec:n4}

Having introduced the CFT setup for the construction of massive string state, let us now turn to explicit vertex operators on the first mass level. We will first of all examine the four-dimensional field content of maximally supersymmetric superstring compactifications to $D=4$ with ${\cal N}=4$ SUSY. This is the dimensional reduction of the ten-dimensional multiplet, so we will again find all the 256 states which have been discussed from the $D=10$ viewpoint in section \ref{sec:10dim}. They form a massive ${\cal N}=4$ multiplet in four dimensions for which we will work out the spin and R-symmetry content as well as the SUSY transformations.

\subsection{NS sector}

With the internal CFT operators from figure \ref{N=4} at hand, the following $h=3/2$ combinations must be considered in the most general NS vertex operator at first mass level:
\begin{align}
V^{(-1)} \eq & \Big( \, \al_{\mu \nu} \, i \pa X^\mu \, \psi^\nu \ + \ e_{\mu \nu \la} \, \psi^\mu \, \psi^\nu \, \psi^\la \ + \ h_\mu \, \pa \psi^\mu \ + \ \be_\mu^m \, i \pa X^\mu \, \Psi_m  \notag \\
&+ \ \ga_\mu^m \, \psi^\mu \, i \pa Z_m \ + \ d_\mu^{mn} \, \psi^\mu \, \Psi_m \, \Psi_n \ + \ Y^m \, \pa \Psi_m \ + \ \om_{\mu \nu}^m \, \psi^\mu \, \psi^\nu \, \Psi_m \notag \\
&+ \ \zeta^{mn} \, i \pa Z_m \, \Psi_n \ + \ \Om^{mnp} \, \Psi_m \, \Psi_n \, \Psi_p
 \, \Big) \, \ee^{-\phi} \, \ee^{ik \cdot X} \ .
\label{12,1}
\end{align}
Requiring vanishing $Q_1$ variation for (\ref{12,1}) implies the following on-shell constraints for the ten wavefunctions above:
\beq \begin{array}{ll}
0 \eq \al_\mu{}^\mu \ + \ k^\mu \, h_\mu \ + \ \zeta_m {}^m  \ \ \ \ \ \ \ \ \ \ \ &0 \eq 2\ap \, Y^m \ + \ k^\mu \, \ga_\mu^m \\
0 \eq \al_{[\mu \nu]} \ + \ 3 \, k^\la \, e_{\la \mu \nu}  \ \ \ \ \ \ \ \ \ \ \  & 0 \eq \be_\mu^m \ - \ \ga_\mu^m \ + \ 2 \, k^\la  \om_{\la \mu}^m \\
0 \eq 2\ap \, \al_{\mu \nu} \, k^\nu \ + \ h_\mu  \ \ \ \ \ \ \ \ \ \ \ &0 \eq k^\mu \, d_\mu^{mn} \ + \ \zeta^{[mn]}
 \end{array} \label{12,2} \eeq
This leaves the following 128 physical solutions
\begin{itemize}
\item one transverse and traceless spin two tensor
\begin{align}
V^{(-1)}_\alpha &\eq \frac{1}{\sqrt{2\ap}} \; \al_{\mu \nu} \, i \pa X^\mu \, \psi^\nu \, \ee^{-\phi} \, \ee^{ik \cdot X} \co k^\mu \, \al_{\mu \nu} = \al_{[\mu \nu]} = \al_\mu{}^\mu = 0 \label{12,3a}
\end{align}
\item 27 transverse vectors (in the vector and two-form representations of the R-symmetry $SO(6)$)
\begin{align}
V^{(-1)}_d &\eq \frac{1}{2} \, d^{mn}_\mu \, \psi^\mu \, \Psi_m \, \Psi_n \,  \ee^{-\phi} \, \ee^{ik \cdot X} \co k^\mu \, d^{mn}_\mu = 0  \label{12,3c} \\
V^{(-1)}_{\beta^{\pm}} &\eq \frac{1}{2 \sqrt{2\ap}} \; \be^{\pm,m}_{\mu} \, \bigl( \, i \pa X^\mu \, \Psi_m \ + \ i \pa Z_m \, \psi^\mu \notag \\
& \hskip3.7cm \pm \ i \ap \, \vep^{\mu \nu \la \rho} \, k_\nu \, \psi_\la \, \psi_\rho \, \Psi_m
\, \big) \, \ee^{-\phi} \, \ee^{ik \cdot X} \co k^\mu \, \be_{\mu}^{\pm,m} =  0 \label{12,3d} \end{align}
\item 42 scalar degrees of freedom (scalars, spin two and and three-form with respect to $SO(6)$)
\begin{align}
V^{(-1)}_{\Phi^{\pm}} &\eq \frac{1}{2\sqrt{2\ap}} \; \Phi^{\pm} \, \Big[ \, (\eta_{\mu \nu} \, + \, 2\ap \, k_\mu \, k_\nu) \, i\pa X^\mu \, \psi^\nu \ + \ 2\ap \, k_\mu \, \pa \psi^\mu \notag \\
& \hskip6cm \pm \ \frac{i\ap}{3} \; \vep_{\mu \nu \la \rho} \, \psi^\mu \, \psi^\nu \, \psi^\la \, k^{\rho} \, \Big] \, \ee^{-\phi} \, \ee^{ik \cdot X} \label{12,3b} \\
V^{(-1)}_\zeta &\eq \frac{1}{\sqrt{2\ap}} \; \zeta^{mn} \, i\pa Z_m \, \Psi_n \, \ee^{-\phi} \, \ee^{ik \cdot X} \co \zeta^{[mn]} = \zeta^m{}_m = 0 \label{12,3f} \\
V^{(-1)}_\Om &\eq \Om^{mnp} \, \Psi_m \, \Psi_n \, \Psi_p  \, \ee^{-\phi} \, \ee^{ik \cdot X} \ . \label{12,3g}
\end{align}
\end{itemize}
The 46 spurious NS sector states from ten dimensions are aligned into six representations of $SO(1,3) \times SO(6)$. They can be constructively obtained as BRST variations of ghost charge -2 objects, see (\ref{4,15aa}):
\begin{align}
V^{(-1)}_{\pi(\te{sp})} &\ \ \sim \ \ \Big[ \, (\pi_\mu \, k_\nu \; + \; k_\mu \, \pi_\nu) \, i \pa X^\mu \, \psi^\nu \ + \ 2 \, \pi_\mu \, \pa \psi^\mu \, \Big] \, \ee^{-\phi} \, \ee^{ik \cdot X} \co k^\mu \, \pi_{\mu}  = 0 \ , \label{12,4a} \\
V^{(-1)}_{\Sigma(\te{sp})} &\ \ \sim \ \ \Big[ \, 2 \, \Si_{[\mu \nu]} \, i \pa X^\mu \, \psi^\nu \ + \ 2\ap \, \Si_{[\mu \nu} \, k_{\la]} \, \psi^\mu \, \psi^\nu \, \psi^\la \, \Big] \, \ee^{-\phi} \, \ee^{ik \cdot X} \co k^\mu \, \Si_{\mu \nu}  = 0 \ , \label{12,4b} \\
V^{(-1)}_{\Lambda_1(\te{sp})} &\ \ \sim \ \ \La_1 \, \Big[ \, (\eta_{\mu \nu} \; + \; 4\ap \, k_\mu \,k_\nu) \, i \pa X^\mu \, \psi^\nu \ + \ 6\ap \, k_\mu \, \pa \psi^\mu \ + \ i \pa Z_m \, \Psi^m \, \Big] \, \ee^{-\phi} \, \ee^{ik \cdot X}  \ , \label{12,4c} \\
V_{\Lambda_2(\te{sp})}^{(-1)} &\ \ \sim \ \ \La_2^m \Bigl( \, k_\mu \, \bigl[ \, i\pa X^\mu \, \Psi_m \ + \ i\pa Z_m\, \psi^\mu \, \bigr] \ + \ 2 \, \pa \Psi_m \, \Bigr) \, \ee^{-\phi} \, \ee^{ik \cdot X} \ , \label{12,4e} \\
V_{\Lambda_3(\te{sp})}^{(-1)} &\ \ \sim \ \ \La_3^{[mn]} \, \bigl[ \,  i\pa Z_m\, \Psi_n  \ + \ \alpha' \, k_{\mu} \, \psi^\mu \, \Psi_m \, \Psi_n \, \bigr] \, \ee^{-\phi} \, \ee^{ik \cdot X} \ , \label{12,4f}  \\
V_{\Lambda_4(\te{sp})}^{(-1)} &\ \ \sim \ \ \La^m _{4 \, \mu} \,  \Bigl(  \, i\pa X^\mu \, \Psi_m \ - \ i\pa Z_m\, \psi^\mu \ - \ 2\alpha' \, k_{\nu}  \, \psi^\mu \, \psi^\nu \, \Psi_m \, \Bigr) \, \ee^{-\phi} \, \ee^{ik \cdot X} \co k^\mu \, \La^m _{4 \, \mu} = 0 \ . \label{12,4d}
\end{align}
Each spurious state corresponds to a gauge freedom. The first one (\ref{12,4a}) admits to gauge away the longitudinal component of the rank two tensor $\al_{\mu \nu}$ whereas the second one (\ref{12,4b}) identifies the antisymmetric part $\al_{[\mu \nu]}$ together with the longitudinal three-form $e_{\mu \nu \la} \sim k_{[\mu} \Si_{\nu \la]}$ as unphysical. Similarly, (\ref{12,4e}), (\ref{12,4f}) and (\ref{12,4d}) eliminate the longitudinal components of $(\be^{m}_\mu + \ga^m_\mu),d^{mn}_\mu$ and $\om^m_{\mu \nu}$ as well as the antisymmetric parts $\be^{m}_\mu - \ga^m_\mu$ and $\zeta_{[mn]}$. The trace of $\al_{\mu \nu}$ can be gauged away using (\ref{12,4c}).

\medskip
Once the three- and two-forms $e_{\mu \nu \la}$ and $\om^k_{\mu \nu}$ are reduced to there transverse part, contraction with $\vep^{\mu \nu \la \rho} k_{\rho}$ dualizes them to a scalar and a vector, respectively. As we will see below, supersymmetry suggests to include these dualized states into the complex combinations (\ref{12,3d}) and (\ref{12,3b}).

\subsection{R sector}

In the R sector, the SCFT operators of appropriate weight give rise to a vertex operator ansatz with six wavefunctions:
\begin{align}
V^{(-\frac{1}{2})} \eq & \Big( \, v_{\mu,I}^a \, i \pa X^\mu \, S_a \, \Si^I \ + \ \bar \rho^\mu_{\dot b,I} \, S_\mu^{\dot b} \, \Si^I \ + \ u^a_I \, \pa S_a \, \Si^I \notag \\
& \ + \ y^a_I \, S_a \, \pa \Si^I \ + \ \bar r_{m,\dot b}^{\bar J} \, i \pa Z^m \, S^{\dot b} \, \bar \Si_{\bar J} \ + \ s_m^{a,\bar J} \, S_a \, \bar \Si^m_{\bar J}  \, \Big) \, \ee^{-\phi/2} \, \ee^{ik\cdot X}\ .
\label{vert}
\end{align}
The same set of states also exists with opposite chiralities with respect to both $SO(1,3)$ and $SO(6)$ (e.g. $v_{\mu,I}^a S_a \Si^I \leftrightarrow \bar v_{\mu,\dot b}^{\bar J} S^{\dot b} \bar \Si_{\bar J}$). However, the BRST constraints for the polarizations in (\ref{vert}) decouple from those of the other chirality sector which we did not display, so the discussion will be limited to the six wavefunctions shown in (\ref{vert}) for the moment. The full list of physical and spurious states follows from doubling the solutions of the on-shell constraints. Imposing invariance under $Q_1$ yields the following three independent constraints:
\beq \begin{array}{l}
0\eq 2 \ap \, v_{I}^{\mu,a} \not \! k_{a \dot b} \ + \ \sqrt{2} \, \bar \rho^\mu_{\dot b,I} \ + \ \tfrac{1}{2} \, u^a_I \, \si^\mu_{a \dot b} \ ,
\\
0 \eq 2\ap \, \bar r_{m,\dot b}^{\bar J} \not \! k^{\dot b a} \ + \ \sqrt{2} \, s_m^{a,\bar J} \ - \ \frac{1}{2} \, y_I^a \, \ga_m^{I \bar J} \ , \\
0 \eq k_\mu \, \bar \rho^\mu_{\dot b,I} \ + \ \frac{1}{2\sqrt{2}} \; \bar r^{\bar J}_{m,\dot b} \, \gab^m_{\bar J I} \ .
\end{array}
\eeq
The first two equations can be further disentangled into a trace and a traceless part with respect to the $\si^\mu$ and $\ga_m$ matrices. Since excited spin fields are $\si$ and $\ga$ traceless, the associated wavefunctions satisfy $\bar \rho^\mu_{\dot b,I} \sib^{\dot b a}_\mu = s_m^{a,\bar J} \gab^m_{\bar J I} = 0$ by construction. Hence, the aforementioned projections simplify the BRST constraints to
\begin{align}
u^a_I &\eq \ap \, v_{\mu,I}^b \, (\! \not \! k \, \sib^\mu)_b{}^a
\notag \\
\bar \rho^\mu_{\dot b,I} &\eq -\, \sqrt{2} \, \ap \, \Bigl( \, v_I^{\mu,a} \not \! k_{a \dot b} \ + \ \tfrac{1}{4} \, v_{\la,I}^a \, (\! \not \! k \, \sib^\la \, \si^\mu)_{a \dot b} \, \Bigr) \notag \\
y^a_I &\eq - \, \frac{2 \ap}{3} \; \bar r_{m,\dot b}^{\bar J} \, \gab^m_{\bar J I} \not \! k^{\dot b a} \\
s_m^{a,\bar J} &\eq  -\, \sqrt{2} \, \ap \, \Bigl( \, \bar r^{\bar J}_{m,\dot b} \not \! k^{\dot b a} \ + \ \tfrac{1}{6} \, \bar r^{\bar I}_{n,\dot b} \, (\gab^n \, \ga_m)_{\bar I}{}^{\bar J} \not \! k^{\dot b a} \, \Bigr) \notag \\
\bar r_{m,\dot b}^{\bar J} \, \gab^m_{\bar J I} &\eq 2\ap \, k_\mu \, v^{\mu,a}_{I} \not \! k_{a \dot b} \ - \ v_{\mu,I}^a \, \si^\mu_{a \dot b} \notag
\end{align}
where $\bar \rho,u,y$ and $s$ are expressed in terms of $v$ and $\bar r$. It turns out that both spin 3/2 and spin 1/2 components of the vector spinors $v_I$ as well as the $\ga$ traceless components of $\bar r$ give rise to an independent physical solution. The former is the $D=4$ analogue of the ten-dimensional spin 3/2 state (\ref{4,13}). But additionally, we find spin 1/2 Dirac fermions $(a^b, \bar r^{\bar I}_{m,a})$ -- both in the fundamental spinor- and in the spin 3/2 representations of the R-symmetry $SO(6)$. To summarize the physical states built from (\ref{vert}) and its opposite chirality counterpart:
\begin{itemize}
\item eight transverse and $\si$ traceless spin 3/2 vector spinors
\begin{align}
V^{(-\frac{1}{2})}_\chi &\eq \frac{1}{\sqrt{2}\ap^{1/4}} \; \chi^a_{\mu,I} \, \Big( \, i \pa X^\mu \, S_a \ - \ \sqrt{2} \, \ap \, \not \! k_{a \dot b} \, S^{\mu \dot b} \, \Big) \, \Si^I \, \ee^{-\phi/2} \, \ee^{ik\cdot X} \ ,\\
V^{(-\frac{1}{2})}_{\bar \chi} &\eq \frac{1}{\sqrt{2}\ap^{1/4}} \; \bar \chi^{\bar I}_{\mu,\dot a} \, \Big( \, i \pa X^\mu \, S^{\dot a} \ - \ \sqrt{2} \, \ap \, \not \! k^{\dot a  b} \, S^{\mu}_{b} \, \Big) \, \bar \Si_{\bar I} \, \ee^{-\phi/2} \, \ee^{ik\cdot X} \ ,\\
0 &\eq k^\mu \, \chi_{\mu,I}^a \eq \chi_{\mu,I}^a \, \si^{\mu}_{a \dot b} \eq k_\mu \, \bar \chi^{\mu, \bar J}_{\dot b} \eq \bar \chi^{\mu,\bar J}_{\dot b} \, \sib_\mu^{\dot b a}
\end{align}
\item 48 spin 1/2 fermions (eight in the fundamental and 40 in spin 3/2 representations of $SO(6)$)
\begin{align}
V^{(-\frac{1}{2})}_a &\eq \frac{\alpha'^{1/4}}{2} \; a^b_I \, \Big( \, (\si_\mu \not \! k)_b{}^a \, S_a \, i \pa X^\mu \ - \ 4 \, \pa S_b \, \Big) \,  \Si^I \, \ee^{-\phi/2} \, \ee^{ik\cdot X} \ ,\\
V^{(-\frac{1}{2})}_{\bar a} &\eq \frac{\alpha'^{1/4}}{2} \; \bar a_{\dot b}^{\bar I} \, \Big( \, (\sib_\mu \not \! k)^{\dot b}{}_{\dot a} \, S^{\dot a} \, i \pa X^\mu \ - \ 4 \, \pa S^{\dot b} \, \Big) \,  \bar \Si_{\bar I} \, \ee^{-\phi/2} \, \ee^{ik\cdot X} \ ,\\
V^{(-\frac{1}{2})}_r &\eq \frac{1}{\sqrt{2}\ap^{1/4}} \; r^a_{m,I} \, \Big( \, i \pa Z^m \, \Si^I \, S_a \ - \ \sqrt{2} \, \ap \, \not \! k_{a \dot b} \, S^{\dot b} \, \Si^{m,I} \, \Big) \, \ee^{-\phi/2} \, \ee^{ik\cdot X} \ , \\
V^{(-\frac{1}{2})}_{\bar r} &\eq \frac{1}{\sqrt{2}\ap^{1/4}} \; \bar r^{\bar I}_{m, \dot a} \, \Big( \, i \pa Z^m \, \bar \Si_{\bar I} \, S^{\dot a} \ - \ \sqrt{2} \, \ap \, \not \! k^{\dot a b} \, S_{ b} \, \bar \Si^{m}_{\bar I} \, \Big) \, \ee^{-\phi/2} \, \ee^{ik\cdot X} \ .
\end{align}
\end{itemize}
The following spurious solutions have been subtracted to remove internal derivatives $\pa \Si^I$ from the vertex operators:
\begin{align}
V^{(-\frac{1}{2})}_{\Theta(\mathrm{sp})} &\ \ \sim \ \ \Theta_{I}^{a} \, \Big[ \, (\slashed k_{a\dot{b}} \, \bar{\sigma}_{\mu}^{\dot{b}b} \; + \; 4k_{\mu}\, \delta_{a}^{b}) \, i\partial X^{\mu} \, S_{b} \, \Sigma^{I} \ - \ 2\sqrt{2} \, \big(\alpha' \, k^{\mu} \, \slashed k_{a\dot{b}} \; + \; \tfrac{1}{4} \, \sigma^\mu_{ a\dot{b}})\, S_{\mu}^{\dot{b}} \, \Sigma^{I}\nonumber \\
 & \qquad+ \ 6 \, \partial S_{a} \, \Sigma^{I} \ + \ 4 \, S_{a} \, \partial\Sigma^{I}\ + \ \slashed k_{a\dot{b}} \, \gamma_{m}^{I\bar{J}} \, i\partial Z^{m} \, S^{\dot{b}} \, \bar{\Sigma}_{\bar{I}} \, \Big] \, \ee^{-{\phi}/{2}} \, \ee^{ik\cdot X} , \\
V^{(-\frac{1}{2})}_{\bar\Theta(\mathrm{sp})} &\ \ \sim \ \ \bar{\Theta}_{\dot{b}}^{\bar{I}} \, \Big[ \, (\slashed k^{\dot{b}a} \,\sigma^\mu_{ a\dot{a}} \; + \; 4k^{\mu} \,\delta_{\dot{a}}^{\dot{b}}) \, i\partial X_{\mu} \, S^{\dot{a}} \, \bar{\Sigma}_{\bar{I}} \ - \ 2\sqrt{2} \, \big(\alpha' \, k_{\mu} \, \slashed k^{\dot{b}a} \; + \; \tfrac{1}{4} \, \bar{\sigma}_{\mu}^{\dot{b}a} \big) \, S_{a}^{\mu} \, \bar{\Sigma}_{\bar{I}}\nonumber \\
 & \qquad+ \ 6 \, \partial S^{\dot{b}} \, \bar{\Sigma}_{\bar{I}} \ + \ 4 \, S^{\dot{b}} \, \partial\bar{\Sigma}_{\bar{I}} \ + \ \slashed k^{\dot{b}a} \, \gamma_{\bar{I}J}^{m} \, i\partial Z_{m} \, S_{a} \, \Sigma^{J} \, \Big] \, \ee^{-{\phi}/{2}} \, \ee^{ik\cdot X} .
\end{align}
They are the dimensional reduction of the ten-dimensional spurious state (\ref{4,15}).

\subsection{SUSY transformations}

Now with all the higher order OPEs and physical spectrum in hands, we are able to compute the SUSY
transformations by acting with the supercharge operators on the physical
states and evaluating the corresponding contour integral.

\medskip
In $\mathcal{N}=4$ SUSY, the SUSY parameters $\eta_{I}^{a},\bar{\eta}_{\dot{a}}^{\bar{I}}$
are chiral spinors of both the $SO(1,3)$ Lorentz group and the internal
$SO(6)$ R-symmetry group. For our convenience, we choose these SUSY
parameters to have mass dimension $[M^{-\frac{1}{2}}]$. As we verify
case by case, action of the supercharges ${\cal Q}_a^I$ and $\bar {\cal Q}^{\dot b}_{\bar J}$ given by (\ref{3,1}) and (\ref{3,2a}), (\ref{3,2b}) takes bosonic (fermionic)
vertex operators exactly into fermionic (bosonic) vertex operators, including their couplings. The polarization wavefunction of the ${\cal Q}$ image state is expressed in terms of $\eta_{I}^{a},\bar{\eta}_{\dot{a}}^{\bar{I}}$ and the pre-image wavefunction.\footnote{In our settings, all the wavefunctions of bosonic fields have mass
dimension 0, and all the wavefunction of fermionic fields have mass
dimension $\frac{1}{2}$, see appendix \ref{app:hel} for their explicit construction in terms of (massive) spinor helicity variables.}

\medskip
Once we perform the SUSY variations, besides
physical fields in the spectrum, we will also get certain spurious states.
As an example, let us consider the anti-supercharge acting on the spin-$\frac{3}{2}$ fermionic
state $\chi_{\mu,I}^{a}$. Evaluating the contour integral yields
\begin{align*}
\big[\bar{\eta}_{\dot{a}}^{\bar{I}}\bar{\mathcal{Q}}_{\bar{I}}^{(-\frac{1}{2}),\dot{a}},V_{\chi}^{(-\frac{1}{2})}\big] & =V_{\alpha}^{(-1)}\Big(\alpha_{\mu\nu}=\frac{1}{\sqrt{2}}\bar{\eta}_{\dot{a}}^{\bar{I}}\big(\bar{\sigma}_{(\mu}^{\dot{a}a}\chi_{\nu),a,I}+\alpha'\slashed k^{\dot{a}a}k_{(\mu}\chi_{\nu),a,I}\big)C_{\bar{I}}^{I}\Big)\\
 & \,+V_{d}^{(-1)}\Big(d_{\mu}^{[mn]}=-\frac{\sqrt{\alpha'}}{4}\bar{\eta}_{\dot{a}}^{\bar{I}}\slashed k^{\dot{a}a}\chi_{\mu,a,J}(\gamma^{[m}\gamma^{n]}C)_{\bar{I}}^{J}\Big)\\
 & \,+V_{\pi(\te{sp})}^{(-1)}\Big(\pi_{\mu}=-\frac{\sqrt{\alpha'}}{4}\bar{\eta}_{\dot{a}}^{\bar{I}}\slashed k^{\dot{a}a}\chi_{\mu,a,I}C_{\bar{I}}^{I}\Big)\\
 & \,+V_{\Sigma(\te{sp})}^{(-1)}\Big(\Sigma_{[\mu\nu]}=-\frac{1}{\sqrt{2}}\bar{\eta}_{\dot{a}}^{\bar{I}}\big(\bar{\sigma}_{[\mu}^{\dot{a}a}\chi_{\nu],a,I}-\alpha'\slashed k^{\dot{a}a}k_{[\mu}\chi_{\nu],a,I}\big)C_{\bar{I}}^{I}\Big) \ .
\end{align*}
As we can see, we obtain two physical states -- a spin two boson $\alpha_{\mu\nu}$
and a vector $d_{\mu}^{[mn]}$ from (\ref{12,3a}) and (\ref{12,3c}), plus two spurious states -- $\pi_{\mu} $
and $\Sigma_{[\mu\nu]} $, see (\ref{12,4a}) and (\ref{12,4b}) for their full vertex operators. We will drop out all these spurious
states in our final results for simplicity.

\medskip
All the physical states form one big supermultiplet of $\mathcal{N}=4$. The structure of the explicit SUSY variations listed in this section is summarized in figure \ref{covN=4} below. This diagram will be refined in section \ref{sec:hel} to take helicity quantum numbers into account.

\begin{figure}[h]
\centerline{
\tikzpicture [xscale=1.05, yscale=1.5,line width=0.30mm]
\draw (-8,-1) node{$\Phi^+$} ;
\draw (-7,-1) node{$\leftrightarrow$} ;
\draw (-6,-1) node{$\bar a^{\bar J}_{\dot b}$} ;
\draw (-5,-1) node{$\leftrightarrow$} ;
\draw (-4,-0.8) node{$\be^{-,m}_\mu $} ;
\draw (-4,-1.2) node{$\Om^{mnp}_-$} ;
\draw (-3,-1) node{$\leftrightarrow$} ;
\draw (-2,-0.8) node{$\chi^a_{\mu,I}$} ;
\draw (-2,-1.2) node{$ \bar r_{\dot b,m}^{\bar J}$} ;
\draw (-1,-1) node{$\leftrightarrow$} ;
\draw (0,-0.6) node{$\al_{\mu  \nu} $} ;
\draw (0,-1) node{$d^{mn}_\mu$} ;
\draw (0,-1.4) node{$ \zeta^{mn}$} ;
\draw (1,-1) node{$\leftrightarrow$} ;
\draw (2,-0.8) node{$\bar \chi_{\dot b}^{\mu,\bar J} $} ;
\draw (2,-1.2) node{$r^{a,m}_{I}$} ;
\draw (3,-1) node{$\leftrightarrow$} ;
\draw (4,-0.8) node{$\be^{+,m}_\mu $} ;
\draw (4,-1.2) node{$ \Om_+^{mnp}$} ;
\draw (5,-1) node{$\leftrightarrow$} ;
\draw (6,-1) node{$a^b_{I}$} ;
\draw (7,-1) node{$\leftrightarrow$} ;
\draw (8,-1) node{$\Phi^-$} ;
\endtikzpicture
}
\caption{${\cal N}=4$ SUSY multiplet: action of the left-handed SUSY charge ${\cal Q}_a^I$ transforms a state into (a combination of) its left neighbors, whereas $\bar {\cal Q}_{\bar J}^{\dot b}$ action maps states into right neighbors.}
\label{covN=4}
\end{figure}

The pattern of SUSY variations depicted in figure \ref{covN=4} justifies the complex combinations (\ref{12,3d}) of vectors and (\ref{12,4a}) of scalars: The complex conjugates appear on widely separated positions of the multiplet (i.e. the $\be^{+}$ and $\be^{-}$ are separated by four ${\cal Q}$ actions whereas $\Phi^{+} \leftrightarrow \Phi^-$ requires eight supercharge applications). Also, the internal scalar $\Omega^{mnp}$ splits into self-dual and anti-self-dual components $\Omega^{mnp}_{\pm}$ which sit at different points of the multiplet.

\medskip
There are group theoretic selection rules for the possible outcome of a physical state's SUSY variations, based on the $SO(1,3) \times SO(6)$ symmetry. Firstly, according to its eigenvalue under diagonal Lorentz currents, ${\cal Q}$ can only change the spin by $\pm \frac{1}{2}$. Secondly, transformations have to compatible with the $SO(6)$ quantum numbers involved. Representation of the $SO(6) \equiv SU(4)$ R-symmetry group are referred to using their Dynkin Labels $[k,p,q]$.\footnote{
Our conventions for the Dynkin labels $[k,p,q]$ are such that $[1,0,0]$ labels the vector representation, and $[0,1,0]$ and $[0,0,1]$ are left- and right-handed spinor. A generic representation with labels $[k,p,q]$ has dimension
\begin{equation}
D_{[k,p,q]}=\frac{1}{12}(k+p+q+3)(k+p+2)(k+q+2)(k+1)(p+1)(q+1) \ ,
\end{equation}
and tensor products act as follows on Dynkin labels:
\begin{align}
[k,p,q] \otimes [0,1,0]
 & =[k,p,q-1]\oplus[k,p+1,q]\oplus[k+1,p-1,q]\oplus[k-1,p,q+1],\\
[k,p,q] \otimes  [0,0,1]
 & =[k,p,q+1]\oplus[k,p-1,q]\oplus[k+1,p,q-1]\oplus[k-1,p+1,q],\\
[k,p,q] \otimes [1,0,0]
 & =[k,p+1,q-1]\oplus[k,p-1,q+1]\oplus[k+1,p,q]\nonumber \\
 & \quad \oplus[k+1,p-1,q-1]\oplus[k-1,p,q]\oplus[k-1,p+1,q+1].
\end{align}
} The SUSY variation of a state $\in [k,p,q]$ aligns into the tensor product with $[0,1,0] \ni {\cal Q}^I$ or $[0,0,1] \ni \bar {\cal Q}_{\bar J}$ of the SUSY charge. Table~\ref{N=4R} gives an overview of the R-symmetry representations involved (see the following subsection for the $\Om^{\pm}$ splitting).
\begin{table}
  \small
  \centering
  \begin{tabular}{|c|c|c||c|c|c|}\hline  Spin &Wavefunctions &$SO(6)$ rep.  &Spin &Wavefunctions &$SO(6)$ rep. \\ \hline \hline
  2 &$\al_{\mu \nu}$  &[0,0,0]  &3/2 &$\chi_{\mu,I}^a$ &[0,1,0]\\\hline
1 &$\be^{\pm,m}_\mu$ &[1,0,0]  &3/2 &$\bar \chi^{\mu,\bar J}_{\dot b}$ &[0,0,1] \\\hline
1 &$d^{[mn]}_\mu$ &[0,1,1]  &1/2 &$r_{m,I}^a$ &[1,1,0] \\\hline
0 &$\zeta^{(mn)}$ &[2,0,0]  &1/2 &$\bar r^{m,\bar J}_{\dot b}$ &[1,0,1] \\\hline
0 &$\Om^+_{mnl}$ &[0,2,0]  &1/2 &$a_{I}^b$ &[0,1,0] \\\hline
0 &$\Om^-_{mnl}$ &[0,0,2]  &1/2 &$\bar a_{\dot b}^{\bar J}$ &[0,0,1] \\\hline
0 &$\Phi^{\pm} $&[0,0,0] & & & \\\hline
\end{tabular}
\caption{\small R-symmetry content of the massive ${\cal N}=4$ multiplet in $SO(6)$ Dynkin label notation}
\label{N=4R}
\end{table}

\subsubsection{SUSY transformation of bosonic states}

In this subsubsection, we will analyze supercharge acting on the bosonic
states. The spin two field $\alpha_{\mu\nu}$ transforms into left- and right-handed spin-$\frac{3}{2}$ fermions $\mathcal{Q}\alpha\rightarrow\chi$ and $\bar{\mathcal{Q}}\alpha\rightarrow\bar{\chi}$ in lines with $[0,0,0]\otimes[0,1,0]\rightarrow[0,1,0]$ for the R-symmetry scalar $\al_{\mu \nu}$. The SUSY variations of this field are parallel to (\ref{D=10,1}) in ten dimensions:
\begin{align}
\big[\eta_{I}^{a}\mathcal{Q}_{a}^{(+\frac{1}{2}),I},V_{\alpha}^{(-1)}\big] & =V_{\chi}^{(-\frac{1}{2})}\Big(\chi_{\mu,I}^{b}=\frac{1}{\sqrt{2}}\eta_{I}^{a}\alpha_{\mu\nu}(\slashed k\bar{\sigma}^{\nu})_{a}^{\phantom{a}b}\Big),\\
\big[\bar{\eta}_{\dot{a}}^{\bar{I}}\bar{\mathcal{Q}}_{\bar{I}}^{(+\frac{1}{2}),\dot{a}},V_{\alpha}^{(-1)}\big] & =V_{\bar{\chi}}^{(-\frac{1}{2})}\Big(\bar{\chi}_{\mu,\dot{a}}^{\bar{I}}=\frac{1}{\sqrt{2}}\bar{\eta}_{\dot{a}}^{\bar{I}}\alpha_{\mu\nu}(\slashed k\sigma^{\nu})_{\phantom{a}\dot{b}}^{\dot{a}}\Big).
\end{align}
The spin one fields fall into vector and two-form representations $[1,0,0]$ and $[0,1,1]$ of the R-symmetry, so their SUSY image belongs to $[0,1,0]\otimes[1,0,0]\rightarrow[1,1,0]\oplus[0,0,1]$ and $[0,1,0]\otimes[0,1,1]\rightarrow[0,1,0]\oplus[0,2,1]\oplus[1,0,1]$, respectively (note that $[0,2,1]$ does not occur in our multiplet). This implies that $\beta_{\mu}^{\pm,m}$ can transform into an internal left-handed fermion $r_{m,I}^{a} \in [1,1,0]$, and right-handed spin-$\frac{3}{2}$ fermions
$\bar{\chi}_{\mu\dot{a}}^{\bar{I}}$ or a spin-$\frac{1}{2}$ fermions
$\bar{a}_{\dot{b}}^{\bar{I}}$, in short: $\mathcal{Q}\beta^{\pm}\rightarrow\bar{\chi}+\bar{a}+r$. For the $SO(6)$ two-form $d^{[mn]}$, we will get the opposite chirality configuration,
$\mathcal{Q}d\rightarrow\chi+a+\bar{r}$. The explicit results for the left-handed ${\cal Q}_a^I$ are given as follows,\footnote{There is a subtlety in these computations (and also for some later ones) related to the fact that gamma matrices associated with spacetime and internal dimensions are anticommuting.}
\begin{align}
\big[\eta_{I}^{b}\mathcal{Q}_{b}^{(+\frac{1}{2}),I},V_{\beta^{+}}^{(-1)}\big] & =V_{\bar{\chi}}^{(-\frac{1}{2})}\Big(\bar{\chi}_{\mu,\dot{b}}^{\bar{I}}=\frac{1}{3\sqrt{2}}\eta_{I}^{b}\big[3\beta_{\mu}^{+,m}\slashed k_{b\dot{b}}-k_{\mu}\slashed\beta_{b\dot{b}}^{+,m}-(\slashed\beta^{+,m}\slashed k\sigma_{\mu})_{b\dot{b}}\big]\gamma_{m}^{I\bar{I}}\Big)\nonumber \\
 & \,+V_{r}^{(-\frac{1}{2})}\Big(r_{n,J}^{c}=-\frac{1}{6\sqrt{2}}\eta_{I}^{b}(\slashed\beta^{+,m}\slashed k)_{b}^{\phantom{b}c}\big[6\delta_{mn}^{(6)}\delta_{J}^{I}+(\gamma_{m}\bar\gamma_{n})_{\phantom{I}J}^{I}\big]\Big),\\
\big[\eta_{I}^{b}\mathcal{Q}_{b}^{(+\frac{1}{2}),I},V_{\beta^{-}}^{(-1)}\big] & =V_{\bar{a}}^{(-\frac{1}{2})}\Big(\bar{a}_{\dot{b}}^{\bar{I}}=-\frac{1}{2\sqrt{\alpha'}}\eta_{I}^{b}\slashed\beta_{b\dot{b}}^{-,m}\gamma_{m}^{I\bar{I}}\Big),\\
\big[\eta_{I}^{b}\mathcal{Q}_{b}^{(+\frac{1}{2}),I},V_{d}^{(-1)}\big] & =V_{\chi}^{(-\frac{1}{2})}\Big(\chi_{\mu,J}^{c}=\frac{1}{6\sqrt{\alpha'}}\eta_{I}^{b}\big[3d_{\mu}^{mn}\delta_{b}^{\phantom{b}c}+(\slashed d^{mn}\bar{\sigma}_{\mu}+\alpha'k_{\mu}\slashed d^{mn}\slashed k)_{b}^{\phantom{b}c}\big](\gamma_{m}\bar\gamma_{n})_{\phantom{I}J}^{I}\Big)\nonumber \\
 & \,+V_{\bar{r}}^{(-\frac{1}{2})}\Big(\bar{r}_{l,\dot{b}}^{\bar{I}}=\frac{1}{6\sqrt{\alpha'}}\eta_{I}^{b}\slashed d_{b\dot{b}}^{mn}\gamma_{n}^{I\bar{I}}\big[6\delta_{ml}^{(6)}\delta_{\bar{I}}^{\phantom{I}\bar{J}}+(\bar\gamma_{m}\gamma_{l})_{\bar{I}}^{\phantom{I}\bar{J}}\big]\Big),
\end{align}
whereas the action of right-handed $\bar {\cal Q}^{\dot b}_{\bar J}$ yields
\begin{align}
\big[\bar{\eta}_{\dot{b}}^{\bar{I}}\bar{\mathcal{Q}}_{\bar{I}}^{(+\frac{1}{2}),\dot{b}},V_{\beta^{+}}^{(-1)}\big] & =V_{a}^{(-\frac{1}{2})}\Big(a_{I}^{b}=-\frac{1}{2\sqrt{\alpha'}}\bar{\eta}_{\dot{b}}^{\bar{I}}\slashed\beta^{+,m,\dot{b}b}\bar\gamma_{m,\bar{I}I}\Big),\\
\big[\bar{\eta}_{\dot{b}}^{\bar{I}}\bar{\mathcal{Q}}_{\bar{I}}^{(+\frac{1}{2}),\dot{b}},V_{\beta^{-}}^{(-1)}\big] & =V_{\chi}^{(-\frac{1}{2})}\Big(\chi_{\mu,I}^{b}=\frac{1}{3\sqrt{2}}\bar{\eta}_{\dot{b}}^{\bar{I}}\big[3\beta_{\mu}^{-,m}\slashed k^{\dot{b}b}-k_{\mu}\slashed\beta^{-,m,\dot{b}b}-(\slashed\beta^{-,m}\slashed k\bar{\sigma}_{\mu})^{\dot{b}b}\big]\bar\gamma_{m,\bar{I}I}\Big)\nonumber \\
 & \,+V_{\bar{r}}^{(-\frac{1}{2})}\Big(\bar{r}_{n,\dot{c}}^{\bar{J}}=-\frac{1}{6\sqrt{2}}\bar{\eta}_{\dot{b}}^{\bar{I}}(\slashed\beta^{-,m}\slashed k)_{\phantom{b}\dot{c}}^{\dot{b}}\big[6\delta_{mn}^{(6)}\delta_{\bar{I}}^{\phantom{I}\bar{J}}+(\bar\gamma_{m}\gamma_{n})_{\bar{I}}^{\phantom{I}\bar{J}}\big]\Big),\\
\big[\bar{\eta}_{\dot{b}}^{\bar{I}}\bar{\mathcal{Q}}_{\bar{I}}^{(+\frac{1}{2}),\dot{b}},V_{d}^{(-1)}\big] & =V_{\bar{\chi}}^{(-\frac{1}{2})}\Big(\bar{\chi}_{\mu,\dot{c}}^{\bar{J}}=\frac{1}{6\sqrt{\alpha'}}\bar{\eta}_{\dot{b}}^{\bar{I}}\big[3d_{\mu}^{mn}\delta_{\phantom{b}\dot{c}}^{\dot{b}}+(\slashed d^{mn}\sigma_{\mu}+\alpha'k_{\mu}\slashed d\slashed k)_{\phantom{b}\dot{c}}^{\dot{b}}\big](\bar\gamma_{m}\gamma_{n})_{\bar{I}}^{\phantom{I}\bar{J}}\Big)\nonumber \\
 & \,+V_{r}^{(-\frac{1}{2})}\Big(r_{J,p}^{b}=\frac{1}{6\sqrt{\alpha'}}\bar{\eta}_{\dot{b}}^{\bar{I}}\slashed d^{mn,\dot{b}b}\bar\gamma_{n,\bar{I}I}\big[6\delta_{mp}^{(6)}\delta_{\phantom{I}J}^{I}+(\gamma_{m}\bar\gamma_{p})_{\phantom{I}J}^{I}\big]\Big).
\end{align}
Then we are left with the $SO(1,3)$ scalar fields $\Phi^{\pm}$, $\zeta^{(mn)}$
and $\Omega_{mnl}$. The internal states $\Omega_{mnl}$ represent
both self-dual and anti-self-dual three-forms of $SO(6)$. We will denote their irreducible
components as $\Omega_{mnl}^{+} \in [0,2,0]$ and $\Omega_{mnl}^{-} \in [0,0,2]$, for the self-dual and anti-self-dual part, respectively. Their defining irreducibility constraint is
\beq
\Omega_{mnl}^{-}(\gamma^{mnl})_{I\bar{I}} \eq \Omega_{mnl}^{+}(\bar\gamma^{mnl})^{\bar{I}I} \eq 0 \ .
\eeq
The $SO(6)$ selection rules constrain ${\cal Q}^I \zeta^{(mn)}\in [0,1,0]\otimes[2,0,0]\rightarrow[2,1,0]\oplus[1,0,1]$ as well as ${\cal Q}^I \Om^+_{mnl} \in [0,1,0]\otimes[0,2,0]\rightarrow[0,3,0]\oplus[1,1,0]$ and ${\cal Q}^I \Om^-_{mnl}  \in [0,1,0]\otimes[0,0,2]\rightarrow[0,1,2]\oplus[0,0,1]$. Thus, we expect the internal spin-$\frac{1}{2}$ fermion
$\bar{r}$ or $r$ by performing the SUSY transformation $\mathcal{Q}\zeta\rightarrow\bar{r}$,
and $\bar{\mathcal{Q}}\zeta\rightarrow r$. Three-forms, on the other hand, are mapped to either $r$ or $\bar{a}$, depending on the self-duality property $\mathcal{Q}\Omega^{+}\rightarrow r$ or $\mathcal{Q}\Omega^{-}\rightarrow\bar{a}$. The supercharges acting on $\Phi^{\pm}$ and $\zeta^{(mn)}$ yield
\begin{gather}
\big[\eta_{I}^{b}\mathcal{Q}_{b}^{(+\frac{1}{2}),I},V_{\Phi^{+}}^{(-1)}\big]=0,\qquad\big[\bar{\eta}_{\dot{b}}^{\bar{I}}\bar{\mathcal{Q}}_{\bar{I}}^{(+\frac{1}{2}),\dot{b}},V_{\Phi^{-}}^{(-1)}\big]=0,\label{N4mulVac}\\
\big[\bar{\eta}_{\dot{b}}^{\bar{I}}\bar{\mathcal{Q}}_{\bar{I}}^{(+\frac{1}{2}),\dot{b}},V_{\Phi^{+}}^{(-1)}\big]=V_{\bar{a}}^{(-\frac{1}{2})}\Big(\bar{a}_{\dot{b}}^{\bar{I}}=-\alpha'^{-\frac{1}{2}}\Phi^{+}\bar{\eta}_{\dot{b}}^{\bar{I}}\Big),\\
\big[\eta_{I}^{b}\mathcal{Q}_{b}^{(+\frac{1}{2}),I},V_{\Phi^{-}}^{(-1)}\big]=V_{a}^{(-\frac{1}{2})}\Big(a_{I}^{b}=-\alpha'^{-\frac{1}{2}}\Phi^{-}\eta_{I}^{b}\Big),
\end{gather}
and
\begin{align}
\big[\eta_{I}^{b}\mathcal{Q}_{b}^{(+\frac{1}{2}),I},V_{\zeta}^{(-1)}\big] & =V_{\bar{r}}^{(-\frac{1}{2})}\Big(\bar{r}_{\dot{b}}^{m,\bar{I}}=\frac{1}{\sqrt{2}}\eta_{I}^{b}\zeta^{(mn)}\slashed k_{b\dot{b}}\gamma_{n}^{I\bar{I}}\Big),\\
\big[\bar{\eta}_{\dot{b}}^{\bar{I}}\bar{\mathcal{Q}}_{\bar{I}}^{(+\frac{1}{2}),\dot{b}},V_{\zeta}^{(-1)}\big] & =V_{r}^{(-\frac{1}{2})}\Big(r_{I}^{m,b}=\frac{1}{\sqrt{2}}\bar{\eta}_{\dot{b}}^{\bar{I}}\zeta^{(mn)}\slashed k^{\dot{b}b}\bar\gamma_{n,\bar{I}I}\Big).
\end{align}
On the three-forms $\Om^{\pm}_{mnl}$, we obtain
\begin{align}
\big[\eta_{I}^{b}\mathcal{Q}_{b}^{(+\frac{1}{2}),I},V_{\Omega^{+}}^{(-1)}\big] & =V_{r}^{(-\frac{1}{2})}\Big(r_{k,J}^{b}=-\frac{1}{4\sqrt{\alpha'}}\eta_{I}^{b}\Omega_{mnl}^{+}(\gamma_{k}\bar\gamma^{mnl})_{\phantom{I}J}^{I}\Big),\\
\big[\bar{\eta}_{\dot{b}}^{\bar{I}}\bar{\mathcal{Q}}_{\bar{I}}^{(+\frac{1}{2}),\dot{b}},V_{\Omega^{+}}^{(-1)}\big] & =V_{a}^{(-\frac{1}{2})}\Big(a_{I}^{b}=\frac{1}{2\sqrt{2}}\bar{\eta}_{\dot{b}}^{\bar{I}}\Omega_{mnl}^{+}\slashed k^{\dot{b}b}(\bar\gamma^{mnl})_{\bar{I}I}\Big),
\end{align}
and
\begin{align}
\big[\eta_{I}^{b}\mathcal{Q}_{b}^{(+\frac{1}{2}),I},V_{\Omega^{-}}^{(-1)}\big] & =V_{\bar{a}}^{(-\frac{1}{2})}\Big(\bar{a}_{\dot{b}}^{\bar{I}}=\frac{1}{2\sqrt{2}}\eta_{I}^{b}\Omega_{mnl}^{-}\slashed k_{b\dot{b}}(\gamma^{mnl})^{I\bar{I}}\Big),\\
\big[\bar{\eta}_{\dot{b}}^{\bar{I}}\bar{\mathcal{Q}}_{\bar{I}}^{(+\frac{1}{2}),\dot{b}},V_{\Omega^{-}}^{(-1)}\big] & =V_{\bar{r}}^{(-\frac{1}{2})}\Big(\bar{r}_{k,\dot{b}}^{\bar{J}}=-\frac{1}{4\sqrt{\alpha'}}\bar{\eta}_{\dot{b}}^{\bar{I}}\Omega_{mnl}^{-}(\bar\gamma_{k}\gamma^{mnl})_{\bar{I}}^{\phantom{I}\bar{J}}\Big).
\end{align}

\subsubsection{SUSY transformation of fermionic states}

In this subsubsection, we investigate the (anti-)supercharge acting on the
fermionic states. Following the strategy outlined before, we first derive a selection rule from group theory and then perform SUSY variations to get the expression
of the bosonic wavefunctions explicitly. All the transformations are symmetric under simultaneous exchange of chiralities on the supercharges and the states (where $\Phi^{+},\beta^{+},\Om^{+} \leftrightarrow \Phi^{-},\beta^{-},\Om^{-}$). We will only comment on one out of two inequivalent cases in the text but also give the formulae for the images under chirality reversal.

\medskip
Since both the spin-$\frac{3}{2}$ fermions $(\chi,\bar \chi)$ and the spin-$\frac{1}{2}$ states $(a,\bar a)$ fall into (anti-)fundamental R-symmetry representations, the $SO(6)$ content of their SUSY variation is $[0,1,0]\otimes[0,1,0]\rightarrow[0,2,0]\oplus[1,0,0]$ and $[0,0,1]\otimes[0,1,0]\rightarrow[0,0,0]\oplus[0,1,1]$. The (anti-)supercharge acting on $\chi_{\mu,I}^{a} (\bar{\chi}_{\mu,\dot{a}}^{\bar{I}})$ will give us vectors $\beta_{\mu}^{\pm, m}$. In the cases ${\cal Q}^I \bar{\chi}_{\mu,\dot{a}}^{\bar{I}}$ and $\bar {\cal Q}_{\bar J} \chi_{\mu,I}^{a}$ of opposite chirality, the spin two field $\alpha_{\mu\nu}$ and the vector $d_{\mu}^{[mn]}$ can emerge. Indeed,
\begin{align}
\big[\eta_{I}^{a}\mathcal{Q}_{a}^{(-\frac{1}{2}),I},V_{\chi}^{(-\frac{1}{2})}\big] & =V_{\beta^{-}}^{(-1)}\Big(\beta_{\mu}^{-,m}=\frac{1}{\sqrt{2}}\eta_{I}^{a}\chi_{\mu,a,J}(\gamma^{m}C)^{IJ}\Big),\\
\big[\bar{\eta}_{\dot{a}}^{\bar{I}}\bar{\mathcal{Q}}_{\bar{I}}^{(-\frac{1}{2}),\dot{a}},V_{\chi}^{(-\frac{1}{2})}\big] & =V_{\alpha}^{(-1)}\Big(\alpha_{\mu\nu}=\frac{1}{\sqrt{2}}\bar{\eta}_{\dot{a}}^{\bar{I}}\big(\bar{\sigma}_{(\mu}^{\dot{a}a}\chi_{\nu),a,I}+\alpha'\slashed k^{\dot{a}a}k_{(\mu}\chi_{\nu),a,I}\big)C_{\bar{I}}^{I}\Big)\nonumber \\
 & \,+V_{d}^{(-1)}\Big(d_{\mu}^{[mn]}=-\frac{\sqrt{\alpha'}}{4}\bar{\eta}_{\dot{a}}^{\bar{I}}\slashed k^{\dot{a}a}\chi_{\mu,a,I}(\bar\gamma^{m n}C)_{\bar{I}}^{\phantom{I}I}\Big),
\end{align}
and\footnote{The notation $M_{\mu_{1}\mu_{2}\cdots(\mu_{i}\cdots\mu_{j-1}|\mu_{j}\cdots\mu_{k}|\mu_{k+1}\cdots\mu_{l})\cdots\mu_{n}}$
indicates we symmetrize over the indices $\mu_{i},\cdots,\mu_{j-1},\mu_{k+1},\cdots,\mu_{l}$, but not over the indices $\mu_{j},\ldots \mu_k$ enclosed between the bars.}
\begin{align}
\big[\eta_{I}^{a}\mathcal{Q}_{a}^{(-\frac{1}{2}),I},V_{\bar{\chi}}^{(-\frac{1}{2})}\big] & =V_{\alpha}^{(-1)}\Big(\alpha_{\mu\nu}=\frac{1}{\sqrt{2}}\eta_{I}^{a}\big(\sigma_{(\mu|a\dot{a}|}\bar{\chi}_{\nu)}^{\dot{a},\bar{I}}+\alpha'\slashed k_{a\dot{a}}k_{(\mu}\bar{\chi}_{\nu)}^{\dot{a},\bar{I}}\big)C_{\bar{I}}^{I}\Big)\nonumber \\
 & \,+V_{d}^{(-1)}\Big(d_{\mu}^{[mn]}=-\frac{\sqrt{\alpha'}}{4}\eta_{I}^{a}\slashed k_{a\dot{a}}\bar{\chi}_{\mu}^{\dot{a},\bar{I}}(\gamma^{m n}C)_{\phantom{I}\bar{I}}^{I}\Big),\\
\big[\bar{\eta}_{\dot{a}}^{\bar{I}}\bar{\mathcal{Q}}_{\bar{I}}^{(-\frac{1}{2}),\dot{a}},V_{\bar{\chi}}^{(-\frac{1}{2})}\big] & =V_{\beta^{+}}^{(-1)}\Big(\beta_{\mu}^{+,m}=\frac{1}{\sqrt{2}}\bar{\eta}_{\dot{a}}^{\bar{I}}\bar{\chi}_{\mu}^{\dot{a},\bar{J}}(\bar\gamma^{m}C)_{\bar{I}\bar{J}}\Big).
\end{align}
The supercharge action on $a_{I}^{b}$ and $\bar{a}_{\dot{b}}^{\bar{I}}$
follows the same selection rules with respect to $SO(6)$ but different ones with respect to spacetime spin. The corresponding SUSY transformations
read
\begin{align}
\big[\eta_{I}^{b}\mathcal{Q}_{b}^{(-\frac{1}{2}),I},V_{a}^{(-\frac{1}{2})}\big] & =V_{\beta^{+}}^{(-1)}\Big(\beta_{\mu}^{+,m}=\frac{\sqrt{\alpha'}}{2}\eta_{I}^{b}\big[k_{\mu}\delta_{b}^{\phantom{b}c}+(\slashed k\bar{\sigma}_{\mu})_{b}^{\phantom{b}c}\big]a_{c,J}(\gamma^{m}C)^{IJ}\Big)\nonumber \\
 & \,+V_{\Omega^{+}}^{(-1)}\Big(\Omega_{mnl}^{+}=\frac{1}{12\sqrt{2}}\eta_{I}^{b}a_{b,J}(\gamma_{mnl}C)^{IJ}\Big),\\
\big[\bar{\eta}_{\dot{b}}^{\bar{I}}\bar{\mathcal{Q}}_{\bar{I}}^{(-\frac{1}{2}),\dot{b}},V_{a}^{(-\frac{1}{2})}\big] & =V_{\Phi^{-}}^{(-1)}\Big(\Phi^-=\sqrt{\alpha'}\bar{\eta}_{\dot{b}}^{\bar{I}}\slashed k^{\dot{b}b}a_{b,I}C_{\bar{I}}^{I}\Big),
\end{align}
and
\begin{align}
\big[\eta_{I}^{b}\mathcal{Q}_{b}^{(-\frac{1}{2}),I},V_{\bar{a}}^{(-\frac{1}{2})}\big] & =V_{\Phi^{+}}^{(-1)}\Big(\Phi^+=\sqrt{\alpha'}\eta_{I}^{b}\slashed k_{b\dot{b}}\bar{a}^{\dot{b},\bar{I}}C_{\bar{I}}^{I}\Big),\\
\big[\bar{\eta}_{\dot{b}}^{\bar{I}}\bar{\mathcal{Q}}_{\bar{I}}^{(-\frac{1}{2}),\dot{b}},V_{\bar{a}}^{(-\frac{1}{2})}\big] & =V_{\beta^{-}}^{(-1)}\Big(\beta_{\mu}^{-,m}=\frac{\sqrt{\alpha'}}{2}\bar{\eta}_{\dot{b}}^{\bar{I}}\big[k_{\mu}\delta_{\phantom{b}\dot{c}}^{\dot{b}}+(\slashed k\sigma_{\mu})_{\phantom{b}\dot{c}}^{\dot{b}}\big]\bar{a}^{\dot{c},\bar{J}}(\bar\gamma^{m}C)_{\bar{I}\bar{J}}\Big)\nonumber \\
 & \,+V_{\Omega^{-}}^{(-1)}\Big(\Omega_{mnl}^{-}=\frac{1}{12\sqrt{2}}\bar{\eta}_{\dot{b}}^{\bar{I}}\bar{a}^{\dot{b},\bar{J}}(\bar\gamma_{mnl}C)_{\bar{I}\bar{J}}\Big).
\end{align}
Notice we do not get a vector $d_{\mu}^{[mn]}$ in the SUSY transformations, although it is allowed by $SO(1,3) \times SO(6)$.

\medskip
Now we are left with the internal spin-$\frac{1}{2}$ fermions $r$
and $\bar{r}$. Group theory admits SUSY variations in $[0,1,0]\otimes[1,1,0]\rightarrow[1,2,0]\oplus[2,0,0]\oplus[0,1,1]$ and $[0,0,1]\otimes[1,1,0]\rightarrow[1,1,1]\oplus[1,0,0]\oplus[0,2,0]$ corresponding to vectors $d_{\mu}^{[mn]}$ and internal scalars $\zeta^{(mn)}$ in the former case and $\bar{\mathcal{Q}}r\rightarrow \beta^{\pm} +\Omega^{+}$ in the latter. The left-handed supercharge yields
\begin{align}
\big[\eta_{I}^{a}\mathcal{Q}_{a}^{(-\frac{1}{2}),I},V_{r}^{(-\frac{1}{2})}\big] & =V_{d}^{(-1)}\Big(d_{\mu}^{[mn]}=\frac{\sqrt{\alpha'}}{2}\eta_{I}^{a}\big[k_{\mu}\delta_{a}^{\phantom{a}b}+(\sigma_{\mu}\slashed k)_{a}^{\phantom{a}b}\big]r_{b,J}^{[m}(\gamma^{n]}C)^{IJ}\Big)\nonumber \\
 & \,+V_{\zeta}^{(-1)}\Big(\zeta^{(mn)}=\frac{1}{\sqrt{2}}\eta_{I}^{a}r_{a,J}^{(m}(\gamma^{n)}C)^{IJ}\Big),\\
\big[\bar{\eta}_{\dot{a}}^{\bar{I}}\bar{\mathcal{Q}}_{\bar{I}}^{(-\frac{1}{2}),\dot{a}},V_{r}^{(-\frac{1}{2})}\big] & =V_{\beta^{+}}^{(-1)}\Big(\beta_{\mu}^{+,m}=\frac{1}{\sqrt{2}}\bar{\eta}_{\dot{a}}^{\bar{I}}\big(\bar{\sigma}_{\mu}^{\dot{a}a}+\alpha'k_{\mu}\slashed k^{\dot{a}a}\big)r_{a,I}^{m}C_{\bar{I}}^{I}\Big)\nonumber \\
 & \,+V_{\Omega_{+}}^{(-1)}\Big(\Omega_{+}^{mnl}=-\frac{\sqrt{\alpha'}}{4}\bar{\eta}_{\dot{a}}^{\bar{I}}\slashed k^{\dot{a}a}r_{a,I}^{[m}(\bar\gamma^{nl]}C)_{\bar{I}}^{\phantom{I}I}\Big),
\end{align}
and the right-handed counterpart reads
\begin{align}
\big[\eta_{I}^{a}\mathcal{Q}_{a}^{(-\frac{1}{2}),I},V_{\bar{r}}^{(-\frac{1}{2})}\big] & =V_{\beta^{-}}^{(-1)}\Big(\beta_{\mu}^{-,m}=\frac{1}{\sqrt{2}}\eta_{I}^{a}\big(\sigma_{\mu a\dot{a}}+\alpha'k_{\mu}\slashed k_{a\dot{a}}\big)\bar{r}_{\bar{I}}^{m,\dot{a}}C_{I}^{\bar{I}}\Big)\nonumber \\
 & \,+V_{\Omega_{-}}^{(-1)}\Big(\Omega_{-}^{mnl}=-\frac{\sqrt{\alpha'}}{4}\eta_{I}^{a}\slashed k_{a\dot{a}}\bar{r}^{[m|,\dot{a},\bar{I}|}(\gamma^{n l]}C)_{\phantom{I}\bar{I}}^{I}\Big),\\
\big[\bar{\eta}_{\dot{a}}^{\bar{I}}\bar{\mathcal{Q}}_{\bar{I}}^{(-\frac{1}{2}),\dot{a}},V_{\bar{r}}^{(-\frac{1}{2})}\big] & =V_{d}^{(-1)}\Big(d_{\mu}^{[mn]}=\frac{\sqrt{\alpha'}}{2}\bar{\eta}_{\dot{a}}^{\bar{I}}\big[k_{\mu}\delta_{\phantom{a}\dot{b}}^{\dot{a}}+(\bar{\sigma}_{\mu}\slashed k)_{\phantom{a}\dot{b}}^{\dot{a}}\big]\bar{r}^{[m|,\dot{b},\bar{J}|}(\bar\gamma^{n]}C)_{\bar{I}\bar{J}}\Big)\nonumber \\
 & \,+V_{\zeta}^{(-1)}\Big(\zeta^{(mn)}=\frac{1}{\sqrt{2}}\bar{\eta}_{\dot{a}}^{\bar{I}}\bar{r}^{(m|,\dot{a},\bar{J}|}(\bar\gamma^{n)}C)_{\bar{I}\bar{J}}\Big).
\end{align}
This completes the list of SUSY transformations within the $\mathcal{N}=4$
multiplet. We will revisit these results from the spinor helicity viewpoint in section \ref{sec:hel}.


\section{Massive supermultiplets for ${\cal N}=1$ SUSY}
\label{sec:n1}

This section is devoted to the universal SUSY multiplets common to all $D=4$ superstring compactifications which preserve ${\cal N}=1$ spacetime SUSY. It was already observed in \cite{Feng:2010yx} that 24 universal states exist, and the reference also investigates their three- and four-point couplings to massless states. We will show that they gather in three multiplets: one spin two representation of 8+8 states and two spin 1/2 representations of 2+2 states each. The first subsections review the construction of these states and the third one contains their SUSY variations.

\subsection{NS sector}

By assembling $h=3/2$ combinations of the conformal fields of figure \ref{N=1}, one arrives at the following general form of an NS state at mass $m^2 = 1/\ap$:
\begin{align}
V^{(-1)} \eq & \Big( \, \al_{\mu \nu} \, i \pa X^\mu \, \psi^\nu \ + \ e_{\mu \nu \la} \, \psi^\mu \, \psi^\nu \, \psi^\la \ + \ h_\mu \, \pa \psi^\mu \ + \ \xi_\mu \, \psi^\mu \, {\cal J} \notag \\
& \ + \ \Om_+ \, {\cal O}^+ \ + \ \Om_- \, {\cal O}^- \ + \ c_+ \, G_{\te{int}}^+ \ + \ c_- \, G_{\te{int}}^- \, \Big) \, \ee^{-\phi} \, \ee^{ik \cdot X} \ .
\label{8,1}
\end{align}
This is BRST invariant if the polarization tensors satisfy
\beq
\begin{array}{ll}
0 \eq \al_{\mu}{}^\mu \ + \ k^\mu \, h_\mu \ + \ \tfrac{3}{2 \sqrt{\ap}} \, (c_+ + c_-) \ \ \ \ \ \ \ \
&0 \eq  \al_{[\mu \nu]} \ + \ 3 \, e_{\mu \nu \la} \, k^\la
\\
 0 \eq k^\mu \, \xi_\mu \ + \ \tfrac{\sqrt{3}}{2 \sqrt{\ap}} \; (c_- - c_+)
 &0 \eq  2\ap \, \al_{\mu \nu} \, k^\nu \ + \  h_\mu
 \end{array}  \label{8,2} \eeq
Twelve physical states solve this system of equations:
\begin{itemize}
\item one transverse and traceless spin two tensor
\begin{align}
V^{(-1)}_\alpha &\eq \frac{1}{\sqrt{2\ap}} \; \al_{\mu \nu} \, i \pa X^\mu \, \psi^\nu \, \ee^{-\phi} \, \ee^{ik \cdot X} \co k^\mu \, \al_{\mu \nu} = \al_{[\mu \nu]} = \al_\mu{}^\mu = 0 \label{8,3}
\end{align}
\item one transverse vector
\begin{align}
V^{(-1)}_d &\eq d_\mu \, \psi^\mu \, {\cal J} \,  \ee^{-\phi} \, \ee^{ik \cdot X} \co k^\mu \, d_\mu = 0
\label{8,5} \end{align}
\item two complex scalars
\begin{align}
V^{(-1)}_{\Phi^{\pm}} &\eq \frac{\Phi^{\pm}}{2\sqrt{2\ap}} \, \Big[ \, (\eta_{\mu \nu} \, + \, 2\ap \, k_\mu \, k_\nu) \, i\pa X^\mu \, \psi^\nu \ + \ 2\ap \, k_\mu \, \pa \psi^\mu \notag \\
& \hskip6cm \pm \ \frac{i\ap}{3} \; \vep_{\mu \nu \la \rho} \, \psi^\mu \, \psi^\nu \, \psi^\la \, k^{\rho} \, \Big] \, \ee^{-\phi} \, \ee^{ik \cdot X} \ , \label{8,4} \\
V^{(-1)}_{\Om^{\pm}} &\eq \Om^{\pm} \, {\cal O}^{\pm} \, \ee^{-\phi} \, \ee^{ik \cdot X} \ .
\label{8,6}
\end{align}
\end{itemize}
In addition, we have spurious solutions to the BRST constraints:
\begin{align}
V^{(-1)}_{\pi(\te{sp})} &\ \ \sim \ \ \Big[ \, (\pi_\mu \, k_\nu \; + \; k_\mu \, \pi_\nu) \, i \pa X^\mu \, \psi^\nu \ + \ 2 \, \pi_\mu \, \pa \psi^\mu \, \Big] \, \ee^{-\phi} \, \ee^{ik \cdot X} \co k^\mu \, \pi_{\mu}  = 0 \ , \label{8,7} \\
V^{(-1)}_{\Sigma(\te{sp})} &\ \ \sim \ \ \Big[ \, 2 \, \Si_{[\mu \nu]} \, i \pa X^\mu \, \psi^\nu \ + \ 2\ap \, \Si_{[\mu \nu} \, k_{\la]} \, \psi^\mu \, \psi^\nu \, \psi^\la \, \Big] \, \ee^{-\phi} \, \ee^{ik \cdot X} \co k^\mu \, \Si_{\mu \nu}  = 0 \ , \label{8,8} \\
V^{(-1)}_{\Lambda_0(\te{sp})} & \ \ \sim \ \ \La_0 \, \Big[ \, (G^+_{\te{int}} \ - \ G^-_{\te{int}}) \ - \ \sqrt{3\ap} \, k_\mu \, \psi^\mu \, {\cal J}  \, \Big] \,  \ee^{-\phi} \, \ee^{ik \cdot X}  \ , \label{8,10} \\
V^{(-1)}_{\Lambda_1(\te{sp})} &\ \ \sim \ \ \La_1 \, \Big[ \, (\eta_{\mu \nu} \; + \; 4\ap \, k_\mu \,k_\nu) \, i \pa X^\mu \, \psi^\nu \ + \ 6\ap \, k_\mu \, \pa \psi^\mu \notag \ , \\
&\hskip6cm + \ 2\sqrt{\ap} \, (G^+_{\te{int}} \ + \ G^-_{\te{int}}) \, \Big] \, \ee^{-\phi} \, \ee^{ik \cdot X}  \ .\label{8,9}
\end{align}
The last two spurious states allow to gauge away both the $c^{\pm}$ scalars and the longitudinal component of the massive vector $\xi_\mu \sim k_\mu$.

\subsection{R sector}

For $D=4$ fermions at mass $m^2 = 1/\ap$, the most general vertex operators built from ${\cal N}=1$ internal SCFT fields reads
\begin{align}
V^{(-\frac{1}{2})} \eq & \Big( \, v_\mu^a \, i \pa X^\mu \, S_a \, \Si^+ \ + \ \bar \rho^\mu_{\dot b} \, S_\mu^{\dot b} \, \Si^+ \ + \ u^a \, \pa S_a \, \Si^+ \notag \\
& \ + \ y^a \, S_a \, \pa \Si^+ \ + \ \bar \om_{\dot b} \, S^{\dot b} \, \tilde \Si^+ \, \Big) \, \ee^{-\phi/2} \, \ee^{ik\cdot X}\ ,
\label{8,11}
\end{align}
see figure \ref{N=1}. Invariance under $Q_1$ yields three independent BRST constraints:
\beq \begin{array}{l}
0 \eq 2 \ap \, v^{\mu,a} \not \! k_{a \dot b} \ + \ \sqrt{2} \, \bar \rho^\mu_{ \dot b} \ + \ \tfrac{1}{2} \, u^a \, \si^\mu_{a \dot b} \ ,\\
0\eq  k_\mu \, \bar \rho^\mu_{\dot b} \ + \ \tfrac{1}{2} \, \sqrt{\tfrac{3}{2\ap} }\, \bar \om_{\dot b} \ ,  \\
0\eq y^a \ + \ 2\,\sqrt{\tfrac{\ap}{3}} \, \bar \om_{\dot b} \not \! k^{\dot b a} \ .
\end{array}
 \label{8,13}
\eeq
They allow to express any wavefunction in terms of $v_\mu^a$
\begin{align}
u^a &\eq \ap \, v_\mu^b \, (\! \not \! k \, \sib^\mu)_b{}^a
\notag \\
\bar \rho_{\mu \dot b} &\eq -\, \sqrt{2} \, \ap \, \Bigl( \, v_\mu^a \not \! k_{a \dot b} \ + \ \tfrac{1}{4} \, v_\la^a \, (\! \not \! k \, \sib^\la \, \si_\mu)_{a \dot b} \, \Bigr)
\label{8,13b} \\
\bar \om_{\dot b}&\eq \sqrt{\tfrac{ \ap}{3}} \; \Bigl( \, 2\ap \, k^\mu \, v_\mu^a \not \! k_{a \dot b} \ - \ v_\mu^a \, \si^{\mu}_{a \dot b} \, \Bigr) \notag \\
y^a &\eq \frac{2\ap}{3} \, \Big( \, v_\mu^b \, (\si^\mu \!  \not \! k)_b{}^a \ - \ 2\, k^\mu \, v_\mu^a \, \Big) \notag \ .
\end{align}
The same set of states exists with opposite $SO(1,3)$ chirality and internal $U(1)$ charge. Including them, we have four physical solutions to (\ref{8,13b}) and four solutions to the conjugate system of equations:
\begin{itemize}
\item two transverse and $\si$ traceless spin 3/2 vector spinors
\begin{align}
V^{(-\frac{1}{2})}_\chi &\eq \frac{ 1 }{ \sqrt{2} \ap^{1/4}} \;  \chi_\mu^a \, \Big( \, i \pa X^\mu \, S_a \ - \ \sqrt{2} \, \ap \not \! k_{a\dot b} \, S^{\mu \dot b} \, \Big)  \, \Si^+\, \ee^{-\phi/2} \, \ee^{ik\cdot X} \ , \label{8,14} \\
V^{(-\frac{1}{2})}_{\bar \chi} &\eq  \frac{ 1 }{\sqrt{2} \ap^{1/4}}\; \bar \chi^\mu_{\dot a} \, \Big( \, i \pa X_\mu \, S^{\dot a} \ - \ \sqrt{2} \, \ap \not \! k^{\dot a b} \, S_{\mu b} \, \Big) \, \Si^-\, \ee^{-\phi/2} \, \ee^{ik\cdot X} \ , \label{8,17}  \\
0 &\eq k^\mu \, \chi_\mu^a \eq \chi^a_\mu \, \si^\mu_{a \dot b} \eq k_\mu \, \bar \chi^\mu_{\dot a} \eq \bar \chi_{\dot a}^\mu \, \sib_\mu^{\dot a b}
\end{align}
\item two spin 1/2 fermions
\begin{align}
V^{(-\frac{1}{2})}_a &\eq \frac{\alpha'^{1/4}}{2} \; a^b \, \Big( \, (\si_\mu \not \! k)_b{}^a \, S_a \, i \pa X^\mu \ - \ 4 \, \pa S_b \, \Big) \,  \Si^+\, \ee^{-\phi/2} \, \ee^{ik\cdot X} \ ,
\label{8,15}\\
V^{(-\frac{1}{2})}_{\bar a} &\eq \frac{\alpha'^{1/4}}{2} \; \bar a_{\dot b} \, \Big( \, (\sib_\mu \not \! k)^{\dot b}{}_{\dot a} \, S^{\dot a} \, i \pa X^\mu \ - \ 4 \, \pa S^{\dot b} \, \Big) \,  \Si^-\, \ee^{-\phi/2} \, \ee^{ik\cdot X} \ .
\label{8,18} \end{align}
\end{itemize}
Spurious solutions can gauge away the internal excitations with wavefunctions $y^a$ and $\bar \om_{\dot b}$:
\begin{align}
V^{(-\frac{1}{2})}_{\Theta(\mathrm{sp})} &\ \ \sim \ \ \Theta^{a} \, \Big[ \, (\slashed k_{a\dot{a}} \, \bar{\sigma}_{\mu}^{\dot{a}b} \; + \; 4k_{\mu} \, \delta_{a}^{b})\, i\partial X^{\mu} \, S_{b} \, \Sigma^{+} \ - \ 2\sqrt{2} \,\big(\alpha' \, k^{\mu} \, \slashed k_{a\dot{b}} \; + \; \tfrac{1}{4} \,\sigma^\mu_{ a\dot{b}} \big) \, S_\mu^{\dot{b}} \, \Sigma^{+}\nonumber \\
 &\ \ \ \ \ \ \ \ \qquad+ \ 6 \, \partial S_{a} \, \Sigma^{+} \  + \ 4 \, S_{a} \, \partial\Sigma^{+} \ - \ 2\sqrt{3\alpha'} \, \slashed k_{a\dot{b}} \, S^{\dot{b}} \, \tilde{\Sigma}^{+} \, \Big] \, \ee^{-\phi/2} \, \ee^{ik\cdot X} ,
\label{8,16} \\
V^{(-\frac{1}{2})}_{\bar\Theta(\mathrm{sp})} &\ \ \sim \ \ \bar{\Theta}_{\dot{b}} \, \Big[ \, (\slashed k^{\dot{b}a} \, \sigma^{\mu}_{ a\dot{a}} \; + \; 4k^{\mu} \, \delta_{\dot{a}}^{\dot{b}}) \, i\partial X_{\mu} \, S^{\dot{a}} \, \Sigma^{-} \ - \ 2\sqrt{2}\, \big(\alpha' \, k_{\mu} \, \slashed k^{\dot{b}a} \; + \; \tfrac{1}{4}\, \bar{\sigma}_{\mu}^{\dot{b}a} \big) \, S_{a}^{\mu} \, \Sigma^{-}\nonumber \\
 &\ \ \ \ \ \ \ \  \qquad+ \ 6 \, \partial S^{\dot{b}} \, \Sigma^{-} \ + \ 4 \, S^{\dot{b}} \, \partial\Sigma^{-} \ - \ 2\sqrt{3\alpha'} \, \slashed k^{\dot{b}a} \, S_{a} \, \tilde{\Sigma}^{-} \, \Big] \, \ee^{-\phi/2} \, \ee^{ik \cdot X} .
\label{8,19}
\end{align}

\subsection{SUSY transformations}

The notation for the $\mathcal{N}=1$ multiplets can be kept lighter because of the Abelian R-symmetry group $U(1)$. The supercharge operators do not carry any R-symmetry indices, only an Abelian charge of $\pm \sqrt{3}/2$. After performing SUSY variation
on all the bosonic and fermionic states in $\mathcal{N}=1$ SUSY,
we find that these states split into three separate massive supermultiplets
-- a spin two multiplet $\{\alpha,\chi,\bar{\chi},d\}$, two spin-$\frac{1}{2}$
multiplets $\{\Phi^{+},\bar{a},\Omega^{-}\}$ and $\{\Omega^{+},a,\Phi^{-}\}$,
see figure \ref{covN=1} below. We will show our results of the SUSY transformations in order.
\begin{figure}[h]
\centerline{
\tikzpicture [scale=1.3,line width=0.30mm]
\draw (-2,0) node{$\Om^+$} ;
\draw (-1,0) node{$\longleftrightarrow$} ;
\draw (0,0) node{$a^b$} ;
\draw (1,0) node{$\longleftrightarrow$} ;
\draw (2,0) node{$\Phi^-$} ;
\draw (-2.5,-1) node{$ \chi_\mu^{a}$} ;
\draw (-1.5,-1) node{$\longleftrightarrow$} ;
\draw (0,-1) node{$\al_{\mu \nu} \, \oplus \, d_\mu$} ;
\draw (1.5,-1) node{$\longleftrightarrow$} ;
\draw (2.5,-1) node{$\bar \chi^\mu_{\dot b}$} ;
\draw (-2,-2) node{$\Phi^+$} ;
\draw (-1,-2) node{$\longleftrightarrow$} ;
\draw (0,-2) node{$\bar a_{\dot b}$} ;
\draw (1,-2) node{$\longleftrightarrow$} ;
\draw (2,-2) node{$\Om^-$} ;
\endtikzpicture
}
\caption{The three disconnected ${\cal N}=1$ SUSY multiplets at the first mass level: As before, ${\cal Q}_a$ ($\bar {\cal Q}^{\dot b}$) action takes states along a left (right) arrow.}
\label{covN=1}
\end{figure}

\subsubsection{SUSY variation of the spin two supermultiplet}

The spin two multiplet includes a spin two boson $\alpha_{\mu\nu}$, a
vector $d_{\mu}$, and two spin-$\frac{3}{2}$ fermions $\chi_{\mu}^{a}$,
$\bar{\chi}_{\mu,\dot{a}}$ with opposite chirality. The SUSY transformation of the bosonic states are:
\begin{align}
\big[\eta^{a}\mathcal{Q}_{a}^{(+\frac{1}{2})},V_{\alpha}^{(-1)}\big] & =V_{\chi}^{(-\frac{1}{2})}\Big(\chi_{\mu}^{b}=\frac{1}{\sqrt{2}}\eta^{a}\alpha_{\mu\nu}(\slashed k\bar{\sigma}^{\nu})_{a}^{\phantom{a}b}\Big),\\
\big[\bar{\eta}_{\dot{a}}\bar{\mathcal{Q}}^{(+\frac{1}{2}),\dot{a}},V_{\alpha}^{(-1)}\big] & =V_{\bar{\chi}}^{(-\frac{1}{2})}\Big(\bar{\chi}_{\mu,\dot{b}}=\frac{1}{\sqrt{2}}\bar{\eta}_{\dot{a}}\alpha_{\mu\nu}(\slashed k\sigma^{\nu})_{\phantom{a}\dot{b}}^{\dot{a}}\Big),\\
\big[\eta^{a}\mathcal{Q}_{a}^{(+\frac{1}{2})},V_{d}^{(-1)}\big] & =V_{\chi}^{(-\frac{1}{2})}\Big(\chi_{\mu}^{b}=\frac{-1}{2\sqrt{3\alpha'}}\eta^{a}\big[3d_{\mu}\delta_{a}^{\phantom{a}b}+(\slashed d\bar{\sigma}_{\mu}+\alpha'k_{\mu}\slashed d\slashed k)_{a}^{\phantom{a}b}\big]\Big), \label{sgn1a} \\
\big[\bar{\eta}_{\dot{a}}\bar{\mathcal{Q}}^{(+\frac{1}{2}),\dot{a}},V_{d}^{(-1)}\big] & =V_{\bar{\chi}}^{(-\frac{1}{2})}\Big(\bar{\chi}_{\mu,\dot{b}}=\frac{1}{2\sqrt{3\alpha'}}\bar{\eta}_{\dot{a}}\big[3d_{\mu}\delta_{\phantom{a}\dot{b}}^{\dot{a}}+(\slashed d\sigma_{\mu}+\alpha'k_{\mu}\slashed d\slashed k)_{\phantom{a}\dot{b}}^{\dot{a}}\big]\Big). \label{sgn1b}
\end{align}
The SUSY transformation of the fermionic states are:
\begin{align}
\big[\eta^{a}\mathcal{Q}_{a}^{(-\frac{1}{2})},V_{\chi}^{(-\frac{1}{2})}\big] & =0,\label{N1spin2mulVac}\\
\big[\bar{\eta}_{\dot{a}}\bar{\mathcal{Q}}^{(-\frac{1}{2}),\dot{a}},V_{\chi}^{(-\frac{1}{2})}\big] & =V_{\alpha}^{(-1)}\Big(\alpha_{\mu\nu}=\frac{1}{\sqrt{2}}\bar{\eta}_{\dot{a}}\big(\bar{\sigma}_{(\mu}^{\dot{a}a}\chi_{\nu)a}+\alpha'\slashed k^{\dot{a}a}k_{(\mu}\chi_{\nu),a}\big)\Big)\nonumber \\
 & \,+V_{d}^{(-1)}\Big(d_{\mu}=\frac{\sqrt{3\alpha'}}{2}\bar{\eta}_{\dot{a}}\slashed k^{\dot{a}a}\chi_{\mu,a}\Big), \label{sgn2a} \\
\big[\eta^{a}\mathcal{Q}_{a}^{(-\frac{1}{2})},V_{\bar{\chi}}^{(-\frac{1}{2})}\big] & =V_{\alpha}^{(-1)}\Big(\alpha_{\mu\nu}=\frac{1}{\sqrt{2}}\eta^{a}\big(\sigma_{(\mu|a\dot{a}|}\bar{\chi}_{\nu)}^{\dot{a}}+\alpha'\slashed k_{a\dot{a}}k_{(\mu}\bar{\chi}_{\nu)}^{\dot{a}}\big)\Big)\nonumber \\
 & \,+V_{d}^{(-1)}\Big(d_{\mu}=-\frac{\sqrt{3\alpha'}}{2}\eta^{a}\slashed k_{a\dot{a}}\bar{\chi}_{\mu}^{\dot{a}}\Big), \label{sgn2b} \\
\big[\bar{\eta}_{\dot{a}}\bar{\mathcal{Q}}^{(-\frac{1}{2}),\dot{a}},V_{\bar{\chi}}^{(-\frac{1}{2})}\big] & =0.
\end{align}
Note that the signs of the SUSY transformations between spin-$\frac{3}{2}$ and spin one are sensitive to the chirality, see the relative signs between (\ref{sgn1a}) and (\ref{sgn1b}) as well as (\ref{sgn2a}) and (\ref{sgn2b}). This is necessary for consistent closure of the SUSY algebra and can be neatly represented by a chirality matrix $\gamma^5$ when passing to Dirac spinor notation.

\subsubsection{SUSY variation of the spin 1/2 supermultiplets}

The first spin-$\frac{1}{2}$ multiplet $\{\Phi^{+},\bar{a},\Omega^{-}\}$
includes a right-handed spin-$\frac{1}{2}$ fermion $\bar{a}$ and two
scalars $\Phi^{+}, \Omega^{-}$. It is governed by
the following SUSY transformations:
\begin{align}
\big[\eta^{b}\mathcal{Q}_{b}^{(+\frac{1}{2})},V_{\Phi^{+}}^{(-1)}\big] & =0,\label{N1spin12mulVac11}\\
\big[\bar{\eta}_{\dot{b}}\bar{\mathcal{Q}}^{(+\frac{1}{2}),\dot{b}},V_{\Phi^{+}}^{(-1)}\big] & =V_{\bar{a}}^{(-\frac{1}{2})}\Big(\bar{a}_{\dot{b}}=-\alpha'^{-\frac{1}{2}}\Phi^{+}\bar{\eta}_{\dot{b}}\Big),\\
\big[\eta^{b}\mathcal{Q}_{b}^{(+\frac{1}{2})},V_{\Omega^{-}}^{(-1)}\big] & =V_{\bar{a}}^{(-\frac{1}{2})}\Big(\bar{a}_{\dot{b}}=\Omega^{-}\eta^{b}\slashed k_{b\dot{b}}\Big),\\
\big[\bar{\eta}_{\dot{b}}\bar{\mathcal{Q}}^{(+\frac{1}{2}),\dot{b}},V_{\Omega^{-}}^{(-1)}\big] & =0,\label{N1spin12mulVac12}
\end{align}
and
\begin{align}
\big[\eta^{b}\mathcal{Q}_{b}^{(-\frac{1}{2})},V_{\bar{a}}^{(-\frac{1}{2})}\big] & =V_{\Phi^{+}}^{(-1)}\Big(\Phi^{+}=\sqrt{\alpha'}\eta^{b}\slashed k_{b\dot{b}}\bar{a}^{\dot{b}}\Big),\\
\big[\bar{\eta}_{\dot{b}}\bar{\mathcal{Q}}^{(-\frac{1}{2}),\dot{b}},V_{\bar{a}}^{(-\frac{1}{2})}\big] & =V_{\Omega^{-}}^{(-1)}\Big(\Omega^{-}=\bar{\eta}_{\dot{b}}\bar{a}^{\dot{b}}\Big).
\end{align}
For $\{\Omega^{+},a,\Phi^{-}\}$ multiplet of opposite R-symmetry charges and fermion chirality, we obtain
\begin{align}
\big[\eta^{b}\mathcal{Q}_{b}^{(+\frac{1}{2})},V_{\Phi^{-}}^{(-1)}\big] & =V_{a}^{(-\frac{1}{2})}\Big(a^{b}=-\alpha'^{-\frac{1}{2}}\Phi^{-}\eta^{b}\Big),\\
\big[\bar{\eta}_{\dot{b}}\bar{\mathcal{Q}}^{(+\frac{1}{2}),\dot{b}},V_{\Phi^{-}}^{(-1)}\big] & =0,\label{N1spin12mulVac22}\\
\big[\eta^{b}\mathcal{Q}_{b}^{(+\frac{1}{2})},V_{\Omega^{+}}^{(-1)}\big] & =0,\label{N1spin12mulVac21}\\
\big[\bar{\eta}_{\dot{b}}\bar{\mathcal{Q}}^{(+\frac{1}{2}),\dot{b}},V_{\Omega^{+}}^{(-1)}\big] & =V_{a}^{(-\frac{1}{2})}\Big(a^{b}=\Omega^{+}\bar{\eta}_{\dot{b}}\slashed k^{\dot{b}b}\Big),
\end{align}
and
\begin{align}
\big[\eta^{b}\mathcal{Q}_{b}^{(-\frac{1}{2})},V_{a}^{(-\frac{1}{2})}\big] & =V_{\Omega^{+}}^{(-1)}\Big(\Omega^{+}=\eta^{b}a_{b}\Big),\\
\big[\bar{\eta}_{\dot{b}}\bar{\mathcal{Q}}^{(-\frac{1}{2}),\dot{b}},V_{a}^{(-\frac{1}{2})}\big] & =V_{\Phi^{-}}^{(-1)}\Big(\Phi^{-}=\sqrt{\alpha'}\bar{\eta}_{\dot{b}}\slashed k^{\dot{b}b}a_{b}\Big).
\end{align}
We will explore the helicity structure of these results in section \ref{sec:hel}.

\section{Massive supermultiplets for ${\cal N}=2$ SUSY}
\label{sec:n2}

In this section, we will show that the first mass level in compactifications with ${\cal N}=2$ spacetime SUSY is populated by 80 universal states which are aligned into one 24+24 state multiplet of highest spin two and two 8+8 state multiplets of maximum spin one.

\subsection{NS sector}

According to the CFT operator content shown in figure \ref{N=2}, we make the following general ansatz for an NS state at the first mass level:\footnote{Recall that we have non-Abelian R-symmetry $SU(2)$ in this setting, and $i,j=1,2$ denote its spinor indices whereas $A=1,2,3$ are adjoint indices.}
\begin{align}
V^{(-1)} \eq \Big( \, &\al_{\mu \nu} \, i \pa X^\mu \, \psi^\nu \ + \ e_{\mu \nu \la} \, \psi^\mu \, \psi^\nu \, \psi^\la \ + \ h_\mu \, \pa \psi^\mu \ + \ Y_+ \, \pa \ee^{iH} \ + \ Y_{-} \, \pa \ee^{-iH} \notag \\
+ \ &\be^+_\mu \, i \pa X^\mu \, \ee^{iH} \ + \ \be^{-}_\mu \, i \pa X^\mu \,\ee^{-iH} \ + \ \ga_\mu^+ \, \psi^\mu \, i \pa Z^+ \ + \ \ga_\mu^{-} \, \psi^\mu \, i \pa Z^- \notag \\
+ \ &\xi_\mu \, \psi^\mu \, i \pa H \ + \ d_\mu^A \, {\cal J}_A \, \psi^\mu \ + \ \Om_+^A \, {\cal J}_A \, \ee^{iH} \ + \ \Om_{-}^A \, {\cal J}_A \, \ee^{-iH}\notag \\
+ \ &\zeta_{++} \, i \pa Z^+ \, \ee^{iH} \ + \ \zeta_{--} \, i \pa Z^- \, \ee^{-iH} \ + \ \zeta_{-+} \, i \pa Z^- \, \ee^{iH} \ + \ \zeta_{+-} \, i \pa Z^+ \, \ee^{-iH} \notag \\
+ \ &\om_{\mu \nu}^+ \, \psi^\mu \, \psi^\nu \, \ee^{iH} \ + \ \om_{\mu \nu}^{-} \, \psi^\mu \, \psi^\nu \, \ee^{-iH} \ + \ c_i{}^j \, \la^i \, g_j \, \Big) \, \ee^{-\phi} \, \ee^{ik \cdot X} \ .
\label{13,1}
\end{align}
Requiring BRST invariance under $Q_1$ yields the following on-shell conditions:
\beq \begin{array}{ll}
0\eq \al_{\mu}{}^\mu \ + \ k^\mu \, h_\mu \ + \ \zeta_{+-} \ + \ \zeta_{-+} \ - \ \ap^{-1/2} \, c_i{}^i  & \\
0\eq 2\ap \, \al_{\mu \nu} \, k^\nu \ + \  h_\mu &0\eq k^\mu \, d^A_\mu \ + \ \tfrac{1}{\sqrt{2\ap}} \, (\tau^A)^i{}_j \, c_i{}^j \\
0\eq \al_{[\mu \nu]} \ + \ 3 \, e_{\mu \nu \la} \, k^\la &0\eq Y_{\pm} \ + \ 2\ap \, \ga^{\pm}_\mu \, k^\mu  \\
0\eq \be_\mu^{\pm} \ - \ \ga_\mu^{\pm} \ + \ 2 \, k^\nu \, \om^{\pm}_{\nu \mu}  &0 \eq k^\mu \, \xi_\mu \ + \ \zeta_{-+} \ - \ \zeta_{+-}
\end{array} \label{13,2} \eeq
These BRST constraints admit 40 physical solutions:
\begin{itemize}
\item one transverse and traceless spin two tensor
\begin{align}
V^{(-1)}_\alpha &\eq \frac{1}{\sqrt{2\ap}} \; \al_{\mu \nu} \, i \pa X^\mu \, \psi^\nu \, \ee^{-\phi} \, \ee^{ik \cdot X} \co k^\mu \, \al_{\mu \nu} = \al_{[\mu \nu]} = \al_\mu{}^\mu = 0 \label{13,3}
\end{align}
\item eight transverse vectors three of which form an R-symmetry triplet (note the sign difference in the pseudovector parts of $\be^{\pm}$ and $\om^{\pm}$)
\begin{align}
V^{(-1)}_\xi &\eq \xi_\mu \, \psi^\mu \, i \pa H \,  \ee^{-\phi} \, \ee^{ik \cdot X} \co k^\mu \, \xi_\mu = 0
\label{13,4} \\
V^{(-1)}_d &\eq d^A_\mu \, \psi^\mu \, {\cal J}_A \,  \ee^{-\phi} \, \ee^{ik \cdot X} \co k^\mu \, d^A_\mu = 0
\label{13,5} \\
V^{(-1)}_{\beta^{\pm}} &\eq \frac{1}{2 \sqrt{2\ap}} \; \be^{\pm}_{\mu} \, \bigl( \, i \pa X^\mu \, \ee^{\pm iH} \ + \ i \pa Z^{\pm} \, \psi^\mu \ \notag \\
& \hskip3.5cm
\pm \ i \ap \, \vep^{\mu \nu \la \rho} \, k_\nu \, \psi_\la \, \psi_\rho \, \ee^{\pm iH}
\, \big) \, \ee^{-\phi} \, \ee^{ik \cdot X} \co k^\mu \, \be_{\mu}^{\pm} =  0 \label{13,6} \\
V^{(-1)}_{\omega^{\pm}} &\eq \frac{1}{2 \sqrt{2\ap}} \; \om^{\pm}_{\mu} \, \bigl( \, i \pa X^\mu \, \ee^{\pm iH} \ + \ i \pa Z^{\pm} \, \psi^\mu \notag \\
& \hskip3.5cm  \mp \ i \ap \, \vep^{\mu \nu \la \rho} \, k_\nu \, \psi_\la \, \psi_\rho \, \ee^{\pm iH}
\, \big) \, \ee^{-\phi} \, \ee^{ik \cdot X} \co k^\mu \, \om_{\mu}^{\pm} =  0 \label{13,8}
\end{align}
\item eleven real scalar degrees of freedom
\begin{align}
V^{(-1)}_{\Phi^{\pm}} &\eq \frac{\Phi^{\pm}}{2\sqrt{2\ap}} \, \Big[ \, (\eta_{\mu \nu} \, + \, 2\ap \, k_\mu \, k_\nu) \, i\pa X^\mu \, \psi^\nu \ + \ 2\ap \, k_\mu \, \pa \psi^\mu \notag \\
& \hskip6cm \pm \ \frac{i\ap}{3} \; \vep_{\mu \nu \la \rho} \, \psi^\mu \, \psi^\nu \, \psi^\la \, k^{\rho} \, \Big] \, \ee^{-\phi} \, \ee^{ik \cdot X} \ , \label{13,10} \\
V^{(-1)}_\phi &\eq \frac{1}{ \sqrt{6\ap}} \; \phi \, \Big[ \, i \pa Z^+ \,  \ee^{-iH} \ + \ i \pa Z^- \, \ee^{iH} \ + \ \sqrt{\ap} \, G^i{}_i \, \Big] \, \ee^{-\phi} \, \ee^{ik \cdot X} \ , \label{13,15} \\
V^{(-1)}_{\Om^{\pm}} &\eq \Om^{\pm}_A \, \ee^{\pm iH} \, {\cal J}^A \, \ee^{-\phi} \, \ee^{ik \cdot X} \ ,
\label{13,11} \\
V^{(-1)}_{\zeta^{\pm}} &\eq \frac{1}{ \sqrt{2\ap}} \; \zeta^{\pm} \, i \pa Z^{\pm} \, \ee^{\pm iH}  \, \ee^{-\phi} \, \ee^{ik \cdot X} \ .
\label{13,13}
\end{align}
\end{itemize}
In addition, we have numerous spurious states:
\begin{align}
V^{(-1)}_{\pi(\te{sp})} &\ \ \sim \ \ \Big[ \, (\pi_\mu \, k_\nu \; + \; k_\mu \, \pi_\nu) \, i \pa X^\mu \, \psi^\nu \ + \ 2 \, \pi_\mu \, \pa \psi^\mu \, \Big] \, \ee^{-\phi} \, \ee^{ik \cdot X} \co k^\mu \, \pi_{\mu}  = 0 \ , \label{13,16} \\
V^{(-1)}_{\Sigma(\te{sp})} &\ \ \sim \ \ \Big[ \, 2 \, \Si_{[\mu \nu]} \, i \pa X^\mu \, \psi^\nu \ + \ 2\ap \, \Si_{[\mu \nu} \, k_{\la]} \, \psi^\mu \, \psi^\nu \, \psi^\la \, \Big] \, \ee^{-\phi} \, \ee^{ik \cdot X} \co k^\mu \, \Si_{\mu \nu}  = 0 \ , \label{13,17} \\
V^{(-1)}_{\Lambda_0(\te{sp})} & \ \ \sim \ \ \La_0^A \, \Big[ \, k_\mu\, \psi^\mu \, {\cal J}_A \ + \ \tfrac{1}{\sqrt{2\ap}} \, (\tau_A)^j{}_i \, G^i{}_j  \, \Big] \,  \ee^{-\phi} \, \ee^{ik \cdot X} \ , \label{13,19}  \\
V^{(-1)}_{\Lambda_1(\te{sp})} &\ \ \sim \ \ \La_1 \, \Big[ \, (\eta_{\mu \nu} \; + \; 4\ap \, k_\mu \,k_\nu) \, i \pa X^\mu \, \psi^\nu \ + \ 6\ap \, k_\mu \, \pa \psi^\mu \notag \\
& \ \ \ \ \ \ \ \ \ \ \ \  \ \ \ \ \ \ \  \ \ \ \ \ \ \  \ \ \ \ \  + \ i \pa Z^+ \, \ee^{-iH} \ + \ i \pa Z^- \, \ee^{iH} \ - \ 2 \sqrt{\ap} \, G^i{}_i \, \Big] \, \ee^{-\phi} \, \ee^{ik \cdot X} \ , \label{13,18} \\
V^{(-1)}_{\Lambda_2^{\pm}(\te{sp})} & \ \ \sim \ \ \La_2^{\pm} \, \Big[ \, k_\mu\, (i \pa Z^{\pm} \, \psi^\mu \, + \, i \pa X^\mu \, \ee^{\pm iH}) \ + \ 2 \, \pa \ee^{\pm iH}  \, \Big] \,  \ee^{-\phi} \, \ee^{ik \cdot X} \ , \label{13,20}  \\
V^{(-1)}_{\Lambda_3(\te{sp})} & \ \ \sim \ \ \La_3 \, \Big[ \, 2\ap \, k_\mu \, \psi^\mu \, i\pa H \ + \ i \pa Z^- \, \ee^{iH} \ - \ i \pa Z^+ \, \ee^{-iH} \, \Big] \,  \ee^{-\phi} \, \ee^{ik \cdot X} \ , \label{13,24}  \\
V^{(-1)}_{\Lambda_4^{\pm}(\te{sp})} & \ \ \sim \ \ \La^{\pm}_{4 \, \mu} \, \Big[ \, 2\ap \, k_\nu \, \psi^\nu \, \psi^\mu \, \ee^{\pm iH} \ + \ i \pa X^\mu \, \ee^{\pm iH} \ - \ i \pa Z^{\pm} \, \psi^\mu  \, \Big] \,  \ee^{-\phi} \, \ee^{ik \cdot X} \co k^\mu \, \La^{\pm}_{4 \, \mu} = 0 \ . \label{13,22}
\end{align}
They allow to eliminate the longitudinal components of six vectors and of the two-forms $\om_{\mu \nu}^{\pm}$. The latter therefore dualize to transverse pseudovectors entering the $\be_{\mu}^{\pm}$ and $\om_\mu^{\pm}$ states. By combining with the $\La_1$ spurious state, one can transform the $\phi$ solution into a form without internal $c=6$ supercurrents:
\begin{align}
V^{(-1)}_\phi \eq \frac{1}{ \sqrt{6\ap}} \; \phi \, \Big[ \, (\eta_{\mu \nu} \; &+ \; 4\ap \, k_\mu \,k_\nu) \, i \pa X^\mu \, \psi^\nu \ + \ 6\ap \, k_\mu \, \pa \psi^\mu \notag \\
&+ \ 3 \, (i \pa Z^+ \, \ee^{-iH} \ + \ i \pa Z^- \, \ee^{iH}) \, \Big] \, \ee^{-\phi} \, \ee^{ik \cdot X} . \label{13,25}
\end{align}

\subsection{R sector}

In the R sector of the first mass level in ${\cal N}=2$ scenarios, the vertex operator ansatz in one chirality sector includes nine SCFT operators:
\begin{align}
V^{(-\frac{1}{2})} \eq & \Big\{ \, v_{\mu i}^a \, i \pa X^\mu \, S_a \, \la^i \, \ee^{iH/2} \ + \ \bar \rho^\mu_{\dot b i} \, S^{\dot b}_\mu \, \la^i \, \ee^{iH/2} \ + \ u_i^a \, \pa S_a \, \la^i \, \ee^{iH/2} \ + \ \bar r_{+\dot b i} \, i \pa Z^+ \, S^{\dot b} \, \la^i \, \ee^{-iH/2}  \notag \\
 & \ + \ \bar r_{- \dot b i} \, i \pa  Z^- \, S^{\dot b} \, \la^i \, \ee^{-iH/2} \ + \ \om_i^a \, S_a \, \la^i \, \pa \ee^{iH/2} \ + \ y_i^a \, S_a \, \pa \la^i \, \ee^{iH/2} \notag \\
 & \ + \ \bar \ell_{\dot b i} \, S^{\dot b} \, g^i \, \ee^{iH/2} \ + \ \psi_i^a \, S_a \, \la^i \, \ee^{-3iH/2} \, \Big\} \, \ee^{-\phi/2} \, \ee^{ik\cdot X} \ .
\label{13,33}
\end{align}
The system of BRST constraints can be reduced to the following independent set:
\beq \begin{array}{l} 0 \eq 2 \, k_\mu \, \bar \rho^\mu_{\dot b i}  \ + \ \bar r_{+\dot b i} \ - \ \sqrt{ \tfrac{1}{\ap} } \, \bar \ell_{\dot b i}
\\
0 \eq 2 \ap \, v_{ i}^{\mu,a} \not \! k_{a \dot b} \ + \ \sqrt{2} \, \bar \rho^\mu_{\dot b i} \ + \ \tfrac{1}{2} \, u^a_i \, \si^\mu_{a \dot b}
\end{array} \co
\begin{array}{l}
0 \eq \om_i^a  \ + \ 2 \sqrt{2} \, \ap \, \bar r_{+\dot b i} \not \! k^{\dot b a} \\
0\eq \psi_i^a  \ + \  \sqrt{2} \, \ap \, \bar r_{-\dot b i} \not \! k^{\dot b a}  \\
0 \eq y_i^a \ - \ \sqrt{2\ap} \, \bar \ell_{\dot b i} \not \! k^{\dot ba} \end{array} \ .
\label{13,34}
\eeq
Adding a sector of opposite chirality and internal charge gives rise to 40 physical solutions. All of them transform in the fundamental representation of the $SU(2)$ R-symmetry:
\begin{itemize}
\item four transverse and $\si$ traceless spin 3/2 vector spinors
\begin{align}
V^{(-\frac{1}{2})}_\chi &\eq \frac{1}{\sqrt{2} \ap^{1/4}}  \; \chi_{\mu ,i}^a \, \Big( \, i \pa X^\mu \, S_a \ - \ \sqrt{2} \, \ap \not \! k_{a\dot b} \, S^{\mu \dot b} \, \Big) \, \la^i \, \ee^{iH/2}  \, \ee^{-\phi/2} \, \ee^{ik\cdot X} \ , \label{13,36} \\
V^{(-\frac{1}{2})}_{\bar \chi} &\eq \frac{1}{\sqrt{2} \ap^{1/4} }\; \bar \chi_{\dot b ,i}^\mu \, \Big( \, i \pa X_\mu \, S^{\dot b} \ - \ \sqrt{2} \, \ap \not \! k^{\dot b a} \, S_{\mu a} \, \Big) \, \la^i \, \ee^{-iH/2} \, \ee^{-\phi/2} \, \ee^{ik\cdot X} \ , \label{13,42} \\
0 &\eq k^\mu \, \chi_{\mu i}^a \eq \chi_{\mu i}^a \, \si^\mu_{a \dot b} \eq k_\mu \, \bar \chi^{\mu}_{\dot b i} \eq \bar \chi^\mu_{\dot b i} \, \sib^{\dot b a}_\mu \end{align}
\item six spin 1/2 fermions:
\begin{align}
V^{(-\frac{1}{2})}_a &\eq \frac{\alpha'^{1/4}}{2} \; a^b_i \, \Big( \, (\si_\mu \not \! k)_b{}^a \, S_a \, i \pa X^\mu \ - \ 4 \, \pa S_b \, \Big) \, \la^i \, \ee^{iH/2}  \, \ee^{-\phi/2} \, \ee^{ik\cdot X} \ ,
\label{13,37} \\
V^{(-\frac{1}{2})}_{\bar a} &\eq \frac{\alpha'^{1/4}}{2} \; \bar a_{\dot a ,i} \, \Big( \, (\sib_\mu \not \! k)^{\dot a}{}_{\dot b} \, S^{\dot b} \, i \pa X^\mu \ - \ 4 \, \pa S^{\dot a} \, \Big) \, \la^i \, \ee^{-iH/2} \, \ee^{-\phi/2} \, \ee^{ik\cdot X} \ ,
\label{13,43} \\
V^{(-\frac{1}{2})}_r &\eq \frac{1}{\sqrt{2} \ap^{1/4}}  \; r^a_{i} \, \Big( \, i \pa  Z^+ \, S_a \, \ee^{iH/2} \ - \ \sqrt{2} \, \alpha' \, \slashed k_{a\dot b} \, S^{\dot b} \, \ee^{3iH/2} \, \Big) \, \la^i \, \ee^{-\phi/2} \, \ee^{ik \cdot X} \ , \label{13,44}  \\
V^{(-\frac{1}{2})}_{\bar r} &\eq \frac{1}{\sqrt{2} \ap^{1/4}}  \, \bar r_{\dot b ,i} \, \Big( \, i \pa Z^- \, S^{\dot b} \, \ee^{-iH/2}  \ - \ \sqrt{2} \, \alpha' \, \slashed k^{\dot b a} \, S_a \, \ee^{-3iH/2}  \, \Big) \, \la^i \, \ee^{-\phi/2} \, \ee^{ik \cdot X} \ , \label{13,38}  \\
V^{(-\frac{1}{2})}_s &\eq \frac{1}{\sqrt{3} \ap^{1/4}}  \; s^a_i \, \Big( \, i \pa Z^- \, S_a \, \la^i \, \ee^{iH/2} \ + \ \sqrt{\ap} \, S_a \, g^i \, \ee^{-iH/2} \notag \\
& \ \ \ \ \ \ \ \ \ \ \ \ + \ \sqrt{2} \, \alpha' \, \slashed k_{a\dot b} \, \big( \, S^{\dot b} \, \pa \la^i \, \ee^{-iH/2} \ - \ 2  \, S^{\dot b}  \, \la^i \, \pa \ee^{-iH/2}  \, \big) \Big) \,  \ee^{-\phi/2} \, \ee^{ik \cdot X} \ , \label{13,45} \\
V^{(-\frac{1}{2})}_{\bar s} &\eq \frac{1}{\sqrt{3} \ap^{1/4}}  \; \bar{s}_{\dot b ,i} \, \Big( \, i \pa Z^+ \, S^{\dot b} \, \la^i \, \ee^{-iH/2}  \ + \ \sqrt{\ap}   \, S^{\dot b} \, g^i \, \ee^{iH/2}  \notag \\
& \ \ \ \ \ \ \ \ \ \ \ \ + \ \sqrt{2} \, \alpha' \, \slashed k^{\dot b a} \big( \, S_a \, \pa \la^i \, \ee^{iH/2}  \ - \ 2 \, S_a \, \la^i \, \pa \ee^{iH/2} \, \big) \Big) \,  \ee^{-\phi/2} \, \ee^{ik \cdot X} \ . \label{13,39}
\end{align}
\end{itemize}
Again, there is a spurious fermion which can be used to remove some internal SCFT fields from the vertex operators:
\begin{align}
V^{(-\frac{1}{2})}_{\Theta(\mathrm{sp})} &\ \ \sim \ \ \Theta_{i}^{a} \, \Big[ \, (\slashed k_{a\dot{b}} \, \bar{\sigma}_{\mu}^{\dot{b}b} \; + \; 4k_{\mu} \, \delta_{a}^{b}) \, i\partial X^{\mu} \, S_{b} \, \lambda^{i} \, \ee^{iH/2} \  - \ 2\sqrt{2} \, \big(\alpha' \, k^{\mu} \, \slashed k_{a\dot{b}} \; + \; \tfrac{1}{4} \, \sigma^\mu_{a\dot{b}} \big) \, S^{\dot{b}}_\mu \, \lambda^{i} \, \ee^{iH/2}\nonumber \\
 & \ \ \ \ \ \ \qquad+ \ 6 \, \partial S_{a} \, \lambda^{i} \, \ee^{iH/2} \ + \ 4 \, S_{a} \,\partial\lambda^{i} \, \ee^{iH/2} \ + \ 4 \, S_{a} \, \lambda^{i} \, \partial \ee^{iH/2}\nonumber \\
 & \ \ \ \ \ \ \qquad+ \ 2\sqrt{2\alpha'} \, \slashed k_{a\dot{b}} \, S^{\dot{b}} \, g^{i} \, \ee^{iH/2} \ - \ \sqrt{2} \, \slashed k_{a\dot{b}} \, i\partial Z^{+} \, S^{\dot{b}}\, \lambda^{i} \, \ee^{-iH/2} \, \Big] \, \ee^{-\phi/2} \, \ee^{ik \cdot X},
\label{13,200} \\
V^{(-\frac{1}{2})}_{\bar\Theta(\mathrm{sp})} &\ \ \sim \ \ \bar{\Theta}_{\dot{b},i} \, \Big[ \, (\slashed k^{\dot{b}a} \, \sigma^\mu_{ a\dot{a}} \;+ \; 4 k^{\mu} \, \delta_{\dot{a}}^{\dot{b}}) \, i\partial X_{\mu} \, S^{\dot{a}} \, \lambda^{i}\, \ee^{-iH/2} \ - \ 2\sqrt{2} \, \big(\alpha' \, k_{\mu} \, \slashed k^{\dot{b}a} \; + \; \tfrac{1}{4} \, \bar{\sigma}_{\mu}^{\dot{b}a} \big) \, S_{a}^{\mu} \, \lambda^{i} \, \ee^{-iH/2}\nonumber \\
 & \ \ \ \ \ \ \qquad+ \ 6 \, \partial S^{\dot{b}} \, \lambda^{i} \, \ee^{-iH/2} \ + \  4 \, S^{\dot{b}} \, \partial\lambda^{i} \, \ee^{-iH/2} \ + \ 4 \, S^{\dot{b}} \, \lambda^{i}\, \partial \ee^{-iH/2}\nonumber \\
 & \ \ \ \ \ \ \qquad+ \ 2\sqrt{2\alpha'} \, \slashed k^{\dot{b}a} \, S_{a} \, g^{i} \, \ee^{-iH/2} \ - \ \sqrt{2} \, \slashed k^{\dot{b}a} \, i\partial Z^{-} \, S_{a} \, \lambda^{i}\ee^{iH/2} \, \Big] \, \ee^{-\phi/2} \, \ee^{ik\cdot X}.
\label{13,201}
\end{align}

\subsection{SUSY transformations}

The charges of $\mathcal{N}=2$ SUSY are spinors of the internal $SU(2)$ R-symmetry and therefore carry an extra index
$i$. In this sector, universal states at the first mass level
split into three separate massive supermultiplets -- a spin two multiplet
$\{\alpha,\chi,\bar{\chi},d,\xi,\beta^{\pm},s,\bar{s},\phi\}$ as well as two
spin one multiplets $\{\omega^{-},\bar{a},\bar{r},\Phi^{+},\zeta^{-},\Omega_{A}^{-}\}$
and $\{\omega^{+},a,r,\Phi^{-},\zeta^{+},\Omega_{A}^{+}\}$, see figure \ref{covN=2} below for their structure.

\begin{figure}[h]
\centerline{
\tikzpicture [scale=1.3,line width=0.30mm]
\draw (-4.5,0) node{$\zeta^+$} ;
\draw (-3.5,0) node{$\longleftrightarrow$} ;
\draw (-2.5,0) node{$r^b_i$} ;
\draw (-1.5,0) node{$\longleftrightarrow$} ;
\draw (0,0) node{$\om^+_\mu \, \oplus \, \Om^+_A$} ;
\draw (1.5,0) node{$\longleftrightarrow$} ;
\draw (2.5,0) node{$a_i^b$} ;
\draw (3.5,0) node{$\longleftrightarrow$} ;
\draw (4.5,0) node{$\Phi^-$} ;
\draw (-5.5,-1) node{$\be^+_\mu$} ;
\draw (-4.5,-1) node{$\longleftrightarrow$} ;
\draw (-3,-1) node{$\chi^a_{\mu,i} \, \oplus \, \bar s_{\dot b,i}$} ;
\draw (-1.5,-1) node{$\longleftrightarrow$} ;
\draw (0,-0.8) node{$\al_{\mu  \nu} \, \oplus \, d^A_\mu$} ;
\draw (0,-1.2) node{$\oplus \, \, \xi_{\mu } \, \oplus \, \phi$} ;
\draw (1.5,-1) node{$\longleftrightarrow$} ;
\draw (3,-1) node{$\bar \chi_{\dot b,i}^\mu \, \oplus \, s^a_{i}$} ;
\draw (4.5,-1) node{$\longleftrightarrow$} ;
\draw (5.5,-1) node{$\be^-_\mu$} ;
\draw (-4.5,-2) node{$\Phi^+$} ;
\draw (-3.5,-2) node{$\longleftrightarrow$} ;
\draw (-2.5,-2) node{$\bar a_{\dot b,i}$} ;
\draw (-1.5,-2) node{$\longleftrightarrow$} ;
\draw (0,-2) node{$\om^-_\mu \, \oplus \, \Om^-_A$} ;
\draw (1.5,-2) node{$\longleftrightarrow$} ;
\draw (2.5,-2) node{$\bar r_{\dot b,i}$} ;
\draw (3.5,-2) node{$\longleftrightarrow$} ;
\draw (4.5,-2) node{$\zeta^-$} ;
\endtikzpicture
}
\caption{Three disconnected ${\cal N}=2$ SUSY multiplets}
\label{covN=2}
\end{figure}

\subsubsection{SUSY variation of the spin two supermultiplet}

The spin two multiplet includes a spin two boson $\alpha_{\mu\nu}$, six
vectors $\xi_{\mu},d_{\mu}^{A=1,2,3},\beta_{\mu}^{\pm}$, one scalar $\phi$, two spin-$\frac{3}{2}$
fermions $\chi_{\mu}^{a}$, $\bar{\chi}_{\mu,\dot{a}}$ and two spin-$\frac{1}{2}$
fermions $s^{a}$, $\bar{s}_{\dot{a}}$. Their
SUSY transformations are:
\begin{align}
\big[\eta_{i}^{a}\mathcal{Q}_{a}^{(+\frac{1}{2}),i},V_{\alpha}^{(-1)}\big] & =V_{\chi}^{(-\frac{1}{2})}\Big(\chi_{\mu,i}^{b}=\frac{1}{\sqrt{2}}\eta_{i}^{a}\alpha_{\mu\nu}(\slashed k\bar{\sigma}^{\nu})_{a}^{\phantom{a}b}\Big),\\
\big[\bar{\eta}_{\dot{a},i}\bar{\mathcal{Q}}^{(+\frac{1}{2}),\dot{a},i},V_{\alpha}^{(-1)}\big] & =V_{\bar{\chi}}^{(-\frac{1}{2})}\Big(\bar{\chi}_{\mu,\dot{a},i}=\frac{1}{\sqrt{2}}\bar{\eta}_{\dot{a},i}\alpha_{\mu\nu}(\slashed k\sigma^{\nu})_{\phantom{a}\dot{b}}^{\dot{a}}\Big).
\end{align}
For the four spin one fields, we have the following results -- the SUSY
variations of $\xi_{\mu}$ field read,
\begin{align}
\big[\eta_{i}^{a}\mathcal{Q}_{a}^{(+\frac{1}{2}),i},V_{\xi}^{(-1)}\big] & =V_{\chi}^{(-\frac{1}{2})}\Big(\chi_{\mu,i}^{b}=-\frac{1}{6\sqrt{\alpha'}}\eta_{i}^{a}\big[3\xi_{\mu}\delta_{a}^{\phantom{a}b}+(\slashed\xi\bar{\sigma}_{\mu}+\alpha'\slashed\xi\slashed kk_{\mu})_{a}^{\phantom{a}b}\big]\Big)\nonumber \\
 & \,+V_{\bar{s}}^{(-\frac{1}{2})}\Big(\bar{s}_{\dot{a},i}=-\frac{1}{\sqrt{3\alpha'}}\eta_{i}^{a}\slashed\xi_{a\dot{a}}\Big),\\
\big[\bar{\eta}_{\dot{a},i}\bar{\mathcal{Q}}^{(+\frac{1}{2}),\dot{a},i},V_{\xi}^{(-1)}\big] & =V_{\bar{\chi}}^{(-\frac{1}{2})}\Big(\bar{\chi}_{\mu,\dot{b},i}=\frac{1}{6\sqrt{\alpha'}}\bar{\eta}_{\dot{a},i}\big[3\xi_{\mu}\delta_{\phantom{a}\dot{b}}^{\dot{a}}+(\slashed\xi\bar{\sigma}_{\mu}+\alpha'\slashed\xi\slashed kk_{\mu})_{\phantom{a}\dot{b}}^{\dot{a}}\big]\Big)\nonumber \\
 & \,+V_{s}^{(-\frac{1}{2})}\Big(s_{i}^{a}=\frac{1}{\sqrt{3\alpha'}}\bar{\eta}_{\dot{a},i}\slashed\xi^{\dot{a}a}\Big),
\end{align}
the $SU(2)$ triplet $d_{\mu}^A$ transforms to,
\begin{align}
\big[\eta_{i}^{a}\mathcal{Q}_{a}^{(+\frac{1}{2}),i},V_{d}^{(-1)}\big] & =V_{\chi}^{(-\frac{1}{2})}\Big(\chi_{\mu,j}^{b}=-\frac{1}{3\sqrt{2\alpha'}}\eta_{i}^{a}\big[3d_{\mu}^{A}\delta_{a}^{\phantom{a}b}+(\slashed d^{A}\bar{\sigma}_{\mu}+\alpha'k_{\mu}\slashed d^{A}\slashed k)_{a}^{\phantom{a}b}\big](\tau_{A})_{\phantom{i}j}^{i}\Big)\nonumber \\
 & \,+V_{\bar{s}}^{(-\frac{1}{2})}\Big(\bar{s}_{\dot{a},j}=\frac{1}{\sqrt{6\alpha'}}\eta_{i}^{a}\slashed d_{a\dot{a}}^{A}(\tau_{A})_{\phantom{i}j}^{i}\Big),\\
\big[\bar{\eta}_{\dot{a},i}\bar{\mathcal{Q}}^{(+\frac{1}{2}),\dot{a},i},V_{d}^{(-1)}\big] & =V_{\bar{\chi}}^{(-\frac{1}{2})}\Big(\bar{\chi}_{\mu,\dot{b},i}=-\frac{1}{3\sqrt{2\alpha'}}\bar{\eta}_{\dot{a},i}\big[3d_{\mu}^{A}\delta_{\phantom{a}\dot{b}}^{\dot{a}}+(\slashed d^{A}\bar{\sigma}_{\mu}+\alpha'k_{\mu}\slashed d^{A}\slashed k)_{\phantom{a}\dot{b}}^{\dot{a}}\big](\tau_{A})_{\phantom{i}j}^{i}\Big)\nonumber \\
 & \,+V_{s}^{(-\frac{1}{2})}\Big(s_{j}^{a}=\frac{1}{\sqrt{6\alpha'}}\bar{\eta}_{\dot{a},i}\slashed d^{A,\dot{a}a}(\tau_{A})_{\phantom{i}j}^{i}\Big),
\end{align}
and the complex vectors $\beta_{\mu}^{\pm}$ are varied to,\footnote{Cocycles would introduce additional minus signs in the computations (and several analogous ones at later points). However, we are able to eliminate these extra minus signs in a consistent way.}
\begin{align}
\big[\eta_{i}^{a}\mathcal{Q}_{a}^{(+\frac{1}{2}),i},V_{\beta^{+}}^{(-1)}\big] & =0,\label{N2spin2mulVac1}\\
\big[\bar{\eta}_{\dot{b},i}\bar{\mathcal{Q}}^{(+\frac{1}{2}),\dot{b},i},V_{\beta^{+}}^{(-1)}\big] & =V_{\chi}^{(-\frac{1}{2})}\Big(\chi_{\mu,i}^{b}=\frac{1}{3}\bar{\eta}_{\dot{b},i}\big[3\beta_{\mu}^{+}\slashed k^{\dot{b}b}-(k_{\mu}\slashed\beta^{+}+\slashed\beta^{+}\slashed k\bar{\sigma}_{\mu})^{\dot{b}b}\big]\Big)\nonumber \\
 & \,+V_{\bar{s}}^{(-\frac{1}{2})}\Big(\bar{s}_{\dot{c},i}=\frac{1}{\sqrt{3}}\bar{\eta}_{\dot{b},i}(\slashed\beta^{+}\slashed k)_{\phantom{b}\dot{c}}^{\dot{b}}\Big),\\
\big[\eta_{i}^{b}\mathcal{Q}_{b}^{(+\frac{1}{2}),i},V_{\beta^{-}}^{(-1)}\big] & =V_{\bar{\chi}}^{(-\frac{1}{2})}\Big(\bar{\chi}_{\mu,\dot{b},i}=\frac{1}{3}\eta_{i}^{b}\big[3\beta_{\mu}^{-}\slashed k_{b\dot{b}}-(k_{\mu}\slashed\beta^{-}+\slashed\beta^{-}\slashed k\sigma_{\mu})_{b\dot{b}}\big]\Big)\nonumber \\
 & \,+V_{s}^{(-\frac{1}{2})}\Big(s_{i}^{c}=\frac{1}{\sqrt{3}}\eta_{i}^{b}(\slashed\beta^{-}\slashed k)_{b}^{\phantom{b}c}\Big),\\
\big[\bar{\eta}_{\dot{a},i}\bar{\mathcal{Q}}^{(+\frac{1}{2}),\dot{a},i},V_{\beta^{-}}^{(-1)}\big] & =0 \ .\label{N2spin2mulVac2}
\end{align}
The SUSY action on the unique scalar field $\phi$ is given by
\begin{align}
\big[\eta_{i}^{a}\mathcal{Q}_{a}^{(+\frac{1}{2}),i},V_{\phi}^{(-1)}\big] & =V_{\bar{s}}^{(-\frac{1}{2})}\Big(\bar{s}_{\dot{a},i}=\frac{1}{\sqrt{2}}\phi\eta_{i}^{a}\slashed k_{a\dot{a}}\Big),\\
\big[\bar{\eta}_{\dot{a},i}\bar{\mathcal{Q}}^{(+\frac{1}{2}),\dot{a},i},V_{\phi}^{(-1)}\big] & =V_{s}^{(-\frac{1}{2})}\Big(s_{i}^{a}=\frac{1}{\sqrt{2}}\phi\bar{\eta}_{\dot{a},i}\slashed k^{\dot{a}a}\Big).
\end{align}
Now we turn to analyze the fermionic states. For $\chi$ and $\bar{\chi}$ at spin-$\frac{3}{2}$, we have SUSY relations,
\begin{align}
\big[\eta_{i}^{a}\mathcal{Q}_{a}^{(-\frac{1}{2}),i},V_{\chi}^{(-\frac{1}{2})}\big] & =V_{\beta^{+}}^{(-1)}\Big(\beta_{\mu}^{+}=\eta_{i}^{a}\chi_{\mu,a,i}\Big),\\
\big[\bar{\eta}_{\dot{a},i}\bar{\mathcal{Q}}^{(-\frac{1}{2}),\dot{a},i},V_{\chi}^{(-\frac{1}{2})}\big] & =V_{\alpha}^{(-1)}\Big(\alpha_{\mu\nu}=\frac{1}{\sqrt{2}}\bar{\eta}_{\dot{a},i}\big(\bar{\sigma}_{(\mu}^{\dot{a}a}\chi_{\nu),a}^{i}+\alpha'\slashed k^{\dot{a}a}\chi_{(\mu|,a|}^{i}k_{\nu)}\big)\Big)\nonumber \\
 & \,+V_{\xi}^{(-1)}\Big(\xi_{\mu}=-\frac{\sqrt{\alpha'}}{2}\bar{\eta}_{\dot{a},i}\slashed k^{\dot{a}a}\chi_{\mu,a}^{i}\Big)\nonumber \\
 & \,+V_{d}^{(-1)}\Big(d_{\mu}^{A}=\sqrt{\frac{\alpha'}{2}}\bar{\eta}_{\dot{a},i}\slashed k^{\dot{a}a}\chi_{\mu,a,j}(\tau^{A}\varepsilon)^{ij}\Big),
\end{align}
and
\begin{align}
\big[\eta_{i}^{a}\mathcal{Q}_{a}^{(-\frac{1}{2}),i},V_{\bar{\chi}}^{(-\frac{1}{2})}\big] & =V_{\alpha}^{(-1)}\Big(\alpha_{\mu\nu}=\frac{1}{\sqrt{2}}\eta_{i}^{a}\big(\sigma_{(\mu|a\dot{a}|}\bar{\chi}_{\nu)}^{\dot{a},i}+\alpha'\slashed k_{a\dot{a}}\bar{\chi}_{(\mu}^{\dot{a},i}k_{\nu)}\big)\Big)\nonumber \\
 & \,+V_{\xi}^{(-1)}\Big(\xi_{\mu}=\frac{\sqrt{\alpha'}}{2}\eta_{i}^{a}\slashed k_{a\dot{a}}\bar{\chi}_{\mu}^{\dot{a},i}\Big)\nonumber \\
 & \,+V_{d}^{(-1)}\Big(d_{\mu}^{A}=\sqrt{\frac{\alpha'}{2}}\eta_{i}^{a}\slashed k_{a\dot{a}}\bar{\chi}_{\mu,j}^{\dot{a}}(\tau^{A}\varepsilon)^{ij}\Big),\\
\big[\bar{\eta}_{\dot{a},i}\bar{\mathcal{Q}}^{(-\frac{1}{2}),\dot{a},i},V_{\bar{\chi}}^{(-\frac{1}{2})}\big] & =V_{\beta^{-}}^{(-1)}\Big(\beta_{\mu}^{-}=\bar{\eta}_{\dot{a},i}\bar{\chi}_{\mu}^{\dot{a},i}\Big).
\end{align}
The spin-$\frac{1}{2}$ states $s$ and $\bar{s}$, on the other hand, transform to
\begin{align}
\big[\eta_{i}^{a}\mathcal{Q}_{a}^{(-\frac{1}{2}),i},V_{s}^{(-\frac{1}{2})}\big] & =V_{\xi}^{(-1)}\Big(\xi_{\mu}=\sqrt{\frac{\alpha'}{3}}\eta_{i}^{a}\big[k_{\mu}\delta_{a}^{\phantom{a}b}+(\sigma_{\mu}\slashed k)_{a}^{\phantom{a}b}\big]s_{b}^{i}\Big)\nonumber \\
 & +V_{d}^{(-1)}\Big(d_{\mu}^{A}=\sqrt{\frac{\alpha'}{6}}\eta_{i}^{a}\big[k_{\mu}\delta_{a}^{\phantom{a}b}+(\sigma_{\mu}\slashed k)_{a}^{\phantom{a}b}\big]s_{b,j}(\tau^{A}\varepsilon)^{ij}\Big)\nonumber \\
 & +V_{\phi}^{(-1)}\Big(\phi=\frac{1}{\sqrt{2}}\eta_{i}^{a}s_{a}^{i}\Big),\\
\big[\bar{\eta}_{\dot{a},i}\bar{\mathcal{Q}}^{(-\frac{1}{2}),\dot{a},i},V_{s}^{(-\frac{1}{2})}\big] & =V_{\beta^{-}}^{(-1)}\Big(\beta_{\mu}^{-}=-\frac{1}{\sqrt{3}}\bar{\eta}_{\dot{a},i}\big(\bar{\sigma}_{\mu}^{\dot{a}a}+\alpha'k_{\mu}\slashed k^{\dot{a}a}\big)s_{a}^{i}\Big),
\end{align}
and
\begin{align}
\big[\eta_{i}^{a}\mathcal{Q}_{a}^{(-\frac{1}{2}),i},V_{\bar{s}}^{(-\frac{1}{2})}\big] & =V_{\beta^{+}}^{(-1)}\Big(\beta_{\mu}^{+}=-\frac{1}{\sqrt{3}}\eta_{i}^{a}\big(\sigma_{\mu a\dot{a}}+\alpha'k_{\mu}\slashed k_{a\dot{a}}\big)\bar{s}^{\dot{a},i}\Big),\\
\big[\bar{\eta}_{\dot{a},i}\bar{\mathcal{Q}}^{(-\frac{1}{2}),\dot{a},i},V_{\bar{s}}^{(-\frac{1}{2})}\big] & =V_{\xi}^{(-1)}\Big(\xi_{\mu}=\sqrt{\frac{\alpha'}{3}}\bar{\eta}_{\dot{a},i}\big[k_{\mu}\delta_{\phantom{a}\dot{b}}^{\dot{a}}+(\bar{\sigma}_{\mu}\slashed k)_{\phantom{a}\dot{b}}^{\dot{a}}\big]\bar{s}^{\dot{b},i}\Big)\nonumber \\
 & +V_{d}^{(-1)}\Big(d_{\mu}^{A}=\sqrt{\frac{\alpha'}{6}}\bar{\eta}_{\dot{a},i}\big[k_{\mu}\delta_{\phantom{a}\dot{b}}^{\dot{a}}+(\bar{\sigma}_{\mu}\slashed k)_{\phantom{a}\dot{b}}^{\dot{a}}\big]\bar{s}^{\dot{b},i}(\tau^{A}\varepsilon)^{ij}\Big)\nonumber \\
 & +V_{\phi}^{(-1)}\Big(\phi=\frac{1}{\sqrt{2}}\bar{\eta}_{\dot{a},i}\bar{r}_{+}^{\dot{a},i}\Big).
\end{align}

\subsubsection{SUSY variation of the spin one supermultiplets}

The first spin one multiplet $\{\omega^{-},\bar{a},\bar{r},\Phi^{+},\zeta^{-},\Omega_{A}^{-}\}$
contains one vector $\omega_{\mu}^{-}$, two right-handed fermions
$\bar{a}_{\dot{b}}$ and $\bar{r}_{\dot{b}}$ of spin 1/2 each, and three scalars $\Phi^{+}$,
$\zeta^{-}$ and $\Omega_{A}^{-}$. The SUSY relations for the spin one $\omega_{\mu}^{-}$ read,
\begin{align}
\big[\eta_{i}^{b}\mathcal{Q}_{b}^{(+\frac{1}{2}),i},V_{\omega^{-}}^{(-1)}\big] & =V_{\bar{a}}^{(-1)}\Big(\bar{a}_{\dot{b},i}=-\frac{1}{\sqrt{2\alpha'}}\eta_{i}^{b}\slashed\omega_{b\dot{b}}^{-}\Big),\\
\big[\bar{\eta}_{\dot{a},i}\bar{\mathcal{Q}}^{(+\frac{1}{2}),\dot{a},i},V_{\omega^{-}}^{(-1)}\big] & =V_{\bar{r}}^{(+\frac{1}{2})}\Big(\bar{r}_{\dot{b},i}=-\frac{1}{\sqrt{2}}\bar{\eta}_{\dot{a},i}(\slashed\omega^{-}\slashed k)_{\phantom{a}\dot{b}}^{\dot{a}}\Big).
\end{align}
For the fermions
$\bar{a}_{\dot{b}}$ and $\bar{r}_{\dot{b}}$, we have,
\begin{align}
\big[\eta_{i}^{b}\mathcal{Q}_{b}^{(-\frac{1}{2}),i},V_{\bar{a}}^{(-\frac{1}{2})}\big] & =V_{\Phi^{+}}^{(-1)}\Big(\Phi^{+}=\sqrt{\alpha'}\eta_{i}^{b}\slashed k_{b\dot{b}}\bar{a}^{\dot{b},i}\Big),\\
\big[\bar{\eta}_{\dot{b},i}\bar{\mathcal{Q}}^{(-\frac{1}{2}),\dot{b},i},V_{\bar{a}}^{(-\frac{1}{2})}\big] & =V_{\omega^{-}}^{(-1)}\Big(\omega_{\mu}^{-}=\sqrt{\frac{\alpha'}{2}}\bar{\eta}_{\dot{b},i}\big[k_{\mu}\delta_{\phantom{b}\dot{c}}^{\dot{b}}+(\slashed k\sigma_{\mu})_{\phantom{b}\dot{c}}^{\dot{b}}\big]\bar{a}^{\dot{c},i}\Big)\nonumber \\
 & \,+V_{\Omega^{-}}^{(-1)}\Big(\Omega_{A}^{-}=-\frac{1}{\sqrt{2}}\bar{\eta}_{\dot{b},i}(\tau_{A}\varepsilon)^{ij}\bar{a}_{j}^{\dot{b}}\Big),
\end{align}
and
\begin{align}
\big[\eta_{i}^{a}\mathcal{Q}_{a}^{(-\frac{1}{2}),i},V_{\bar{r}}^{(-\frac{1}{2})}\big] & =V_{\omega^{-}}^{(-1)}\Big(\omega_{\mu}^{-}=\frac{1}{\sqrt{2}}\eta_{i}^{a}\big(\sigma_{\mu a\dot{a}}+\alpha'k_{\mu}\slashed k_{a\dot{a}}\big)\bar{r}^{\dot{a},i}\Big)\nonumber \\
 & \,+V_{\Omega^{-}}^{(-1)}\Big(\Omega_{A}^{-}=\sqrt{\frac{\alpha'}{2}}\eta_{i}^{a}\slashed k_{a\dot{a}}\bar{r}_{j}^{\dot{a}}(\tau_{A}\varepsilon)^{ij}\Big),\\
\big[\bar{\eta}_{\dot{a},i}\bar{\mathcal{Q}}^{(-\frac{1}{2}),\dot{a},i},V_{\bar{r}}^{(-\frac{1}{2})}\big] & =V_{\zeta^{-}}^{(-1)}\Big(\zeta^{-}=\bar{\eta}_{\dot{a},i}\bar{r}^{\dot{a},i}\Big).
\end{align}
The results for the scalar fields are:
\begin{align}
\big[\eta_{i}^{b}\mathcal{Q}_{b}^{(+\frac{1}{2}),i},V_{\Phi^{+}}^{(-1)}\big] & =0,\label{N2spin1mulVac1}\\
\big[\bar{\eta}_{\dot{b},i}\bar{\mathcal{Q}}^{(+\frac{1}{2}),\dot{b},i},V_{\Phi^{+}}^{(-1)}\big] & =V_{\bar{a}}^{(-\frac{1}{2})}\Big(\bar{a}_{\dot{b},i}=-\alpha'^{-\frac{1}{2}}\Phi^{+}\bar{\eta}_{\dot{b},i}\Big),
\end{align}
and
\begin{align}
\big[\eta_{i}^{a}\mathcal{Q}_{a}^{(+\frac{1}{2}),i},V_{\zeta^{-}}^{(-1)}\big] & =V_{\bar{r}}^{(-\frac{1}{2})}\Big(\bar{r}_{\dot{a},i}=\zeta^{-}\eta_{i}^{a}\slashed k_{a\dot{a}}\Big),\\
\big[\bar{\eta}_{\dot{a},i}\bar{\mathcal{Q}}^{(+\frac{1}{2}),\dot{a},i},V_{\zeta^{-}}^{(-1)}\big] & =0,
\end{align}
and
\begin{align}
\big[\eta_{i}^{b}\mathcal{Q}_{b}^{(+\frac{1}{2}),i},V_{\Omega^{-}}^{(-1)}\big] & =V_{\bar{a}}^{(-\frac{1}{2})}\Big(\bar{a}_{\dot{b},j}=-\frac{1}{\sqrt{2}}\eta_{i}^{b}\slashed k_{b\dot{b}}\Omega_{A}^{-}(\tau^{A})_{\phantom{i}j}^{i}\Big),\\
\big[\bar{\eta}_{\dot{a},i}\bar{\mathcal{Q}}^{(+\frac{1}{2}),\dot{a},i},V_{\Omega^{-}}^{(-1)}\big] & =V_{\bar{r}}^{(-\frac{1}{2})}\Big(\bar{r}_{\dot{a},j}=\frac{1}{\sqrt{2\alpha'}}\bar{\eta}_{\dot{a},i}\Omega_{A}^{-}(\tau^{A})_{\phantom{i}j}^{i}\Big).
\end{align}
The second spin one multiplet $\{\omega^{+},a,r,\Phi^{-},\zeta^{+},\Omega_{A}^{+}\}$ is just the complex conjugate of the former, so let us simply list the analogous SUSY transformations:
\begin{align}
\big[\eta_{i}^{a}\mathcal{Q}_{a}^{(+\frac{1}{2}),i},V_{\omega^{+}}^{(-1)}\big] & =V_{r}^{(+\frac{1}{2})}\Big(r_{i}^{b}=-\frac{1}{\sqrt{2}}\eta_{i}^{a}(\slashed\omega^{+}\slashed k)_{a}^{\phantom{a}b}\Big),\\
\big[\bar{\eta}_{\dot{b},i}\bar{\mathcal{Q}}^{(+\frac{1}{2}),\dot{b},i},V_{\omega^{+}}^{(-1)}\big] & =V_{a}^{(-1)}\Big(a_{i}^{b}=-\frac{1}{\sqrt{2\alpha'}}\bar{\eta}_{\dot{b},i}\slashed\omega^{+,\dot{b}b}\Big), \\
\big[\eta_{i}^{b}\mathcal{Q}_{b}^{(-\frac{1}{2}),i},V_{a}^{(-\frac{1}{2})}\big] & =V_{\omega^{+}}^{(-1)}\Big(\omega_{\mu}^{+}=\sqrt{\frac{\alpha'}{2}}\eta_{i}^{b}\big[k_{\mu}\delta_{b}^{\phantom{b}c}+(\slashed k\bar{\sigma}_{\mu})_{b}^{\phantom{b}c}\big]a_{c}^{i}\Big)\nonumber \\
 & \,+V_{\Omega^{+}}^{(-1)}\Big(\Omega_{A}^{+}=-\frac{1}{\sqrt{2}}\eta_{i}^{b}(\tau_{A}\varepsilon)^{ij}a_{b,j}\Big),\\
\big[\bar{\eta}_{\dot{b},i}\bar{\mathcal{Q}}^{(-\frac{1}{2}),\dot{b},i},V_{a}^{(-\frac{1}{2})}\big] & =V_{\Phi^{-}}^{(-1)}\Big(\Phi^{-}=\sqrt{\alpha'}\bar{\eta}_{\dot{b},i}\slashed k^{\dot{b}b}a_{b}^{i}\Big), \\
\big[\eta_{i}^{a}\mathcal{Q}_{a}^{(-\frac{1}{2}),i},V_{r}^{(-\frac{1}{2})}\big] & =V_{\zeta^{+}}^{(-1)}\Big(\zeta^{+}=\eta_{i}^{a}r_{a}^{i}\Big),\\
\big[\bar{\eta}_{\dot{a},i}\bar{\mathcal{Q}}^{(-\frac{1}{2}),\dot{a},i},V_{r}^{(-\frac{1}{2})}\big] & =V_{\omega^{+}}^{(-1)}\Big(\omega_{\mu}^{+}=\frac{1}{\sqrt{2}}\bar{\eta}_{\dot{a},i}\big(\bar{\sigma}_{\mu}^{\dot{a}a}+\alpha'k_{\mu}\slashed k^{\dot{a}a}\big)r_{a}^{i}\Big)\nonumber \\
 & \,+V_{\Omega^{+}}^{(-1)}\Big(\Omega_{A}^{+}=\sqrt{\frac{\alpha'}{2}}\bar{\eta}_{\dot{a},i}\slashed k^{\dot{a}a}r_{a,j}(\tau_{A}\varepsilon)^{ij}\Big) , \\
\big[\eta_{i}^{b}\mathcal{Q}_{b}^{(+\frac{1}{2}),i},V_{\Phi^{-}}^{(-1)}\big] & =V_{a}^{(-\frac{1}{2})}\Big(a_{i}^{b}=-\alpha'^{-\frac{1}{2}}\Phi^{-}\eta_{i}^{b}\Big),\\
\big[\bar{\eta}_{\dot{b},i}\bar{\mathcal{Q}}^{(+\frac{1}{2}),\dot{b},i},V_{\Phi^{-}}^{(-1)}\big] & =0, \\
\big[\eta_{i}^{a}\mathcal{Q}_{a}^{(+\frac{1}{2}),i},V_{\zeta^{+}}^{(-1)}\big] & =0,\label{N2spin1mulVac2}\\
\big[\bar{\eta}_{\dot{a},i}\bar{\mathcal{Q}}^{(+\frac{1}{2}),\dot{a},i},V_{\zeta^{+}}^{(-1)}\big] & =V_{r}^{(-\frac{1}{2})}\Big(r_{i}^{a}=\zeta^{+}\bar{\eta}_{\dot{a},i}\slashed k^{\dot{a}a}\Big),
 \\ %
\big[\eta_{i}^{a}\mathcal{Q}_{a}^{(+\frac{1}{2}),i},V_{\Omega^{+}}^{(-1)}\big] & =V_{r}^{(-\frac{1}{2})}\Big(r_{j}^{a}=\frac{1}{\sqrt{2\alpha'}}\eta_{i}^{a}\Omega_{A}^{+}(\tau^{A})_{\phantom{i}j}^{i}\Big),\\
\big[\bar{\eta}_{\dot{b},i}\bar{\mathcal{Q}}^{(+\frac{1}{2}),\dot{b},i},V_{\Omega^{+}}^{(-1)}\big] & =V_{a}^{(-\frac{1}{2})}\Big(a_{j}^{b}=-\frac{1}{\sqrt{2}}\bar{\eta}_{\dot{b},i}\slashed k^{\dot{b}b}\Omega_{A}^{+}(\tau^{A})_{\phantom{i}j}^{i}\Big).
\end{align}

\section{Helicity structure of massive on-shell multiplets}
\label{sec:hel}

In this section, we apply the massive version of the spinor helicity formalism \cite{HWF1, HWF2} to obtain a refined understanding of the structure of the previously constructed SUSY multiplets. A brief summary of the spinor techniques is collected in appendix \ref{app:hel}, including the explicit form of massive wavefunctions associated with different spin components. The spin quantization axis is chosen covariantly by decomposing the timelike momentum
$k$ into two arbitrary light-like reference momenta $p$ and $q$:
\begin{equation}
k^{\mu}=p^{\mu}+q^{\mu},\qquad k^{2}=-m^{2}=2pq,\qquad p^{2}=q^{2}=0.
\end{equation}
As was explained in detail in \cite{Boels:2011zz}, the supercharges
can be expanded in the basis of the momentum spinors $p_{a},p^{* \dot a}$ and $q_{a},q^{* \dot a}$ defined by $p_\mu \si^\mu_{a \dot a} = -p_{a}p^{*}_{ \dot a}$ and $q_\mu \si^\mu_{a \dot a} = -q_{a}q^{*}_{ \dot a}$:
\begin{align}
\mathcal{Q}_{a} & =\frac{[q\mathcal{Q}]}{[qp]}p_{a}+\frac{[p\mathcal{Q}]}{[pq]}q_{a}=\mathcal{Q}_{+}p_{a}+\mathcal{Q}_{-}q_{a}, \label{Qdecomp} \\
\bar{\mathcal{Q}}^{\dot{a}} & =\frac{\langle p\bar{\mathcal{Q}}\rangle}{\langle pq\rangle}q^{* \dot{a}}+\frac{\langle q\bar{\mathcal{Q}}\rangle}{\langle qp\rangle}p^{* \dot{a}}=\bar{\mathcal{Q}}_{+}q^{* \dot{a}}+\bar{\mathcal{Q}}_{-}p^{* \dot{a}}.\label{barQdecomp}
\end{align}
This defines the supercharge components ${\cal Q}_{\pm}$ and $\bar {\cal Q}_{\pm}$ to be
\begin{gather}
\mathcal{Q}_{+}\equiv\frac{[q\mathcal{Q}]}{[qp]},\qquad \mathcal{Q}_{-}\equiv\frac{[p\mathcal{Q}]}{[pq]},\\
\bar{\mathcal{Q}}_{+}\equiv\frac{\langle p\bar{\mathcal{Q}}\rangle}{\langle pq\rangle},\qquad\bar{\mathcal{Q}}_{-}\equiv\frac{\langle q\bar{\mathcal{Q}}\rangle}{\langle qp\rangle}.\label{N1SUSYChargeCom}
\end{gather}
The $\mathcal{Q}_{+}$ and $\bar{\mathcal{Q}}_{+}$ raise the spin quantum $j_z$ number along the quantization axis by $1/2$, while $\mathcal{Q}_{-}$ and $\bar{\mathcal{Q}}_{-}$ lower it by $1/2$. The corresponding Lorentz generator which is diagonalized with eigenvalues $j_z$ reads
\beq
J_z = \frac{1}{m^2} \; \vep^{\mu \nu \la \rho} \, P_\mu \, q_\nu \, M_{\la \rho}\ ,
\eeq
where $P_\mu$ denotes the translation operator and $M_{\la \rho}$ an $SO(1,3)$ rotation.

\medskip
A convenient way of organizing representations of the super Poincar\'e group is to pick a highest weight state which is annihilated by half the supercharges -- either the left-handed ${\cal Q}_a$ or the right-handed $\bar {\cal Q}^{\dot b}$. States with this property are referred to as (anti-)Clifford vacua, and we shall use the vacuum eliminated by the left-handed ${\cal Q}_a$ by convention. The rest of the supermultiplet is then constructed by applying the nontrivially acting $\bar {\cal Q}_{+}$ and $\bar {\cal Q}_-$, see the figures in this section. In our notation, each
diamond shaped diagram represents one supermultiplet. The dashed lines connecting bosonic and fermionic states indicate $\mathcal{Q}_{\pm}$ and $\bar{\mathcal{Q}}_{\pm}$ applications, and we assign the following directions:
\begin{gather}
\nearrow\;\equiv\bar{\mathcal{Q}}_{+},\qquad\searrow\;\equiv\bar{\mathcal{Q}}_{-} \ \ \ \ \ \ \te{and} \ \ \ \ \ \
\nwarrow\;\equiv \mathcal{Q}_{+},\qquad\swarrow\;\equiv \mathcal{Q}_{-}.
\end{gather}
The Clifford vacuum state being annihilated by the left-handed ${\cal Q}_{\pm}$ is located on the far left of the diamond, and we can construct the full supermultiplet by repeated action of $\bar {\cal Q}_{\pm}$.\footnote{Alternatively, we can also construct this supermultiplet starting from the anti-Clifford vacuum
state on the right side of this diamond, which is eliminated by the anti-supercharge $\bar {\cal Q}_{\pm}$, and the remaining states follow by acting $\mathcal{Q}_{\pm}$ on it.} In this section, we will show how $\bar{\mathcal{Q}}_{\pm}$ transform all the states in the multiplet from the left side of the diamond all the way to the right. The SUSY algebras $\{ {\cal Q}_{\pm} , \bar  {\cal Q}_{\mp} \} = 1$ and $\{ {\cal Q}_{\pm} , \bar  {\cal Q}_{\pm} \} = 0$ \footnote{To show this, we simply plug the supercharge decompositions (\ref{Qdecomp}) and
(\ref{barQdecomp}) into the $\mathcal{N}=1$ SUSY algebra (\ref{N=1SUSYalg}), and obtain,
\begin{align*}
\big\{\mathcal{Q}_{a}^{(+\frac{1}{2})},\mathcal{\bar{Q}}^{(-\frac{1}{2}),\dot{a}}\big\} & =p_{a}p^{*\dot{a}}\big\{\mathcal{Q}_{+},\mathcal{\bar{Q}}_{-}\big\}+p_{a}q^{*\dot{a}}\big\{\mathcal{Q}_{+},\mathcal{\bar{Q}}_{+}\big\}+q_{a}q^{*\dot{a}}\big\{\mathcal{Q}_{-},\mathcal{\bar{Q}}_{+}\big\}+q_{a}p^{*\dot{a}}\big\{\mathcal{Q}_{-},\mathcal{\bar{Q}}_{-}\big\}\\
 & =(\sigma^{\mu}\varepsilon)_{a}^{\phantom{a}\dot{a}}P_{\mu}\sim(\sigma^{\mu}\varepsilon)_{a}^{\phantom{a}\dot{a}}k_{\mu}=p_{a}p^{*\dot{a}}+q_{a}q^{*\dot{a}}.
\end{align*}
Thus we arrive at,
\[
\left(\begin{array}{cc}
\big\{\mathcal{Q}_{+},\mathcal{\bar{Q}}_{-}\big\} & \big\{\mathcal{Q}_{+},\mathcal{\bar{Q}}_{+}\big\}\\
\big\{\mathcal{Q}_{-},\mathcal{\bar{Q}}_{-}\big\} & \big\{\mathcal{Q}_{-},\mathcal{\bar{Q}}_{+}\big\}
\end{array}\right)=\left(\begin{array}{cc}
1 & 0\\
0 & 1
\end{array}\right).
\]
} imply that $\mathcal{Q}_{\pm}$ undoes $\bar{\mathcal{Q}}_{\pm}$ applications and transforms states from right to left in the diamond.

\medskip
This section starts with the ${\cal N}=1$ situation to illustrate the methods, and the additional features of extended SUSY are explained in the later subsections on ${\cal N}=2,4$ supermultiplets. To make everything simple and clear, instead of using our old notation of vertex operators in the previous sections, we will use the ``ket'' notation to express the states inside the diamonds. For example, the spin two boson with $j_z=+2$ is expressed by
\beq
|\alpha,+2\rangle \ \equiv \ V_{\al}^{(-1)}\Big( \, \al^{\mu \nu}= \frac{1}{2m^{2}}\,\bar{\sigma}^{\mu\dot{a}a}\bar{\sigma}^{\nu\dot{b}b}p_{\dot{a}}^{*}q_{a}p_{\dot{b}}^{*}q_{b} \, \Big) \ ,
\eeq
and a combined state $\{\alpha,d\}$ with $j_z=+1$ is expressed by $|\alpha\oplus d,+1\rangle$. The commutators of ${\cal Q}_a$ and $\bar {\cal Q}^{\dot b}$ with vertex operators are replaced by SUSY transformations acting directly on the states.


%

\subsection{${\cal N}=1$ supermultiplets}

According to the strategy outlined above, it suffices to evaluate the anti-supercharge components $\bar {\cal Q}_{\pm}$ on the helicity states in the $\mathcal{N}=1$ supermultiplets. The decomposition $\bar{\mathcal{Q}}^{\dot{a}} = \bar{\mathcal{Q}}_{+}q^{\dot{a}*}+\bar{\mathcal{Q}}_{-}p^{\dot{a}*}$ corresponds to the mass dimension $[M^{-\frac{1}{2}}]$ choices for $\bar \eta$:
\begin{equation}
\bar {\cal Q}_{\pm} = \bar \eta_{\bar a}^{\pm} \, {\cal Q}^{\bar a} \ \ \ \longleftrightarrow \ \ \
\bar{\eta}_{\dot{a}}^{+} = \frac{p_{\dot{a}}^{*}}{\langle pq\rangle} ,\ \  \bar{\eta}_{\dot{a}}^{-} = \frac{q_{\dot{a}}^{*}}{\langle qp\rangle}.
\end{equation}

\subsubsection{Spin one half supermultiplets}

\begin{figure}[h]
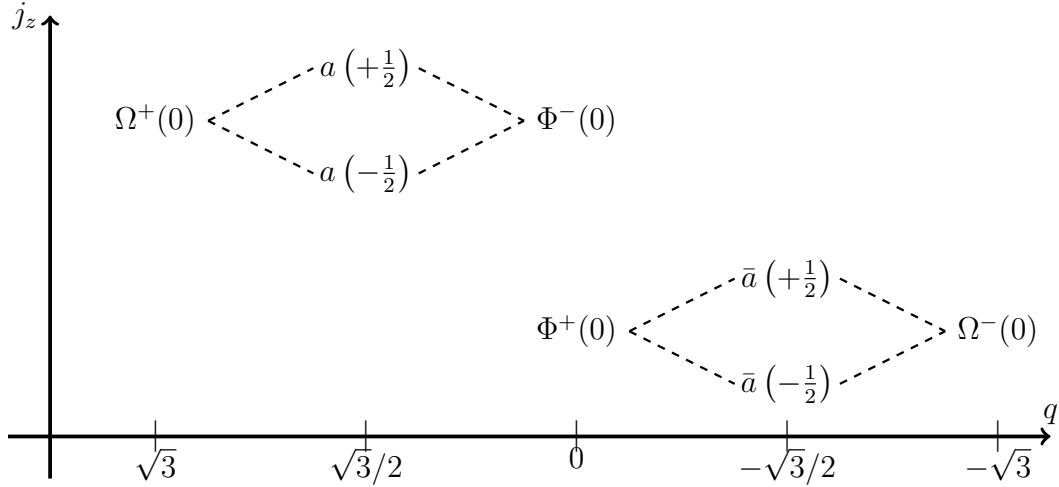

\centerline{
\tikzpicture [scale=1.4,line width=0.30mm]
\draw[line width=0.50mm,->] (1,-0.4) -- (1,4) node[left]{$j_z$};
\draw[line width=0.50mm,->] (0.6,0) -- (10.5,0) node[above]{$q$};
\draw (2,0)node[below]{$\sqrt{3}$} node{$|$};
\draw (4,0)node[below]{$\sqrt{3}/2$} node{$|$};
\draw (6,0)node[below]{$0$} node{$|$};
\draw (8,0)node[below]{$-\sqrt{3}/2$} node{$|$};
\draw (10,0)node[below]{$-\sqrt{3}$} node{$|$};
\draw (2,3) node{$\Om^+(0)$} ;
\draw (4,3.5) node{$a\left( +\tfrac{1}{2} \right)$} ;
\draw (4,2.5) node{$a\left( -\tfrac{1}{2} \right)$} ;
\draw (6,3) node{$\Phi^-(0)$} ;
\draw (6,1) node{$\Phi^+(0)$} ;
\draw (8,1.5) node{$\bar a\left( +\tfrac{1}{2} \right)$} ;
\draw (8,0.5) node{$\bar a\left( -\tfrac{1}{2} \right)$} ;
\draw (10,1) node{$\Om^-(0)$} ;
\draw[dashed] (2.5,3) -- (3.5,3.5) ;
\draw[dashed] (4.5,3.5) -- (5.5,3) ;
\draw[dashed] (2.5,3) -- (3.5,2.5) ;
\draw[dashed] (4.5,2.5) -- (5.5,3) ;
\draw[dashed] (6.5,1) -- (7.5,1.5) ;
\draw[dashed] (8.5,1.5) -- (9.5,1) ;
\draw[dashed] (6.5,1) -- (7.5,0.5) ;
\draw[dashed] (8.5,0.5) -- (9.5,1) ;
\endtikzpicture
}
\caption{${\cal N}=1$ SUSY multiplets with scalar Clifford vacuum: In ${\cal N}=1$ scenarios, the $U(1)$ charge $q$ with respect to the internal current ${\cal J}$ is plotted along the horizontal axis. The SUSY charges have eigenvalue $\pm \sqrt{3}/2$ under ${\cal J}$ and therefore change $q$ by a fixed offset.}
\label{N=1spin12}
\end{figure}

We firstly consider the $\{\Phi^{+},\bar{a},\Omega^{-}\}$ multiplet of highest spin 1/2 whose scalar Clifford vacuum $|\Phi^{+} \rangle$ is eliminated
by the supercharge ${\cal Q}^{a}$. By repeated actions of the anti-supercharge $\bar{\mathcal{Q}}_{\pm}$ on $\Phi^{+}$, we can construct the remainder of the multiplet, see figure \ref{N=1spin12}.

The spin-$\frac{1}{2}$ multiplet is the minimal massive representation of the ${\cal N}=1$ SUSY algebra, since it only contains
four states. Very straightly, we obtain, up to a phase,
\beq
\bar{\mathcal{Q}}_{\pm} \ |\Phi^+,0\rangle = |\bar a,\pm \tfrac{1}{2}\rangle,
\eeq
and
\begin{gather}
\bar{\mathcal{Q}}_{\pm}\ |{\bar{a},\mp \tfrac{1}{2}}\rangle = |\Omega^{-},0\rangle,\qquad \bar{\mathcal{Q}}_{\pm}\ |{\bar{a},\pm \tfrac{1}{2}}\rangle =0.
\end{gather}
The anti-Clifford vacuum $|\Om^-\rangle$ is then annihilated by $\bar {\cal Q}_{\pm}$ action,
\beq
\bar {\cal Q}_{\pm}\ |\Om^-,0\rangle = 0 .
\eeq
Secondly, we consider the mirror multiplet $\{\Omega^{+},a,\Phi^{-}\}$ which is also summarized in figure \ref{N=1spin12}. Starting from the Clifford vacuum $|\Omega^{+} \rangle $, c.f. (\ref{N1spin12mulVac21}),
we obtain,
\beq
\bar {\cal Q}_{\pm} \ |\Omega^{+},0\rangle =  | a,\pm \tfrac{1}{2}\rangle,
\eeq
and
\beq
\bar {\cal Q}_{\pm} \ | a,\mp \tfrac{1}{2}\rangle = |\Phi^-,0\rangle ,\qquad \bar {\cal Q}_{\pm} \ | a,\pm \tfrac{1}{2}\rangle = 0.
\eeq

\subsubsection{Spin two supermultiplet}

\begin{figure}[h]
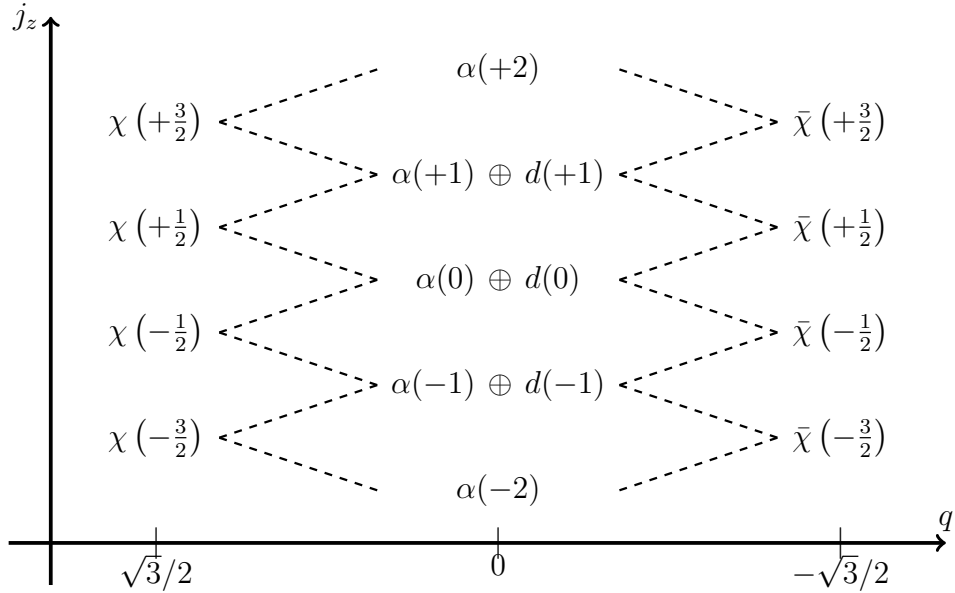

\centerline{
\tikzpicture [scale=1.4,line width=0.30mm]
\draw[line width=0.50mm,->] (1,-0.4) -- (1,5) node[left]{$j_z$};
\draw[line width=0.50mm,->] (0.6,0) -- (9.5,0) node[above]{$q$};
\draw (2,0)node[below]{$\sqrt{3}/2$} node{$|$};
\draw (5.25,0)node[below]{$0$} node{$|$};
\draw (8.5,0)node[below]{$-\sqrt{3}/2$} node{$|$};
\draw (2,4) node{$\chi \left( +\tfrac{3}{2} \right)$} ;
\draw (2,3) node{$\chi \left( +\tfrac{1}{2} \right)$} ;
\draw (2,2) node{$\chi \left( -\tfrac{1}{2} \right)$} ;
\draw (2,1) node{$\chi \left( -\tfrac{3}{2} \right)$} ;
\draw (8.5,4) node{$\bar \chi \left( +\tfrac{3}{2} \right)$} ;
\draw (8.5,3) node{$\bar \chi \left( +\tfrac{1}{2} \right)$} ;
\draw (8.5,2) node{$\bar \chi \left( -\tfrac{1}{2} \right)$} ;
\draw (8.5,1) node{$\bar \chi \left( -\tfrac{3}{2} \right)$} ;
\draw (5.25,3.5) node{$\al(+1) \, \oplus \, d(+1)$} ;
\draw (5.25,2.5) node{$\al(0) \, \oplus \, d(0)$} ;
\draw (5.25,1.5) node{$\al(-1) \, \oplus \, d(-1)$} ;
\draw (5.25,4.5) node{$\al(+2)$} ;
\draw (5.25,0.5) node{$\al(-2)$} ;
\draw[dashed] (2.6,4) -- (4.1,4.5) ;
\draw[dashed] (2.6,3) -- (4.1,3.5) ;
\draw[dashed] (2.6,2) -- (4.1,2.5) ;
\draw[dashed] (2.6,1) -- (4.1,1.5) ;
\draw[dashed] (2.6,4) -- (4.1,3.5) ;
\draw[dashed] (2.6,3) -- (4.1,2.5) ;
\draw[dashed] (2.6,2) -- (4.1,1.5) ;
\draw[dashed] (2.6,1) -- (4.1,0.5) ;
\draw[dashed] (7.9,4) -- (6.4,4.5) ;
\draw[dashed] (7.9,3) -- (6.4,3.5) ;
\draw[dashed] (7.9,2) -- (6.4,2.5) ;
\draw[dashed] (7.9,1) -- (6.4,1.5) ;
\draw[dashed] (7.9,4) -- (6.4,3.5) ;
\draw[dashed] (7.9,3) -- (6.4,2.5) ;
\draw[dashed] (7.9,2) -- (6.4,1.5) ;
\draw[dashed] (7.9,1) -- (6.4,0.5) ;
\endtikzpicture
}
\caption{${\cal N}=1$ SUSY multiplets with spin 3/2 Clifford vacuum}
\label{N=1spin2mult}
\end{figure}

In addition to the two minimal spin 1/2 multiplets, there is a larger multiplet $\{\alpha,\chi,\bar{\chi},d\}$ with spins up to $j_z=2$ in each ${\cal N}=1$ scenario. All the left-handed spin 3/2 states $| \chi,j_z \rangle$ with $-3/2 \leq j_z \leq +3/2$ are annihilated by ${\cal Q}_a$, c.f. (\ref{N1spin2mulVac}). Hence, the Clifford vacuum transforms in a nontrivial $SO(1,3)$ representation. Starting from the four states $|\chi ,j_z\rangle$, we build the full spin two multiplet by $\bar {\cal Q}_{\pm}$ application, see figure \ref{N=1spin2mult}. The spin-$\frac{3}{2}$ states with wavefunction $\bar \chi^\mu_{\dot a}$ of opposite chirality are obtained by $|\bar \chi,j_z \rangle =\bar {\cal Q}_{+}\bar {\cal Q}_{-} |\chi,j_z \rangle$, so they form the anti-Clifford vacua.

The helicity SUSY transformations are such that normalized states are either mapped to equally normalized states or annihilated. This becomes particularly interesting at the intersection points $\bar{\mathcal{Q}}_{-}\,|\chi,j_z \rangle \leftrightarrow \bar{\mathcal{Q}}_{+}\,|\chi, j_z-1\rangle$ within the diamond where combination states of type $|\al \oplus d \rangle$ arise. From the $j_z =\pm \frac{3}{2}$ components, we obtain
\begin{align}
\bar{\mathcal{Q}}_{\pm}\,|\chi,\pm\tfrac{3}{2}\rangle & =|\alpha,\pm2\rangle, \label{begin} \\
\bar{\mathcal{Q}}_{\mp}\,|\chi,\pm\tfrac{3}{2}\rangle & =\frac{1}{2}|\alpha,\pm1\rangle\pm\frac{\sqrt{3}}{2}|d,\pm1\rangle\equiv|\alpha\pm d,\pm1\rangle,
\end{align}
whereas $\bar {\cal Q}_{\pm}$ action on $j_z =\pm \frac{1}{2}$ components yields
\begin{align}
\bar{\mathcal{Q}}_{\pm}\,|\chi,\pm\tfrac{1}{2}\rangle & =\frac{\sqrt{3}}{2}|\alpha,\pm1\rangle\mp\frac{1}{2}|d,\pm1\rangle\equiv|\alpha\mp d,\pm1\rangle,\\
\bar{\mathcal{Q}}_{\mp}\,|\chi,\pm\tfrac{1}{2}\rangle & =\frac{1}{\sqrt{2}}|\alpha,0\rangle\pm\frac{1}{\sqrt{2}}|d,0\rangle\equiv|\alpha\pm d,0\rangle \label{end} .
\end{align}
We use canonical normalization conventions for vertex operators
as well as helicity wavefunctions: Let $| \psi, j_z\rangle$ denote some physical state with polarization tensor $\psi$ and spin component $j_z$ along the quantization axis. Then, $|\psi,+j_z\rangle$ has unit scalar product with $|\psi,-j_z\rangle$ and is orthogonal to all states whose wavefunction belongs to a different $SO(3)$ representation. We can see from above results that all the states on the right-hand sides of (\ref{begin}) to (\ref{end}) have unit norm. Furthermore, we find that the combined states $|\alpha\pm d, \pm 1 \rangle$ obtained from $\bar{\mathcal{Q}}_{\mp}\,|\chi,\pm\frac{3}{2}\rangle $ are orthogonal to
$|\alpha\mp d,\pm 1\rangle$ from distinct Clifford vacuum components $\bar{\mathcal{Q}}_{\pm}\,|\chi,\pm\frac{1}{2}\rangle$, as expected.

\medskip
To complete the other half of the diamond, we have,
\begin{gather}
\bar{\mathcal{Q}}_{\pm}\,|\alpha,+2\rangle=0,\qquad\bar{\mathcal{Q}}_{\mp}\,|\alpha,+2\rangle=|\bar{\chi},\pm\tfrac{3}{2}\rangle,
\end{gather}
and\textbf{
\begin{gather}
\bar{\mathcal{Q}}_{\pm}\,|\alpha\pm d,\pm1\rangle=|\bar{\chi},\pm\tfrac{3}{2}\rangle,\qquad\bar{\mathcal{Q}}_{\mp}\,|\alpha\pm d,\pm1\rangle=0,\\
\bar{\mathcal{Q}}_{\mp}\,|\alpha\mp d,\pm1\rangle=|\bar{\chi},\pm\tfrac{1}{2}\rangle,\qquad\bar{\mathcal{Q}}_{\pm}\,|\alpha\mp d,\pm1\rangle=0,
\end{gather}
}and\textbf{
\begin{gather}
\bar{\mathcal{Q}}_{\pm}\,|\alpha\pm d,0\rangle=|\bar{\chi},\pm\tfrac{1}{2}\rangle,\qquad\bar{\mathcal{Q}}_{\pm}\,|\alpha\mp d,0\rangle=0.
\end{gather}}The diamond is symmetric about the $j_z=0$ line. In other words, once we obtained all the transformations for the states in its upper half, the lower half can be filled up by interchanging
momentum spinors $p\leftrightarrow q$. This holds by the construction of the massive helicity wavefunctions in appendix \ref{app:hel}, see also \cite{Feng:2010yx} and \cite{Feng:2011qc}.

\subsection{${\cal N}=2$ supermultiplets}

The new feature of extended $\mathcal{N}=2$ SUSY is the non-Abelian $SU(2)$ R-symmetry group. The supercharges are spinors with respect to this $SU(2)$ and therefore carry fundamental indices $i$. That is why we have to introduce a bookkeeping Grassmann variable $\eta_i$ which decouples from the spacetime spinor index structure. In other words, this $\eta_i$ is a spinor of the R-symmetry but a scalar with respect to the spacetime $SO(1,3)$. We define supercharge components $\bar {\cal Q}_{\pm}(\eta)$ which are associated with the choices $\bar \eta_{\dot a,i}^{+} = \eta_i p^{\ast}_{\dot a} / \langle pq \rangle$ and $\bar \eta_{\dot a,i}^{-} = \eta_i q^{\ast}_{\dot a} / \langle qp \rangle$:
\begin{align}
\bar {\cal Q}_{+}(\eta)
& =  \eta_{i} \;  \frac{ p^{\ast}_{\dot a} }{ \langle pq \rangle } \; \bar {\cal Q}^{\dot a,i}, \\
\bar {\cal Q}_{-}(\eta)
& =  \eta_{i} \;  \frac{ q^{\ast}_{\dot a} }{ \langle qp \rangle } \; \bar {\cal Q}^{\dot a,i}.
\end{align}
In the construction of ${\cal N}=2$ supermultiplets from their Clifford vacua, we obtain states in nontrivial representations of the $SU(2)$ R-symmetry.\footnote{In fact, it is a peculiar feature of the first mass level that its Clifford vacua are R-symmetry scalars.} Their $SU(2)$ tensor structures will be displayed inside the ket vectors, right after the $J_z$ eigenvalue, separated by a semicolon.\footnote{In the literature, on-shell supersymmetry is usually described by the notion of supercharge eigenstates -- Grassmann coherent states, firstly in \cite{Nair}, and recently in \cite{Boels:2011zz} and also \cite{Coherent}. Our presentation of SUSY transformations including internal wavefunctions (carrying the R-symmetry quantum numbers) are an equivalent way of expressing their information content.}

\subsubsection{Spin one supermultiplets}

Again, we start our presentation with the smaller multiplets of lower spin. The universal sector due to $\mathcal{N}=2$ SUSY
encompasses two spin one multiplets with scalar Clifford vacua, see figures \ref{N=2spin1amult} and \ref{N=2spin1bmult} below.
\begin{figure}[h]
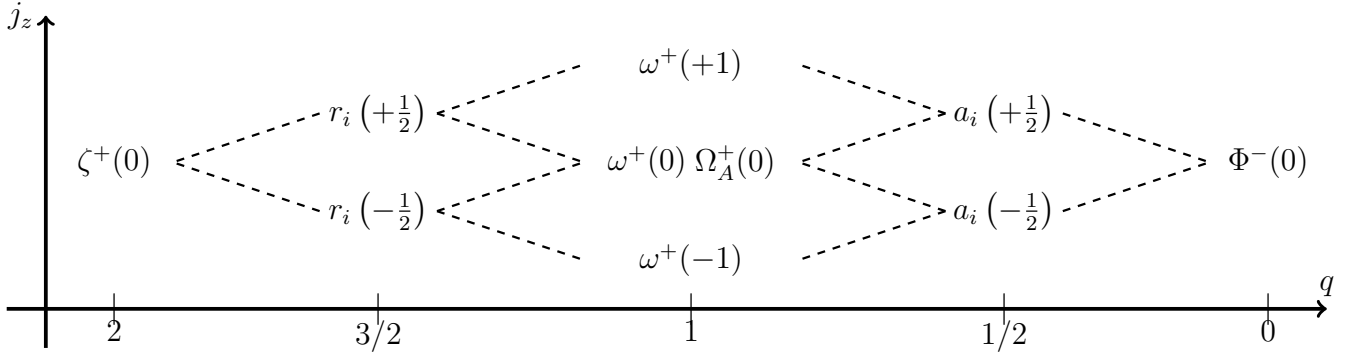

\centerline{
\tikzpicture [scale=1.3,line width=0.30mm]
\draw[line width=0.50mm,->] (-1.4,1.6) -- (-1.4,5) node[left]{$j_z$};
\draw[line width=0.50mm,->] (-1.8,2) -- (11.7,2) node[above]{$q$};
\draw (-0.7,2)node[below]{$2$} node{$|$};
\draw (2,2)node[below]{$3/2$} node{$|$};
\draw (5.2,2)node[below]{$1$} node{$|$};
\draw (8.4,2)node[below]{$1/2$} node{$|$};
\draw (11.1,2)node[below]{$0$} node{$|$};
\draw (2,4) node{$r_i \left( +\tfrac{1}{2} \right)$} ;
\draw (2,3) node{$r_i \left( -\tfrac{1}{2} \right)$} ;
\draw (8.4,4) node{$a_i \left( +\tfrac{1}{2} \right)$} ;
\draw (8.4,3) node{$a_i \left( -\tfrac{1}{2} \right)$} ;
\draw (5.2,4.5) node{$\om^+(+1)$} ;
\draw (4.75,3.5) node{$\om^+(0)$} ;
\draw (5.65,3.5) node{$\Om^+_A(0)$} ;
\draw (5.2,2.5) node{$\om^+(-1)$} ;
\draw[dashed] (2.6,4) -- (4.1,4.5) ;
\draw[dashed] (2.6,3) -- (4.1,3.5) ;
\draw[dashed] (1.4,4) -- (-0.1,3.5) ;
\draw[dashed] (1.4,3) -- (-0.1,3.5) ;
\draw (-0.7,3.5) node{$\zeta^+(0)$} ;
\draw[dashed] (2.6,4) -- (4.1,3.5) ;
\draw[dashed] (2.6,3) -- (4.1,2.5) ;
\draw[dashed] (7.8,4) -- (6.3,4.5) ;
\draw[dashed] (7.8,3) -- (6.3,3.5) ;
\draw[dashed] (7.8,4) -- (6.3,3.5) ;
\draw[dashed] (7.8,3) -- (6.3,2.5) ;
\draw (11.1,3.5) node{$\Phi^-(0)$} ;
\draw[dashed] (9,4) -- (10.5,3.5) ;
\draw[dashed] (9,3) -- (10.5,3.5) ;
\endtikzpicture
}
\caption{${\cal N}=2$ SUSY multiplet with scalar Clifford vacuum: In ${\cal N}=2$ scenarios, the $U(1)$ charge $q$ with respect to the internal toroidal directions is plotted along the horizontal axis. Since the world-sheet fields $i\pa Z^{\pm}$ and $\ee^{iqH}$ have charge $\pm1$ and $q$, respectively, the SUSY generators built from $\ee^{\pm iH/2}$ and $i \pa Z^{\pm} \ee^{\mp iH/2}$ change $q$ by the fixed offset $\pm 1/2$.}
\label{N=2spin1amult}
\end{figure}
\begin{figure}[h]
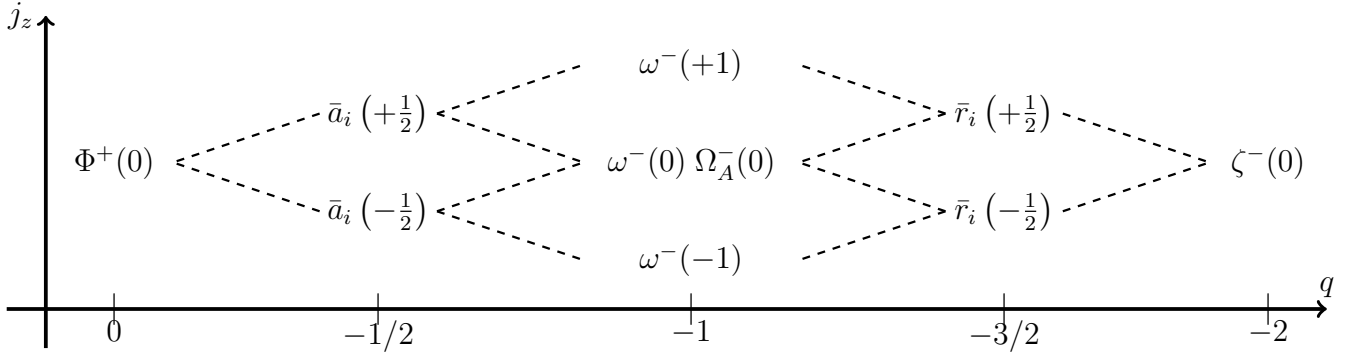

\centerline{
\tikzpicture [scale=1.3,line width=0.30mm]
\draw[line width=0.50mm,->] (-1.4,1.6) -- (-1.4,5) node[left]{$j_z$};
\draw[line width=0.50mm,->] (-1.8,2) -- (11.7,2) node[above]{$q$};
\draw (-0.7,2)node[below]{$0$} node{$|$};
\draw (2,2)node[below]{$-1/2$} node{$|$};
\draw (5.2,2)node[below]{$-1$} node{$|$};
\draw (8.4,2)node[below]{$-3/2$} node{$|$};
\draw (11.1,2)node[below]{$-2$} node{$|$};
\draw (2,4) node{$\bar a_i \left( +\tfrac{1}{2} \right)$} ;
\draw (2,3) node{$\bar a_i \left( -\tfrac{1}{2} \right)$} ;
\draw (8.4,4) node{$\bar r_i \left( +\tfrac{1}{2} \right)$} ;
\draw (8.4,3) node{$\bar r_i \left( -\tfrac{1}{2} \right)$} ;
\draw (5.2,4.5) node{$\om^-(+1)$} ;
\draw (4.75,3.5) node{$\om^-(0)$} ;
\draw (5.65,3.5) node{$\Om^-_A(0)$} ;
\draw (5.2,2.5) node{$\om^-(-1)$} ;
\draw[dashed] (2.6,4) -- (4.1,4.5) ;
\draw[dashed] (2.6,3) -- (4.1,3.5) ;
\draw[dashed] (1.4,4) -- (-0.1,3.5) ;
\draw[dashed] (1.4,3) -- (-0.1,3.5) ;
\draw (-0.7,3.5) node{$\Phi^+(0)$} ;
\draw[dashed] (2.6,4) -- (4.1,3.5) ;
\draw[dashed] (2.6,3) -- (4.1,2.5) ;
\draw[dashed] (7.8,4) -- (6.3,4.5) ;
\draw[dashed] (7.8,3) -- (6.3,3.5) ;
\draw[dashed] (7.8,4) -- (6.3,3.5) ;
\draw[dashed] (7.8,3) -- (6.3,2.5) ;
\draw (11.1,3.5) node{$\zeta^-(0)$} ;
\draw[dashed] (9,4) -- (10.5,3.5) ;
\draw[dashed] (9,3) -- (10.5,3.5) ;
\endtikzpicture
}
\caption{conjugate ${\cal N}=2$ SUSY multiplet with scalar Clifford vacuum}
\label{N=2spin1bmult}
\end{figure}

The first multiplet $\{\omega^{+},a,r,\Phi^{-},\zeta^{+},\Omega_{A}^{+}\}$
is constructed from a scalar Clifford vacuum $\zeta^{+}$, c.f. (\ref{N2spin1mulVac2}).
Omitting all the vanishing results, we obtain
\begin{equation}
\bar{\mathcal{Q}}_{\pm}(\eta_{i})\,|\Phi^{+},0;1\rangle=|\bar{a},\pm\tfrac{1}{2};\eta_{i}\rangle,
\end{equation}
and
\begin{align}
\bar{\mathcal{Q}}_{\pm}(\epsilon_{j})\,|\bar{a},\pm\tfrac{1}{2},\eta_{i}\rangle & =|\omega^{-},\pm1;(\epsilon\eta)\rangle,\\
\bar{\mathcal{Q}}_{\mp}(\epsilon_{j})\,|\bar{a},\pm\tfrac{1}{2},\eta_{i}\rangle & =\frac{1}{\sqrt{2}}|\omega^{-},0;(\epsilon \eta)\rangle\pm\frac{1}{\sqrt{2}}|\Omega^{-},0;\epsilon_{j}(\tau_{A}\varepsilon)^{ji}\eta_{i}\rangle\nonumber \\
 & \equiv|\omega^{-}\pm\Omega_{A}^{-},0\rangle,
\end{align}
where $(\epsilon\eta)=\epsilon_{j}\varepsilon^{ji}\eta_{i}$. The $\om^-$ and $\Om^-$ states in the center of the diamond transform to\textbf{
\begin{gather}
\bar{\mathcal{Q}}_{\mp}(\eta_{i})\,|\omega^{-},\pm1;(\epsilon\eta)\rangle=\bar{\mathcal{Q}}_{\pm}(\eta_{i})\,|\omega^{-}\pm\Omega_{A}^{-},0\rangle=|\bar{r},\pm\tfrac{1}{2};(\epsilon\eta)\eta_{i}\rangle,\\
\bar{\mathcal{Q}}_{\mp}(\eta_{i})\,|\omega^{-}\pm\Omega_{A}^{-},0\rangle=0,
\end{gather}
}and
\begin{equation}
\bar{\mathcal{Q}}_{\mp}(\epsilon_{j})\,|\bar{r},\pm\tfrac{1}{2};(\epsilon\eta)\eta_{i}\rangle=|\zeta^{-},0;(\epsilon\eta)^{2}\rangle.
\end{equation}
Similar results are obtained for the mirror spin one multiplet $\{\omega^{-},\bar{a},\bar{r},\Phi^{+},\zeta^{-},\Omega_{A}^{-}\}$,
which is constructed from the scalar Clifford vacuum. The helicity SUSY transformations are
\begin{equation}
\bar{\mathcal{Q}}_{\pm}(\eta_{i})\,|\zeta^{+},0;1\rangle=|r,\pm\tfrac{1}{2};\eta_{i}\rangle,
\end{equation}
and
\begin{align}
\bar{\mathcal{Q}}_{\pm}(\epsilon_{j})\,|r,\pm\tfrac{1}{2};\eta_{i}\rangle & =|\omega^{+},\pm1;(\epsilon\eta)\rangle,\\
\bar{\mathcal{Q}}_{\mp}(\epsilon_{j})\,|r,\pm\tfrac{1}{2};\eta_{i}\rangle & =\frac{1}{\sqrt{2}}|\omega^{+},0;(\epsilon\eta)\rangle\pm\frac{1}{\sqrt{2}}|\Omega^{+},0;\epsilon_{j}(\tau_{A}\varepsilon)^{ji}\eta_{i}\rangle\nonumber \\
 & \equiv|\omega^{+}\pm\Omega^{+},0\rangle,
\end{align}
and\textbf{
\begin{gather}
\bar{\mathcal{Q}}_{\mp}(\eta_{i})\,|\omega^{+},\pm1;(\epsilon\eta)\rangle=\bar{\mathcal{Q}}_{\pm}(\eta_{i})\,|\omega^{+}\pm\Omega^{+},0\rangle=|a,\pm\tfrac{1}{2};(\epsilon\eta)\eta_{i}\rangle,\\
\bar{\mathcal{Q}}_{\mp}(\eta_{i})\,|\omega^{+}\pm\Omega^{+},0\rangle=0,
\end{gather}
}and
\begin{equation}
\bar{\mathcal{Q}}_{\mp}(\epsilon_{j})\,|a,\pm\tfrac{1}{2};(\epsilon\eta)\eta_{i}\rangle=|\Phi^{-},0;(\epsilon\eta)^{2}\rangle.
\end{equation}

\subsubsection{Spin two supermultiplet}

\begin{figure}[h]
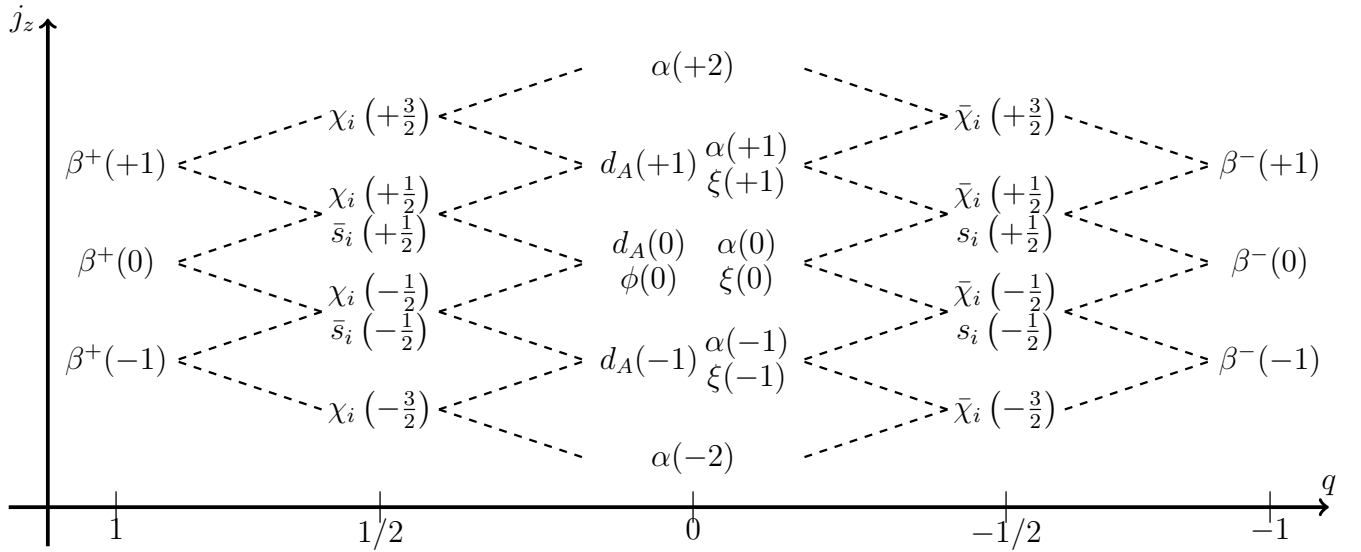

\centerline{
\tikzpicture [scale=1.3,line width=0.30mm]
\draw[line width=0.50mm,->] (-1.4,-0.4) -- (-1.4,5) node[left]{$j_z$};
\draw[line width=0.50mm,->] (-1.8,0) -- (11.7,0) node[above]{$q$};
\draw (-0.7,0)node[below]{$1$} node{$|$};
\draw (2,0)node[below]{$1/2$} node{$|$};
\draw (5.2,0)node[below]{$0$} node{$|$};
\draw (8.4,0)node[below]{$-1/2$} node{$|$};
\draw (11.1,0)node[below]{$-1$} node{$|$};
\draw (2,4) node{$\chi_i \left( +\tfrac{3}{2} \right)$} ;
\draw (2,3.2) node{$\chi_i \left( +\tfrac{1}{2} \right)$} ;
\draw (2,2.8) node{$\bar s_i \left( +\tfrac{1}{2} \right)$} ;
\draw (2,2.2) node{$\chi_i \left( -\tfrac{1}{2} \right)$} ;
\draw (2,1.8) node{$\bar s_i \left( -\tfrac{1}{2} \right)$} ;
\draw (2,1) node{$\chi_i \left( -\tfrac{3}{2} \right)$} ;
\draw (8.4,4) node{$\bar \chi_i \left( +\tfrac{3}{2} \right)$} ;
\draw (8.4,3.2) node{$\bar \chi_i \left( +\tfrac{1}{2} \right)$} ;
\draw (8.4,2.8) node{$s_i \left( +\tfrac{1}{2} \right)$} ;
\draw (8.4,2.2) node{$\bar \chi_i \left( -\tfrac{1}{2} \right)$} ;
\draw (8.4,1.8) node{$s_i \left( -\tfrac{1}{2} \right)$} ;
\draw (8.4,1) node{$\bar \chi_i \left( -\tfrac{3}{2} \right)$} ;
\draw (4.75,3.5) node{$d_A(+1)$} ;
\draw (5.75,3.68) node{$ \al(+1)$} ;
\draw (5.75,3.32) node{$ \xi(+1)$} ;
\draw (4.75,2.68) node{$d_A(0)$} ;
\draw (4.75,2.32) node{$\phi(0)$} ;
\draw (5.75,2.68) node{$ \al(0)$} ;
\draw (5.75,2.32) node{$ \xi(0)$} ;
\draw (4.75,1.5) node{$d_A(-1)$} ;
\draw (5.75,1.68) node{$ \al(-1)$} ;
\draw (5.75,1.32) node{$ \xi(-1)$} ;
\draw (5.2,4.5) node{$\al(+2)$} ;
\draw (5.2,0.5) node{$\al(-2)$} ;
\draw[dashed] (2.6,4) -- (4.1,4.5) ;
\draw[dashed] (2.6,3) -- (4.1,3.5) ;
\draw[dashed] (2.6,2) -- (4.1,2.5) ;
\draw[dashed] (2.6,1) -- (4.1,1.5) ;
\draw[dashed] (1.4,4) -- (-0.1,3.5) ;
\draw[dashed] (1.4,3) -- (-0.1,2.5) ;
\draw[dashed] (1.4,2) -- (-0.1,1.5) ;
\draw[dashed] (1.4,3) -- (-0.1,3.5) ;
\draw[dashed] (1.4,2) -- (-0.1,2.5) ;
\draw[dashed] (1.4,1) -- (-0.1,1.5) ;
\draw (-0.7,3.5) node{$\be^+(+1)$} ;
\draw (-0.7,2.5) node{$\be^+(0)$} ;
\draw (-0.7,1.5) node{$\be^+(-1)$} ;
\draw[dashed] (2.6,4) -- (4.1,3.5) ;
\draw[dashed] (2.6,3) -- (4.1,2.5) ;
\draw[dashed] (2.6,2) -- (4.1,1.5) ;
\draw[dashed] (2.6,1) -- (4.1,0.5) ;
\draw[dashed] (7.8,4) -- (6.3,4.5) ;
\draw[dashed] (7.8,3) -- (6.3,3.5) ;
\draw[dashed] (7.8,2) -- (6.3,2.5) ;
\draw[dashed] (7.8,1) -- (6.3,1.5) ;
\draw[dashed] (7.8,4) -- (6.3,3.5) ;
\draw[dashed] (7.8,3) -- (6.3,2.5) ;
\draw[dashed] (7.8,2) -- (6.3,1.5) ;
\draw[dashed] (7.8,1) -- (6.3,0.5) ;
\draw (11.1,3.5) node{$\be^-(+1)$} ;
\draw (11.1,2.5) node{$\be^-(0)$} ;
\draw (11.1,1.5) node{$\be^-(-1)$} ;
\draw[dashed] (9,4) -- (10.5,3.5) ;
\draw[dashed] (9,3) -- (10.5,2.5) ;
\draw[dashed] (9,2) -- (10.5,1.5) ;
\draw[dashed] (9,3) -- (10.5,3.5) ;
\draw[dashed] (9,2) -- (10.5,2.5) ;
\draw[dashed] (9,1) -- (10.5,1.5) ;
\endtikzpicture
}
\caption{${\cal N}=2$ SUSY multiplet with vector Clifford vacuum}
\label{N=2spin2mult}
\end{figure}

The highest spin state of the first mass level populate a spin two multiplet $\{\alpha,\chi,\bar{\chi},d,\xi,\beta^{\pm},s,\bar{s},\phi\}$
(see figure \ref{N=2spin2mult}), which is built from a vector Clifford vacuum
$\beta_{\mu}^{+}$ state, c.f. (\ref{N2spin2mulVac1}).

The supermultiplet structure is more complicated here due to intersection points in the diamond like $\bar{\mathcal{Q}}_{-}(\eta_{i})\,|\beta^{+},+1;1\rangle \leftrightarrow  \bar{\mathcal{Q}}_{+}(\eta_{i})\,|\beta^{+},0;1\rangle$. Since $j_z \mapsto -j_z$ reflection can be implemented by $p \leftrightarrow q$ exchange, we will only show the transformations for the upper half of the diamond. Omitting all the trivial
relations, we obtain
\begin{align}
\bar{\mathcal{Q}}_{+}(\eta_{i})\,|\beta^{+},+1;1\rangle & =|\chi,+\tfrac{3}{2};\eta_{i}\rangle,\\
\bar{\mathcal{Q}}_{-}(\eta_{i})\,|\beta^{+},+1;1\rangle & =\frac{1}{\sqrt{3}}|\chi,+\tfrac{1}{2};\eta_{i}\rangle+\frac{\sqrt{2}}{\sqrt{3}}|\bar{s},+\tfrac{1}{2};\eta_{i}\rangle\equiv|\chi\oplus\bar{s},+\tfrac{1}{2}\rangle_{1},\\
\bar{\mathcal{Q}}_{+}(\eta_{i})\,|\beta^{+},0;1\rangle & =\frac{\sqrt{2}}{\sqrt{3}}|\chi,+\tfrac{1}{2};\eta_{i}\rangle-\frac{1}{\sqrt{3}}|\bar{s},+\tfrac{1}{2};\eta_{i}\rangle\equiv|\chi\oplus\bar{s},+\tfrac{1}{2}\rangle_{2},
\end{align}
where $|\chi\oplus\bar{s},+\frac{1}{2}\rangle_{1}$ is orthogonal
to $|\chi\oplus\bar{s},+\frac{1}{2}\rangle_{2}$. For the helicity
SUSY transformation of the second column of figure \ref{N=2spin2mult}, we have
\begin{align}
\bar{\mathcal{Q}}_{+}(\epsilon_{j})\,|\chi,+\tfrac{3}{2};\eta_{i}\rangle & =|\alpha,+2;(\epsilon\eta)\rangle,\\
\bar{\mathcal{Q}}_{-}(\epsilon_{j})\,|\chi,+\tfrac{3}{2};\eta_{i}\rangle & =-\frac{1}{2}|\alpha,+1;(\epsilon\eta)\rangle+\frac{1}{2}|\xi,+1;(\epsilon\eta)\rangle-\frac{1}{\sqrt{2}}|d,+1;(\epsilon\eta)\rangle\nonumber \\
 & \equiv|\alpha\oplus\xi\oplus d,+1\rangle_{1},\\
\bar{\mathcal{Q}}_{+}(\epsilon_{j})\,|\chi\oplus\bar{s},+\tfrac{1}{2}\rangle_{1} & =-\frac{1}{2}|\alpha,+1;(\epsilon\eta)\rangle+\frac{1}{2}|\xi,+1;(\epsilon\eta)\rangle+\frac{1}{\sqrt{2}}|d,+1;(\epsilon\eta)\rangle\nonumber \\
 & \equiv|\alpha\oplus\xi\oplus d,+1\rangle_{2},\\
\bar{\mathcal{Q}}_{+}(\epsilon_{j})\,|\chi\oplus\bar{s},+\tfrac{1}{2}\rangle_{2} & =-\frac{1}{\sqrt{2}}|\alpha,+1;(\epsilon\eta)\rangle-\frac{1}{\sqrt{2}}|\xi,+1;(\epsilon\eta)\rangle\nonumber \\
 & \equiv|\alpha\oplus\xi\oplus d,+1\rangle_{3}.
\end{align}
One can easily check that the three states $|\alpha\oplus\xi\oplus d,+1\rangle_{1,2,3}$ are orthonormal. Moreover,
\begin{align}
\bar{\mathcal{Q}}_{-}(\epsilon_{j})\,|\chi\oplus\bar{s},+\tfrac{1}{2}\rangle_{1} & =-\frac{1}{\sqrt{6}}|\alpha,0;(\epsilon\eta)\rangle+\frac{1}{\sqrt{2}}|\xi,0;(\epsilon\eta)\rangle+\frac{1}{\sqrt{3}}|\phi,0;(\epsilon\eta)\rangle\nonumber \\
 & \equiv|\alpha\oplus\xi\oplus d\oplus\phi\rangle_{1},\\
\bar{\mathcal{Q}}_{-}(\epsilon_{j})\,|\chi\oplus\bar{s},+\tfrac{1}{2}\rangle_{2} & =-\frac{1}{\sqrt{3}}|\alpha,0;(\epsilon\eta)\rangle-\frac{1}{\sqrt{2}}|d,0;(\epsilon\eta)\rangle-\frac{1}{\sqrt{6}}|\phi,0;(\epsilon\eta)\rangle\nonumber \\
 & \equiv|\alpha\oplus\xi\oplus d\oplus\phi\rangle_{2}.
\end{align}
By interchanging $p\leftrightarrow q$ we get the states
\begin{align}
|\alpha\oplus\xi\oplus d\oplus\phi\rangle'_{1} & =|\alpha\oplus\xi\oplus d\oplus\phi\rangle_{1}(p\leftrightarrow q)=-\frac{1}{\sqrt{6}}|\alpha,0;(\epsilon\eta)\rangle-\frac{1}{\sqrt{2}}|\xi,0;(\epsilon\eta)\rangle+\frac{1}{\sqrt{3}}|\phi,0;(\epsilon\eta)\rangle,\\
|\alpha\oplus\xi\oplus d\oplus\phi\rangle'_{2} & =|\alpha\oplus\xi\oplus d\oplus\phi\rangle_{2}(p\leftrightarrow q)=-\frac{1}{\sqrt{3}}|\alpha,0;(\epsilon\eta)\rangle+\frac{1}{\sqrt{2}}|d,0;(\epsilon\eta)\rangle-\frac{1}{\sqrt{6}}|\phi,0;(\epsilon\eta)\rangle,
\end{align}
which are the results obtained from $\bar{\mathcal{Q}}_{+}(\eta_{j})\,|\chi\oplus\bar{s},-\frac{1}{2}\rangle$.
Clearly, $|\alpha\oplus\xi\oplus d\oplus\phi\rangle_{1(2)}$ is orthogonal
to $|\alpha\oplus\xi\oplus d\oplus\phi\rangle'_{1(2)}$. The helicity
SUSY transformations of this column are
\begin{gather}
\bar{\mathcal{Q}}_{-}(\eta_{i})\,|\alpha,+2;(\epsilon\eta)\rangle=\bar{\mathcal{Q}}_{+}(\eta_{i})\,|\alpha\oplus\xi\oplus d,+1\rangle_{1}=|\bar{\chi},+\tfrac{3}{2};(\epsilon\eta)\eta_{i}\rangle,\\
\bar{\mathcal{Q}}_{+}(\eta_{i})\,|\alpha\oplus\xi\oplus d,+1\rangle_{2}=\bar{\mathcal{Q}}_{+}(\eta_{i})\,|\alpha\oplus\xi\oplus d,+1\rangle_{3}=0,
\end{gather}
and
\begin{align}
\bar{\mathcal{Q}}_{-}(\eta_{i})\,|\alpha\oplus\xi\oplus d,+1\rangle_{1} & =0,\\
\bar{\mathcal{Q}}_{-}(\eta_{i})\,|\alpha\oplus\xi\oplus d,+1\rangle_{2} & =\frac{1}{\sqrt{3}}|\bar{\chi},+\tfrac{1}{2};(\epsilon\eta)\eta_{i}\rangle-\frac{\sqrt{2}}{\sqrt{3}}|s,+\tfrac{1}{2};(\epsilon\eta)\eta_{i}\rangle\equiv|\bar{\chi}\oplus s,+\tfrac{1}{2}\rangle_{1},\\
\bar{\mathcal{Q}}_{-}(\eta_{i})\,|\alpha\oplus\xi\oplus d,+1\rangle_{3} & =\frac{\sqrt{2}}{\sqrt{3}}|\bar{\chi},+\tfrac{1}{2};(\epsilon\eta)\eta_{i}\rangle+\frac{1}{\sqrt{3}}|s,+\tfrac{1}{2};(\epsilon\eta)\eta_{i}\rangle\equiv|\bar{\chi}\oplus s,+\tfrac{1}{2}\rangle_{2}.
\end{align}
States in the center of the diamond transform as
\begin{gather}
\bar{\mathcal{Q}}_{+}(\eta_{i})\,|\alpha\oplus\xi\oplus d\oplus\phi\rangle'_{1}=\bar{\mathcal{Q}}_{+}(\eta_{i})\,|\alpha\oplus\xi\oplus d\oplus\phi\rangle'_{2}=0,\\
\bar{\mathcal{Q}}_{+}(\eta_{i})\,|\alpha\oplus\xi\oplus d\oplus\phi\rangle_{1}=|\bar{\chi}\oplus s,+\tfrac{1}{2}\rangle_{1},\\
\bar{\mathcal{Q}}_{+}(\eta_{i})\,|\alpha\oplus\xi\oplus d\oplus\phi\rangle_{2}=|\bar{\chi}\oplus s,+\tfrac{1}{2}\rangle_{2},
\end{gather}
where $|\bar{\chi}\oplus s,+\frac{1}{2}\rangle_{1}$ and $|\bar{\chi}\oplus s,+\frac{1}{2}\rangle_{2}$
are orthogonal to each other. Now we are left with the transformations to the anti-Clifford vacuum states $|\beta^- \rangle$ in last column of the diamond:
\begin{gather}
\bar{\mathcal{Q}}_{-}(\epsilon_{j})\,|\bar{\chi},+\tfrac{3}{2};(\epsilon\eta)\eta_{i}\rangle=\bar{\mathcal{Q}}_{+}(\epsilon_{j})\,|\bar{\chi}\oplus s,+\tfrac{1}{2}\rangle_{1}=|\beta^{-},+1;(\epsilon\eta)^{2}\rangle,\\
\bar{\mathcal{Q}}_{+}(\epsilon_{j})\,|\bar{\chi}\oplus s,+\tfrac{1}{2}\rangle_{2}=\bar{\mathcal{Q}}_{-}(\epsilon_{j})\,|\bar{\chi}\oplus s,+\tfrac{1}{2}\rangle_{1}=0,\\
\bar{\mathcal{Q}}_{-}(\epsilon_{j})\,|\bar{\chi}\oplus s,+\tfrac{1}{2}\rangle_{2}=|\beta^{-},0;(\epsilon\eta)^{2}\rangle.
\end{gather}
This completes the helicity SUSY transformations for the upper half of
the diamond representing the spin two supermultiplet of $\mathcal{N}=2$.

\subsection{${\cal N}=4$ supermultiplet}

In $\mathcal{N}=4$ SUSY, the supercharges carry internal $SO(6)\equiv SU(4)$
spinor indices $I$ or $\bar{I}$. Similar to $\mathcal{N}=2$ case,
we introduce the internal spinors $\eta_{I}$ and $\bar{\eta}^{\bar{I}}$.
Then the components of the (right-handed) anti-supercharge can be written as
\begin{gather}
\bar{\mathcal{Q}}_{+}=\bar{\eta}^{\bar{I}}\frac{p_{\dot{a}}^{*}}{\langle pq\rangle}\bar{\mathcal{Q}}_{\bar{I}}^{\dot{a}},\qquad\bar{\mathcal{Q}}_{-}=\bar{\eta}^{\bar{I}}\frac{q_{\dot{a}}^{*}}{\langle qp\rangle}\bar{\mathcal{Q}}_{\bar{I}}^{\dot{a}}.
\end{gather}
We only have one big spin two supermultiplet in $\mathcal{N}=4$, see figure
13. Starting from the Clifford vacuum $\Phi^{+}$, c.f. (\ref{N4mulVac}), the remainder of the multiplet is filled by $\bar{\mathcal{Q}}_{\pm}$ application. Following the symmetry argument of the last subsections,
we will only show the helicity SUSY transformation of the states in
the upper half $j_z \geq 0$ of the diamond. And again, the internal wavefunctions
of the physical states are displayed right behind the semicolon in
the ket.
\begin{center}
\begin{sidewaysfigure}
\tikzpicture [xscale=0.97,yscale=2.5,line width=0.30mm]
\draw[line width=0.50mm,->] (-1.6,0.6) -- (-1.6,6) node[left]{$j_z$} ;
\draw[line width=0.50mm,->] (-2.5,1) -- (21.5,1) ;
\draw (9.5,0.2) node{Figure 13: ${\cal N}=4$ SUSY multiplet with scalar Clifford vacuum};
\draw (-0.7,3.5) node{$\Phi^+(0)$} ;
\draw[dashed] (1.1,4) -- (0,3.5) ;
\draw[dashed] (1.1,3) -- (0,3.5) ;
\draw (2,4) node{$\bar a^{\bar I} \left( +\tfrac{1}{2} \right)$} ;
\draw (2,3) node{$\bar a^{\bar I} \left( -\tfrac{1}{2} \right)$} ;
\draw[dashed] (2.8,4) -- (3.9,4.5) ;
\draw[dashed] (2.8,3) -- (3.9,3.5) ;
\draw[dashed] (2.8,4) -- (3.9,3.5) ;
\draw[dashed] (2.8,3) -- (3.9,2.5) ;
\draw (4.7,4.5) node{$\be_m^-(+1)$} ;
\draw (4.7,3.62) node{$\be_m^-(0)$} ;
\draw (4.7,3.38) node{$\Om_{mnp}^-(0)$} ;
\draw (4.7,2.5) node{$\be_m^-(-1)$} ;
\draw[dashed] (6.6,4) -- (5.5,4.5) ;
\draw[dashed] (6.6,3) -- (5.5,3.5) ;
\draw[dashed] (6.6,2) -- (5.5,2.5) ;
\draw[dashed] (6.6,5) -- (5.5,4.5) ;
\draw[dashed] (6.6,4) -- (5.5,3.5) ;
\draw[dashed] (6.6,3) -- (5.5,2.5) ;
\draw (7.4,3.88) node{$\bar r_m^{\bar I} \left( +\tfrac{1}{2} \right)$} ;
\draw (7.4,2.88) node{$\bar r_m^{\bar I} \left( -\tfrac{1}{2} \right)$} ;
\draw (7.4,4.12) node{$ \chi_I \left( +\tfrac{1}{2} \right)$} ;
\draw (7.4,3.12) node{$  \chi_I \left( -\tfrac{1}{2} \right)$} ;
\draw (7.4,5) node{$ \chi_I \left( +\tfrac{3}{2} \right)$} ;
\draw (7.4,2) node{$  \chi_I \left( -\tfrac{3}{2} \right)$} ;
\draw[dashed] (8.2,5) -- (9.3,5.5) ;
\draw[dashed] (8.2,4) -- (9.3,4.5) ;
\draw[dashed] (8.2,3) -- (9.3,3.5) ;
\draw[dashed] (8.2,2) -- (9.3,2.5) ;
\draw[dashed] (8.2,5) -- (9.3,4.5) ;
\draw[dashed] (8.2,4) -- (9.3,3.5) ;
\draw[dashed] (8.2,3) -- (9.3,2.5) ;
\draw[dashed] (8.2,2) -- (9.3,1.5) ;
\draw (10.1,5.5) node{$\al(+2)$} ;
\draw (10.1,4.5) node{$\al(+1)$} ;
\draw (10.1,3.5) node{$\al(0)$} ;
\draw (10.1,2.5) node{$\al(-1)$} ;
\draw (10.1,1.5) node{$\al(-2)$} ;
\draw (10.1,4.74) node{$d_{mn}(+1)$} ;
\draw (10.1,2.26) node{$d_{mn}(-1)$} ;
\draw (10.1,3.74) node{$d_{mn}(0)$} ;
\draw (10.1,3.26) node{$\zeta_{mn}(0)$} ;
\draw[dashed] (12,5) -- (10.9,5.5) ;
\draw[dashed] (12,4) -- (10.9,4.5) ;
\draw[dashed] (12,3) -- (10.9,3.5) ;
\draw[dashed] (12,2) -- (10.9,2.5) ;
\draw[dashed] (12,5) -- (10.9,4.5) ;
\draw[dashed] (12,4) -- (10.9,3.5) ;
\draw[dashed] (12,3) -- (10.9,2.5) ;
\draw[dashed] (12,2) -- (10.9,1.5) ;
\draw (12.8,3.88) node{$ r^m_I \left( +\tfrac{1}{2} \right)$} ;
\draw (12.8,2.88) node{$ r^m_I \left( -\tfrac{1}{2} \right)$} ;
\draw (12.8,4.12) node{$\bar \chi^{\bar I} \left( +\tfrac{1}{2} \right)$} ;
\draw (12.8,3.12) node{$\bar \chi^{\bar I} \left( -\tfrac{1}{2} \right)$} ;
\draw (12.8,5) node{$\bar \chi^{\bar I} \left( +\tfrac{3}{2} \right)$} ;
\draw (12.8,2) node{$\bar \chi^{\bar I} \left( -\tfrac{3}{2} \right)$} ;
\draw[dashed] (13.6,4) -- (14.7,4.5) ;
\draw[dashed] (13.6,3) -- (14.7,3.5) ;
\draw[dashed] (13.6,2) -- (14.7,2.5) ;
\draw[dashed] (13.6,5) -- (14.7,4.5) ;
\draw[dashed] (13.6,4) -- (14.7,3.5) ;
\draw[dashed] (13.6,3) -- (14.7,2.5) ;
\draw (15.5,4.5) node{$\be_m^+(+1)$} ;
\draw (15.5,3.62) node{$\be_m^+(0)$} ;
\draw (15.5,3.38) node{$\Om_{mnp}^+(0)$} ;
\draw (15.5,2.5) node{$\be_m^+(-1)$} ;
\draw[dashed] (17.4,4) -- (16.3,4.5) ;
\draw[dashed] (17.4,3) -- (16.3,3.5) ;
\draw[dashed] (17.4,4) -- (16.3,3.5) ;
\draw[dashed] (17.4,3) -- (16.3,2.5) ;
\draw (18.3,4) node{$a_{I} \left( +\tfrac{1}{2} \right)$} ;
\draw (18.3,3) node{$a_I \left( -\tfrac{1}{2} \right)$} ;
\draw[dashed] (19.1,4) -- (20.2,3.5) ;
\draw[dashed] (19.1,3) -- (20.2,3.5) ;
\draw (21,3.5) node{$\Phi^-(0)$} ;
\endtikzpicture
\end{sidewaysfigure}
\end{center}

We start from Clifford vacuum state $|\Phi^{+},0;1\rangle$ located
at the far left of the diamond. The helicity SUSY transformations
read
\begin{equation}
\bar{\mathcal{Q}}_{+}(\bar{\eta}^{\bar{I}})\,|\Phi^{+},0;1\rangle=|\bar{a},+\tfrac{1}{2};\bar{\eta}^{\bar{I}}\rangle,
\end{equation}
and
\begin{align}
\bar{\mathcal{Q}}_{+}(\bar{\epsilon}^{\bar{J}})\,|\bar{a},+\tfrac{1}{2};\bar{\eta}^{\bar{I}}\rangle & =|\beta^{-},+1;\tfrac{1}{\sqrt{2}}\bar{\epsilon}^{\bar{J}}(\bar\gamma_{m}C)_{\bar{J}\bar{I}}\bar{\eta}^{\bar{I}}\rangle,\\
\bar{\mathcal{Q}}_{-}(\bar{\epsilon}^{\bar{J}})\,|\bar{a},+\tfrac{1}{2};\bar{\eta}^{\bar{I}}\rangle & =-\frac{1}{\sqrt{2}}|\beta^{-},0;\tfrac{1}{\sqrt{2}}\bar{\epsilon}^{\bar{J}}(\bar\gamma_{m}C)_{\bar{J}\bar{I}}\bar{\eta}^{\bar{I}}\rangle+\frac{1}{\sqrt{2}}|\Omega^{-},0;\tfrac{1}{12}\bar{\epsilon}^{\bar{J}}(\bar{\gamma}_{mnl}C)_{\bar{J}\bar{I}}\bar{\eta}^{\bar{I}}\rangle\nonumber \\
 & \equiv|\beta^{-}\oplus\Omega^{-},0\rangle,
\end{align}
and
\begin{align}
\bar{\mathcal{Q}}_{+}(\bar{\xi}^{\bar{K}})\,|\beta^{-},+1;\tfrac{1}{\sqrt{2}}\bar{\epsilon}^{\bar{J}}(\bar\gamma_{m}C)_{\bar{J}\bar{I}}\bar{\eta}^{\bar{I}}\rangle & =|\chi,+\tfrac{3}{2};\varepsilon_{\bar I \bar J \bar K \bar L} \bar{\eta}^{\bar{I}} \bar{\epsilon}^{\bar{J}} \bar{\xi}^{\bar{K}} C^{\bar L}_L \rangle,\\
\bar{\mathcal{Q}}_{-}(\bar{\xi}^{\bar{K}})\,|\beta^{-},+1;\tfrac{1}{\sqrt{2}}\bar{\epsilon}^{\bar{J}}(\bar\gamma_{m}C)_{\bar{J}\bar{I}}\bar{\eta}^{\bar{I}}\rangle & =-\frac{1}{\sqrt{3}}|\chi,+\tfrac{1}{2};\varepsilon_{\bar I \bar J \bar K \bar L} \bar{\eta}^{\bar{I}} \bar{\epsilon}^{\bar{J}} \bar{\xi}^{\bar{K}} C^{\bar L}_L\rangle+\frac{\sqrt{2}}{\sqrt{3}}|\bar{r},+\tfrac{1}{2};\bar{r}_{\beta}\rangle\nonumber \\
 & \equiv|\chi\oplus\bar{r},+\tfrac{1}{2}\rangle_{1},\\
\bar{\mathcal{Q}}_{+}(\bar{\xi}^{\bar{K}})\,|\beta^{-}\oplus\Omega^{-},0\rangle & =\frac{1}{\sqrt{3}}|\chi,+\tfrac{1}{2};\varepsilon_{\bar I \bar J \bar K \bar L} \bar{\eta}^{\bar{I}} \bar{\epsilon}^{\bar{J}} \bar{\xi}^{\bar{K}} C^{\bar L}_L\rangle+\frac{1}{\sqrt{6}}|\bar{r},+\tfrac{1}{2};\bar{r}_{\beta}\rangle+\frac{1}{\sqrt{2}}|\bar{r},+\tfrac{1}{2};\bar{r}_{\Omega}\rangle\nonumber \\
 & \equiv|\chi\oplus\bar{r},+\tfrac{1}{2}\rangle_{2},
\end{align}
where
\begin{align}
\bar{r}_{\beta} & =\frac{\sqrt{3}}{2}\bar{\epsilon}^{\bar{J}}(\bar\gamma^{m}C)_{\bar{J}\bar{I}}\bar{\eta}^{\bar{I}}\bar{\xi}^{\bar{K}}\big(\delta_{mn}^{(6)}\delta_{\bar{K}}^{\phantom{K}\bar{L}}+\frac{1}{6}(\bar{\gamma}_{m}\gamma_n)_{\bar{K}}^{\phantom{K}\bar{L}}\big),\\
\bar{r}_{\Omega} & =\frac{1}{48}\bar{\epsilon}^{\bar{J}}(\bar{\gamma}_{mnl}C)_{\bar{J}\bar{I}}\bar{\eta}^{\bar{I}}\bar{\xi}^{\bar{K}}(\bar{\gamma}_{k}\gamma^{mnl})_{\bar{K}}^{\phantom{K}\bar{L}}.
\end{align}
Note that $\bar{r}_{\beta}$ and $\bar{r}_{\Omega}$ represent different and mutually orthogonal internal wavefunctions of $\bar{r}$.

\medskip
The left-handed spin 3/2 states in the third column of the ${\cal N}=4$ diamond transform to
\begin{align}
\bar{\mathcal{Q}}_{+}(\bar{\theta}^{\bar{M}})\,|\chi,+\tfrac{3}{2};\varepsilon_{\bar I \bar J \bar K \bar L} \bar{\eta}^{\bar{I}} \bar{\epsilon}^{\bar{J}} \bar{\xi}^{\bar{K}} C^{\bar L}_L\rangle & =|\alpha,+2;\varepsilon(\bar \eta \bar \epsilon \bar \xi \bar \theta)\rangle,\\
\bar{\mathcal{Q}}_{-}(\bar{\theta}^{\bar{M}})\,|\chi,+\tfrac{1}{2};\varepsilon_{\bar I \bar J \bar K \bar L} \bar{\eta}^{\bar{I}} \bar{\epsilon}^{\bar{J}} \bar{\xi}^{\bar{K}} C^{\bar L}_L\rangle & =-\frac{1}{2}|\alpha,+1;\varepsilon(\bar \eta \bar \epsilon \bar \xi \bar \theta) \rangle+\frac{\sqrt{3}}{2}|d,+1;d_{\chi}\rangle \equiv|\alpha\oplus d,+1\rangle_{1},\\
\bar{\mathcal{Q}}_{+}(\bar{\theta}^{\bar{M}})\,|\chi\oplus\bar{r},+\tfrac{1}{2}\rangle_{1} & =-\frac{1}{2}|\alpha,+1;\varepsilon(\bar \eta \bar \epsilon \bar \xi \bar \theta)\rangle-\frac{1}{2\sqrt{3}}|d,+1;d_{\chi}\rangle+\frac{\sqrt{2}}{\sqrt{3}}|d,+1;d_{\bar{r}_{\beta}}\rangle\nonumber \\
 & \equiv|\alpha\oplus d,+1\rangle_{2},\\
\bar{\mathcal{Q}}_{+}(\bar{\theta}^{\bar{M}})\,|\chi\oplus\bar{r},+\tfrac{1}{2}\rangle_{2} & =\frac{1}{2}|\alpha,+1;\varepsilon(\bar \eta \bar \epsilon \bar \xi \bar \theta) \rangle+\frac{1}{2\sqrt{3}}|d,+1;d_{\chi}\rangle+\frac{1}{\sqrt{6}}|d,+1;d_{\bar{r}_{\beta}}\rangle\nonumber \\
 & +\frac{1}{\sqrt{2}}|d,+1;d_{\bar{r}_{\Omega}}\rangle \equiv|\alpha\oplus d,+1\rangle_{3},\\
\bar{\mathcal{Q}}_{-}(\bar{\theta}^{\bar{M}})\,|\chi\oplus\bar{r},+\tfrac{1}{2}\rangle_{1} & =\frac{1}{\sqrt{6}}|\alpha,0;\varepsilon(\bar \eta \bar \epsilon \bar \xi \bar \theta) \rangle-\frac{1}{\sqrt{6}}|d,0;d_{\chi}\rangle+\frac{1}{\sqrt{3}}|d,0;d_{\bar{r}_{\beta}}\rangle+\frac{1}{\sqrt{3}}|\zeta,0;\zeta_{\bar{r}_{\beta}}\rangle\nonumber \\
 & \equiv|\alpha\oplus d\oplus\zeta,0\rangle_{1},\\
\bar{\mathcal{Q}}_{-}(\bar{\theta}^{\bar{M}})\,|\chi\oplus\bar{r},+\tfrac{1}{2}\rangle_{2} & =-\frac{1}{\sqrt{6}}|\alpha,0;\varepsilon(\bar \eta \bar \epsilon \bar \xi \bar \theta)\rangle+\frac{1}{\sqrt{6}}|d,0;d_{\chi}\rangle+\frac{1}{2\sqrt{3}}|d,0;d_{\bar{r}_{\beta}}\rangle\nonumber \\
 & \,+\frac{1}{2}|d,0;d_{\bar{r}_{\Omega}}\rangle+\frac{1}{2\sqrt{3}}|\zeta,0;\zeta_{\bar{r}_{\beta}}\rangle+\frac{1}{2}|\zeta,0;\zeta_{\bar{r}_{\Omega}}\rangle\nonumber \\
 & \equiv|\alpha\oplus d\oplus\zeta,0\rangle_{2},
\end{align}
where we have used the following abbreviations:
\begin{gather}
\varepsilon(\bar{\eta}\bar{\epsilon}\bar{\xi}\bar{\theta})=\varepsilon_{\bar{I}\bar{J}\bar{K}\bar{L}}\bar{\eta}^{\bar{I}}\bar{\epsilon}^{\bar{J}}\bar{\xi}^{\bar{K}}\bar{\theta}^{\bar{L}},\\
d_{\chi}=\frac{1}{2\sqrt{3}}\bar{\theta}^{\bar{M}}(\bar{\gamma}^{[m}\gamma^{n]})_{\bar{M}}^{\phantom{M}\bar{L}}\varepsilon_{\bar{I}\bar{J}\bar{K}\bar{L}}\bar{\eta}^{\bar{I}}\bar{\epsilon}^{\bar{J}}\bar{\xi}^{\bar{K}},\\
d_{\bar{r}_{\beta}}=\frac{1}{\sqrt{2}}\bar{\theta}^{\bar{M}}\bar{r}_{\beta}^{[m|,\bar{L}|}(\bar{\gamma}^{n]}C)_{\bar{M}\bar{L}},\qquad d_{\bar{r}_{\Omega}}=\frac{1}{\sqrt{2}}\bar{\theta}^{\bar{M}}\bar{r}_{\Omega}^{[m|,\bar{L}|}(\bar{\gamma}^{n]}C)_{\bar{M}\bar{L}},\\
\zeta_{\bar{r}_{\beta}}=\frac{1}{\sqrt{2}}\bar{\theta}^{\bar{M}}\bar{r}_{\beta}^{(m|,\bar{L}|}(\bar{\gamma}^{n)}C)_{\bar{M}\bar{L}},\qquad\zeta_{\bar{r}_{\Omega}}=\frac{1}{\sqrt{2}}   \bar{\theta}^{\bar{M}}\bar{r}_{\Omega}^{(m|,\bar{L}|}(\bar{\gamma}^{n)}C)_{\bar{M}\bar{L}}.
\end{gather}
Similarly, $d_{\chi}$, $d_{\bar{r}_{\beta}}$, $d_{\bar{r}_{\Omega}}$
and $\zeta_{\bar{r}_{\beta}}$, $\zeta_{\bar{r}_{\Omega}}$ are two
pairs of orthogonal states with respect to the internal R-symmetry.
Thus, the explicit computation confirms that different states located
at the same point inside the diamond (with the same $j_{z}$) are
orthogonal to each other.

\medskip
Now we are left with the helicity SUSY transformations for the right
half of the diamond. After some manipulations, we obtain
\begin{gather}
\bar{\mathcal{Q}}_{-}(\bar{\eta}^{\bar{I}})\,|\alpha,+2;\varepsilon(\bar{\eta}\bar{\epsilon}\bar{\xi}\bar{\theta})\rangle=\bar{\mathcal{Q}}_{+}(\bar{\eta}^{\bar{I}})\,|\alpha\oplus d,+1\rangle_{1}=|\bar{\chi},+\tfrac{3}{2};\varepsilon(\bar{\eta}\bar{\epsilon}\bar{\xi}\bar{\theta})\bar{\eta}^{\bar{I}}\rangle,\qquad\qquad \\
\bar{\mathcal{Q}}_{-}(\bar{\eta}^{\bar{I}})\,|\alpha\oplus d,+1\rangle_{2}=\bar{\mathcal{Q}}_{+}(\bar{\eta}^{\bar{I}})\,|\alpha\oplus d\oplus\zeta,0\rangle_{1}\qquad\qquad\qquad\qquad\qquad\qquad\nonumber \\
\qquad\qquad=\frac{1}{\sqrt{3}}|\bar{\chi},+\tfrac{1}{2};\varepsilon(\bar{\eta}\bar{\epsilon}\bar{\xi}\bar{\theta})\bar{\eta}^{\bar{I}}\rangle-\frac{\sqrt{2}}{\sqrt{3}}|r,+\tfrac{1}{2};\varepsilon(\bar{\eta}\bar{\epsilon}\bar{\xi}\bar{\theta})\bar{\eta}^{\bar{I}}\bar{\gamma}_{\bar{I}I}^{m}\big(\delta_{mn}^{(6)}\delta_{\phantom{I}J}^{I}+(\gamma_{m}\bar{\gamma}_{n})_{\phantom{I}J}^{I}\big)\rangle\nonumber \\
\equiv|\bar{\chi}\oplus r,+\tfrac{1}{2}\rangle_{1},\qquad\qquad\qquad\qquad\qquad\qquad\qquad\qquad\qquad\qquad\;\\
\bar{\mathcal{Q}}_{-}(\bar{\eta}^{\bar{I}})\,|\alpha\oplus d,+1\rangle_{3}=\bar{\mathcal{Q}}_{+}(\bar{\eta}^{\bar{I}})\,|\alpha\oplus d\oplus\zeta,0\rangle_{2}\qquad\qquad\qquad\qquad\qquad\qquad\nonumber \\
\qquad\qquad=\frac{1}{\sqrt{3}}|\bar{\chi},+\tfrac{1}{2};\varepsilon(\bar{\eta}\bar{\epsilon}\bar{\xi}\bar{\theta})\bar{\eta}^{\bar{I}}\rangle+\frac{1}{\sqrt{6}}|r,+\tfrac{1}{2};\varepsilon(\bar{\eta}\bar{\epsilon}\bar{\xi}\bar{\theta})\bar{\eta}^{\bar{I}}\bar{\gamma}_{\bar{I}I}^{m}\big(\delta_{mn}^{(6)}\delta_{\phantom{I}J}^{I}+(\gamma_{m}\bar{\gamma}_{n})_{\phantom{I}J}^{I}\big)\rangle\nonumber \\
+\frac{1}{\sqrt{2}}|r,+\tfrac{1}{2};\varepsilon(\bar{\eta}\bar{\epsilon}\bar{\xi}\bar{\theta})\bar{\eta}^{\bar{I}}\bar{\gamma}_{\bar{I}I}^{mnl}(\gamma_{k}\bar{\gamma}_{mnl})_{\phantom{I}J}^{I}\rangle\qquad\qquad\qquad\qquad\qquad\nonumber \\
\equiv|\bar{\chi}\oplus r,+\tfrac{1}{2}\rangle_{2},\qquad\qquad\qquad\qquad\qquad\qquad\qquad\qquad\qquad\qquad\;
\end{gather}
and
\begin{gather}
\bar{\mathcal{Q}}_{-}(\bar{\epsilon}^{\bar{J}})\,|\bar{\chi},+\tfrac{3}{2};\varepsilon(\bar{\eta}\bar{\epsilon}\bar{\xi}\bar{\theta})\bar{\eta}^{\bar{I}}\rangle=\bar{\mathcal{Q}}_{+}(\bar{\epsilon}^{\bar{J}})\,|\bar{\chi}\oplus r,+\tfrac{1}{2}\rangle_{1}=|\beta^{+},+1;\tfrac{1}{\sqrt{2}}\varepsilon(\bar{\eta}\bar{\epsilon}\bar{\xi}\bar{\theta})\bar{\epsilon}^{\bar{J}}(\bar{\gamma}_{m}C)_{\bar{J}\bar{I}}\bar{\eta}^{\bar{I}}\rangle,\\
\bar{\mathcal{Q}}_{-}(\bar{\epsilon}^{\bar{J}})\,|\bar{\chi}\oplus r,+\tfrac{1}{2}\rangle_{2}=-\frac{1}{\sqrt{2}}|\beta^{+},0;\tfrac{1}{\sqrt{2}}\varepsilon(\bar{\eta}\bar{\epsilon}\bar{\xi}\bar{\theta})\bar{\epsilon}^{\bar{J}}(\bar{\gamma}_{m}C)_{\bar{J}\bar{I}}\bar{\eta}^{\bar{I}}\rangle\qquad\qquad\qquad\qquad\;\, \nonumber \\
\qquad\qquad\qquad\qquad\qquad\quad\; +\frac{1}{\sqrt{2}}|\Omega^{+},0;\tfrac{1}{12}\varepsilon(\bar{\eta}\bar{\epsilon}\bar{\xi}\bar{\theta})\bar{\epsilon}^{\bar{J}}(C\gamma_{mnl})_{\bar{J}\bar{I}}\bar{\eta}^{\bar{I}}\rangle\equiv|\beta^{+}\oplus\Omega^{+},0\rangle,
\end{gather}
and
\begin{equation}
\bar{\mathcal{Q}}_{-}(\bar{\xi}^{\bar{K}})\,|\beta^{+},+1;\tfrac{1}{\sqrt{2}}\varepsilon(\bar{\eta}\bar{\epsilon}\bar{\xi}\bar{\theta})\bar{\epsilon}^{\bar{J}}(\bar{\gamma}_{m}C)_{\bar{J}\bar{I}}\bar{\eta}^{\bar{I}}\rangle=\bar{\mathcal{Q}}_{+}(\bar{\xi}^{\bar{K}})\,|\beta^{+}\oplus\Omega^{+},0\rangle=|a,+\tfrac{1}{2};\varepsilon(\bar{\eta}\bar{\epsilon}\bar{\xi}\bar{\theta})\varepsilon_{\bar{I}\bar{J}\bar{K}\bar{L}}\bar{\eta}^{\bar{I}}\bar{\epsilon}^{\bar{J}}\bar{\xi}^{\bar{K}}C_{L}^{\bar{L}}\rangle,
\end{equation}
and finally we have
\begin{equation}
\bar{\mathcal{Q}}_{-}(\bar{\theta}^{\bar{L}})\,|a,+\tfrac{1}{2};\varepsilon(\bar{\eta}\bar{\epsilon}\bar{\xi}\bar{\theta})\varepsilon_{\bar{I}\bar{J}\bar{K}\bar{L}}\bar{\eta}^{\bar{I}}\bar{\epsilon}^{\bar{J}}\bar{\xi}^{\bar{K}}C_{L}^{\bar{L}}\rangle=|\Phi^{+},0;[\varepsilon(\bar{\eta}\bar{\epsilon}\bar{\xi}\bar{\theta})]^{2}\rangle.
\end{equation}
This completes the chain of transformations that take the Clifford vacuum $| \Phi^+ \rangle$ into its anti-Clifford counterpart $| \Phi^- \rangle$.

\section{Conclusions \& Outlook}

The main purpose of this paper is the explicit construction of vertex operators and SUSY transformation of universal multiplets of the first mass level. In sections \ref{sec:n4}, \ref{sec:n1} and \ref{sec:n2}, we have identified the $\ap m^2 = 1$ particle content of superstring compactifications to four dimensions whose presence is implied by ${\cal N}=4,1,2$ SUSY respectively. The universality arguments are based on Ramond sector SCFT operators which necessarily enter the SUSY charges and are available for building vertex operators, see section \ref{sec:SCFTuniv} and \cite{BD1,BD2,BD3}. Then, using subleading terms of OPEs, we have explicitly evaluated all the SUSY transformations and found that the 24 (80) first mass level states in ${\cal N}=1$ (${\cal N}=2$) scenarios are aligned into three supermultiplets. This has to be contrasted with the maximally supersymmetric case where the 256 states form one single ${\cal N}=4$ multiplet.

\medskip
The multiplet structure of all the cases is investigated using spinor helicity methods and the results are summarized in figures \ref{N=1spin12}, \ref{N=1spin2mult} for ${\cal N}=1$, figures \ref{N=2spin1amult}, \ref{N=2spin1bmult}, \ref{N=2spin2mult} for ${\cal N}=2$ and figure 13 for ${\cal N}=4$. We worked out the transformation properties of helicity eigenstates along a covariantly chosen quantization axis, see section \ref{sec:hel} for the main results and appendix \ref{app:hel} for some background information on spinor helicity methods.

\medskip
This work motivates a lot of further studies. It would be desirable to determine the universal particle content at higher mass levels, i.e. to classify the $SO(3)$ and R-symmetry quantum numbers of universal ${\cal N}=1$ and ${\cal N}=2$ SUSY multiplets of any mass level along the lines of the $SO(9)$ analysis in \cite{Hanany:2010da}. Explicit vertex operators on the second mass level are available in ten \cite{Bianchi:2010es} and four dimensions \cite{Feng:2011qc}. These results suggest an investigation of subleading Regge trajectories, i.e. closed form expressions for vertex operators of non-maximal spin $n,n-1,\ldots$ at mass level $n$.

\medskip
Having a good control over vertex operators is necessary to gain further insight into the S matrix of massive string excitations. The leading Regge trajectory is a good example where cubic and quartic interactions could be discussed for all mass levels, see \cite{Sagnotti:2010at} for bosonic string theory and \cite{Schlotterer:2010kk} for the superstring. The simple structure of the $N$-point open superstring disk amplitude of massless states \cite{Mafra:2011nv, Mafra:2011nw} suggests that also the amplitudes of heavy vibration modes enjoy a hidden harmony. It would be desirable to work out the kinematic building blocks and the most natural basis of world-sheet integrals for their tree- and loop amplitudes.

\medskip
Supersymmetry is certainly a key ingredient for investigating scattering amplitudes of massive states. An efficient way of constraining (or in some cases even determining) massive superamplitudes via supersymmetric Ward identities is explained in \cite{Boels:2012ie, Boels:2012if}. For the purpose of a full-fledged superstring computation, the pure spinor formalism \cite{Berkovits:2000fe} is a very useful approach to take advantage of manifest supersymmetry. Unfortunately, the only explicitly known vertex operators in pure spinor superspace are at mass levels zero and one \cite{Berkovits:2002qx, Park:2011if}, so determining and applying their higher mass counterparts is an open challenge. Also, there exists a manifestly ${\cal N}=1$ supersymmetric approach in four dimensions known as the hybrid formalism \cite{Berkovits:1996bf}, see \cite{Berkovits:1997M1,Berkovits:1998M2} for a treatment of the first mass level in this framework. In any case, understanding the super Poincar\'e multiplet structure of massive states is the indispensable first step to exploit the power of SUSY for scattering amplitude, this was a key motivation for the present article.

\vskip1cm
\goodbreak
\centerline{\noindent{\bf Acknowledgments} }\vskip 2mm
This work is supported by the DFG Transregional Collaborative Research Centre TRR 33 and the DFG cluster of excellence ``Origin and Structure of the Universe".
WZ.F. is also supported in part by the National Science Foundation Grant PHY-0757959. Any opinions, findings, and conclusions or recommendations expressed in this material are those of the author and do not necessarily reflect the views of the National Science Foundation.
WZ.F. is grateful to the Max-Planck-Institut f\"ur Physik (Werner-Heisenberg-Institut) for its kind hospitality while this work was initiated. He also would like to thank Cheng Peng for very helpful discussions at the final stage of this work. D.L. likes to thank the Simons Center for Geometry and Physics for its hospitality. O.S. would like to thank Northeastern University for hospitality during the time of preparation and Rutger Boels, Stefan Hohenegger, Stephan Stieberger and Tomasz Taylor for enlightening discussions. We are also very grateful to Robert Richter for carefully reading our draft and giving us multiple valuable suggestions.

\appendix

\section{Notation and convention}
\label{appCONV}

Various types of indices appear in this article, so it is essential to keep the notation as clear and unambiguous as possible. Here is a list of occurring index classes together with the preferably used alphabets and letters:
\begin{itemize}
\item In ten dimensions, vector indices of $SO(1,9)$ are taken from the middle of the Latin alphabet $m,n,p,...$. The corresponding Weyl spinor indices are Greek letters from the beginning of the alphabet, $\al,\be,\ga,...$ for left-handed spinors, and their dotted version $\dal,\dbe,\dga,...$ for the right-handed counterparts.
\item Vectors in four-dimensional Minkowski spacetime have indices from the middle of the Greek alphabet $\mu,\nu,\la,\rho,...$. Spinor indices of $SO(1,3)$ are lower case Latin letters $a,b,c,...$ for left-handed Weyl spinors and upper case $\dot a,\dot b, \dot c$ for right-handed Weyl spinors.
\item The R-symmetry group of ${\cal N}=4$ spacetime SUSY is $SO(6) \equiv SU(4)$. We will use $m,n,p\ldots$ as vector indices and $I,J,K$ ($\bar I,\bar J, \bar K$) as left-handed (right-handed) spinor indices. Confusions with the $D=10$ vector indices are excluded by the context.
\item In case of ${\cal N}=2$ spacetime SUSY, we denote the fundamental indices of the $SU(2)$ R-symmetry by $i,j,k$ and the corresponding adjoint indices by $A,B,C$.
\item Chan-Paton generators carrying the color degrees of freedom of the vertex operator are suppressed throughout this work since they are the same for all members of the SUSY multiplet.
\item Also, the coupling $g_{\te{A}} = \sqrt{2\ap} g_{\te{YM}} $ of vertex operators is suppressed, i.e. set to unity.
\end{itemize}
All these symmetry groups involve their metrics $\eta^{mn},\eta^{\mu \nu},\de_{mn}^{(6)}$ as well as gamma matrices and charge conjugation matrices as Clebsch-Gordan coefficients:
\begin{itemize}
\item $\ga^m_{\al \dbe} , \bar \ga_m^{\dal \be}$ and $C_\al{}^{\dbe},C^{\dal}{}_\be$ in $D=10$
\item $\si^\mu_{a \dot b},\sib_\mu^{\dot a b}$ and $\vep_{ab} ,\vep^{\dot a \dot b}$ in $D=4$
\item $\ga_m^{I \bar J}, \bar \ga^m_{\bar I J}$ and $C^{I}{}_{\bar J}, C_{\bar I}{}^{J}$ for the internal $SO(6)$ of ${\cal N}=4$ SUSY
\item standard Pauli matrices $\tau_A{}^i{}_j$ and $\vep^{ij}$ for the $SU(2)$ R-symmetry of ${\cal N}=2$ SUSY
\end{itemize}
Our conventions for the slash notation is
\beq \begin{array}{rl}
\not \! k_{\al \dbe} \eq k_m \, \ga^m_{\al \dbe} \co  \not \! k^{ \dbe \al} \eq k^m \, \gab_m^{ \dbe \al } \ \ \ \ \ \ &\te{in \ $D=10$}  \\
 \not \! k_{a \dot b} \eq k_\mu \, \si^\mu_{a \dot b} \co \ \,  \not \! k^{ \dot b a} \eq k^\mu \, \sib_\mu^{ \dot b a } \ \ \ \ \ \ \ &\te{in \ $D=4$}  \end{array} \ .
\eeq
The totally antisymmetric $\vep$ tensors are normalized to having nonzero $\pm 1$, e.g. $\vep^{\mu \nu \la \rho}$ for $D=4$ vectors and $\vep_{ABC}$ for the adjoint representation of $SU(2)$.

\medskip
The signature of the Dirac algebras is negative in lines with the Wess \& Bagger conventions:
\begin{align}
\ga^m_{\al \dbe} \, \bar \ga^{n \dbe \ga} \ + \ \ga^n_{\al \dbe} \, \bar \ga^{m \dbe \ga} &\eq - \, 2 \, \eta^{mn} \, \de_\al^\ga  \\
\si^{\mu}_{a \dot b} \, \sib^{\nu \dot b c} \ + \ \si^{\nu}_{a \dot b} \, \sib^{\mu \dot b c} &\eq - \, 2 \, \eta^{\mu \nu} \, \de_{a}^c \\
 \ga_m^{I\bar J} \, \gab_{n \bar J K} \ + \ \ga_n^{I\bar J} \, \gab_{m \bar J K}  &\eq - \, 2 \, \de_{mn}^{(6)} \, \de^I_K \ .
\end{align}
On the other hand, the $SU(2)$ Pauli matrices obey the multiplication rule
\beq
(\tau_A){}^i{}_j \,  (\tau_B){}^j{}_k \eq \de_{AB} \, \de^{i}_k \ + \ i \vep_{ABC} \, (\tau^{C})^{i}{}_k
\eeq
Useful material on spinors in various spacetime dimensions can be found in \cite{gamma1, gamma2, gamma3}, the present conventions closely follow \cite{ Haertl:2009yf, Hartl:2010ks,Thesis}.


\section{Operator product expansions}

This appendix gathers the operator product expansions needed to evaluate the BRST constraints and SUSY variations. Before taking a closer look at the interacting SCFTs, let us display the free field OPEs for the sake of completeness, namely
\begin{align}
i\pa X^\mu(z) \,  \ee^{ik \cdot X}(w) \ \ &\sim \ \ \left[ \,  \frac{ 2\al' \, k^\mu }{z-w} \ + \ i\pa X^\mu (w) \ + \ (z-w) \, i \pa^2 X^\mu(w) \ + \ \ldots \, \right] \, \ee^{ik \cdot X}(w)  \label{dX1} \\
i\pa X^\mu(z) \, i\pa X^{\nu}(w) \, \ee^{ik \cdot X}(w) \ \ &\sim \ \ \left[ \, \frac{2\al' \, \eta^{\mu \nu}}{(z-w)^2} \ + \ \frac{ 2\al' \, k^\mu \, i\pa X^\nu(w)}{z-w} \ + \ i\pa X^\mu \, i\pa X^{\nu}(w) \ + \ \ldots \, \right] \, \ee^{ik \cdot X}(w)  \label{dX2}
\end{align}
as well as
\beq
\psi^\mu(z) \, \psi^\nu(w) \ \ \sim \ \ \frac{ \eta^{\mu \nu} }{z-w} \ + \ \psi^\mu \, \psi^\nu(w) \ + \ (z-w) \, \pa \psi^\mu \, \psi^\nu(w) \ + \ \ldots \label{5,1}
\eeq
They are valid in any number of compactification dimensions. Another universal feature is the superghost CFT, governed by
\begin{align}
\ee^{ q_1 \phi(z)} \, \ee^{q_2 \phi(w)} \eq &(z-w)^{-q_1 q_2} \, \Bigl[ \, \ee^{(q_1 + q_2) \,\phi(w)} \ + \ q_1 \, (z-w) \, \pa \phi \, \ee^{(q_1 + q_2) \,\phi(w)} \Bigr. \notag \\
& \ \ \Bigl. \ + \ \tfrac{1}{2} \; (z-w)^2 \, \bigl[ \, q_1 \, \pa^2 \phi \; + \; q_1^2 \, (\pa \phi)^2 \, \bigr] \, \ee^{(q_1 + q_2) \,\phi(w)} \ +\ \ldots \, \Bigr] \ .
\label{5,4}
\end{align}
The following subsections consider the interacting RNS CFT of the $\psi$ fermion and its spin fields $S$ as well as its excited versions. The OPEs were pioneered in \cite{Kostelecky:1986xg} and can be checked by means of correlation functions gathered in \cite{Haertl:2009yf, Hartl:2010ks,Thesis}, and a broader discussion of the RNS operator algebra in various dimensions will be given in \cite{progress}.

\subsection{Spacetime CFT in $D=10$}
\label{appD=10cft}

Evaluating the BRST conditions on the most general fermion vertex operator at the first mass level requires OPEs
\begin{align}
\psi^m(z) \, S_\al(w) \ \ &\sim \ \ \frac{ \ga^m_{\al \dbe} \, S^{\dbe}(w) }{\sqrt{2} \, (z-w)^{1/2}}  \ + \ (z-w)^{1/2} \, \left[ \, S^m_{\al}(w) \ + \ \frac{ 2 \,  \ga^m_{\al \dbe}  }{\sqrt{2} \, 5 } \; \pa S^{\dbe}(w) \, \right]  \ + \ \ldots
\label{5,2X1} \\
%
%
\psi^m(z) \, \pa S_\al(w) \ \ &\sim \ \ \frac{ \ga^m_{\al \dbe} \, S^{\dbe}(w) }{2\sqrt{2} \, (z-w)^{3/2}} \ - \ \frac{ S^m_\al(w) }{2 \, (z-w)^{1/2}} \ + \ \frac{4 \, \ga^m_{\al \dbe} \, \pa S^{\dbe}(w) }{5 \, \sqrt{2}  \, (z-w)^{1/2}} \ + \ \ldots
\label{5,2X3} \\
%
%
 \psi_m(z) \, S_n^{\dbe}(w) \ \ &\sim \ \ \frac{   \eta_{mn} \, S^{\dbe}(w) }{  (z-w)^{3/2}}  \ + \ \frac{ \gab_m^{\dbe \al} \, S_{n \al}(w) }{\sqrt{2} \, (z-w)^{1/2}} \ - \ \frac{ 2 \, \eta_{mn} \, \pa S^{\dbe}(w)}{ 5 \, (z-w)^{1/2} }  \ + \ \ldots
 \label{5,2X5} 
 %
\end{align}
in $D=10$. The corresponding SUSY variations are computed by means of
\begin{align}
S^{\dbe}(z) \, \psi_m(w)  \ \ &\sim \ \ \frac{ \gab_m^{\dbe \al} \, S_{\al}(w) }{\sqrt{2} \, (z-w)^{1/2}} \ + \ (z-w)^{1/2} \, \left[ \, S_m^{\dbe}(w) \ + \ \frac{3 \, \gab_m^{\dbe \al} \, \pa S_{\al}(w)}{5 \, \sqrt{2} }  \, \right]  \ + \ \ldots
\label{5,95X1} \\
S^{\dbe}(z) \, \psi_{m} \, \psi_{n} \, \psi_{p}(w) \ \ &\sim \ \ \frac{- \, 1 }{2 \sqrt{2} \, (z-w)^{3/2}} \; \gab_{mnp}^{ \dbe \al } \, S_{\al}(w) \ - \ \frac{3}{2 \, (z-w)^{1/2}} \; (\gab_{[mn})^{\dbe}{}_{\dal} \, S_{p]}^{\dal}(w)  \notag \\
& \hskip2.5cm+ \  \frac{ 1}{10 \,\sqrt{2} \,  (z-w)^{1/2} } \; \gab_{mnp}^{\dbe \al} \, \pa S_\al(w)  \ + \ \ldots
\label{5,95X2} \\
S^{\dbe}(z) \, \pa \psi_m(w) \ \ &\sim \ \ \frac{ \gab_m^{ \dbe \al } \, S_{\al}(w) }{2\sqrt{2} \, (z-w)^{3/2}} \ - \ \frac{ S_m^{\dbe}(w) }{2 \, (z-w)^{1/2}} \ + \ \frac{7 \, \gab_m^{\dbe \al} \, \pa S_{\al}(w) }{10 \, \sqrt{2} \, (z-w)^{1/2}} \ + \ \ldots \label{5,95X3}
\end{align}
for the NS sector and
\begin{align}
S_\al(z) \, S_\be(w) \ \ &\sim \ \ \frac{(\ga^m C)_{\al \be} \, \psi_m(w)}{\sqrt{2} \, (z-w)^{3/4}} \ + \  (z-w)^{1/4} \,  \frac{(\ga^m C )_{\al \be} \, \pa \psi_m(w)}{2 \sqrt{2}} \notag \\
& \hskip2.5cm - \ (z-w)^{1/4} \, \frac{ (\ga^{mnp} C)_{\al \be} \, \psi_m \, \psi_n \, \psi_p(w)}{12 \sqrt{2}}   \ + \ \ldots \label{5,96X1} \\
S_\al(z) \, S_m^{\dbe}(w) \ \ &\sim \ \ \frac{  C_\al{}^{\dbe} \, \psi_m(w)  }{ (z-w)^{ 7/4}} \  - \ \frac{C_\al{}^{\dbe} \, \pa \psi_m(w) }{2 \, (z-w)^{ 3/4}} \ -  \ \frac{(\ga^{np} \, C)_{\al}{}^{\dbe} \, \psi_m \, \psi_n \, \psi_p(w)}{4 \, (z-w)^{ 3/4}}  \ + \ \ldots \label{5,96X2} \\
S_\al(z) \, \pa S_\be(w) \ \ &\sim \ \ \frac{3 \, (\ga^m C)_{\al \be} \, \psi_m(w)}{4\sqrt{2} \, (z-w)^{7/4}} \ + \ \frac{7 \, (\ga^m C )_{\al \be} \, \pa \psi_m(w)}{8 \sqrt{2} \, (z-w)^{3/4}} \notag \\
&\hskip2.5cm + \ \frac{(\ga^{mnp} C)_{\al \be} \, \psi_m \, \psi_n \, \psi_p(w)}{48\sqrt{2} \, (z-w)^{3/4}} \ + \ \ldots \label{5,96X3}
\end{align}
for the R sector.

\subsection{Spacetime CFT in $D=4$}
\label{appD=4cft}

In $D=4$ spacetime dimensions, $h=\frac{1}{4}$ spin fields $S_a, S^{\dot b}$ of both chiralities are present. The OPEs between spinors and vectors or $p$-forms treat both chiralities on equal footing, e.g.
\begin{align}
\psi^\mu(z) \, S_a(w) \ \ &\sim \ \ \frac{ \si^\mu_{a \dot b} \, S^{\dot b}(w) }{\sqrt{2} \, (z-w)^{1/2}}  \ + \ (z-w)^{1/2} \, \left[ \, S^\mu_{a}(w) \ + \ \frac{ \si^\mu_{a \dot b}  }{\sqrt{2}} \; \pa S^{\dot b}(w) \, \right]  \ + \ \ldots
\label{5,3X1} \\
\psi_\mu(z) \, S^{\dot b}(w) \ \ &\sim \ \ \frac{ \sib_\mu^{ \dot ba} \, S_a(w) }{\sqrt{2} \, (z-w)^{1/2}}  \ + \ (z-w)^{1/2} \, \left[ \, S_\mu^{\dot b}(w) \ + \ \frac{ \sib_\mu^{\dot b a}  }{\sqrt{2}} \; \pa S_{a}(w) \, \right]  \ + \ \ldots
\label{5,3X2}
\end{align}
that is why we only display one chiral half of further OPEs:
\begin{align}
\psi^\mu(z) \, \pa S_a(w) \ \ &\sim \ \ \frac{ \si^\mu_{a \dot b} \, S^{\dot b}(w) }{2\sqrt{2} \, (z-w)^{3/2}} \ - \ \frac{ S^\mu_a(w) }{2 \, (z-w)^{1/2}} \ + \ \frac{\si^\mu_{a \dot b} \, \pa S^{\dot b}(w) }{2\sqrt{2}  \, (z-w)^{1/2}} \ + \ \ldots
\label{5,3X3} \\
%
%
 \psi_\mu(z) \, S_\nu^{\dot b}(w) \ \ &\sim \ \ \frac{   \eta_{\mu \nu} \, S^{\dot b}(w) }{  (z-w)^{3/2}}  \ + \ \frac{ \sib_{\mu}^{\dot b a} \, S_{\nu a}(w) }{\sqrt{2} \, (z-w)^{1/2}} \ - \ \frac{ \eta_{\mu \nu} \, \pa S^{\dot b}(w)}{ (z-w)^{1/2} }  \ + \ \ldots
 \label{5,3X5} 
 %
\end{align}
Four-dimensional SUSY variations of NS operators require
\begin{align}
S^{\dot b}(z) \, \psi_\mu(w)  \ \ &\sim \ \ \frac{ \sib_\mu^{\dot b a} \, S_{a}(w) }{\sqrt{2} \, (z-w)^{1/2}} \ + \ (z-w)^{1/2} \,  S_\mu^{\dot b}(w)  \ + \ \ldots
\label{5,97X1} \\
S^{\dot b}(z) \, \psi_\mu \, \psi_\nu(w) \ \ &\sim \ \ \frac{ -\, (\sib_{\mu \nu})^{\dot b}{}_{\dot a} \, S^{\dot a}(w)}{2 \, (z-w)} \ + \ \sqrt{2} \, \sib^{[\nu | \dot b a} \, S^{|\mu]}_a(w) \ + \ \frac{1}{2} \; (\sib_{\mu \nu})^{\dot b}{}_{\dot a} \, \pa S^{\dot a}(w) \ + \ \ldots \\
S^{\dot b}(z) \, \psi_{\mu} \, \psi_{\nu} \, \psi_{\la}(w) \ \ &\sim \ \ \frac{- \, 1 }{2 \sqrt{2} \, (z-w)^{3/2}} \; \sib_{\mu \nu \la }^{ \dot b a } \, S_{a}(w) \ - \ \frac{3}{2 \, (z-w)^{1/2}} \; (\sib_{[\mu \nu})^{\dot b}{}_{\dot a} \, S_{\la]}^{\dot a}(w)  \notag \\
& \hskip2.5cm+ \  \frac{ 1}{\sqrt{2} \,  (z-w)^{1/2} } \; \sib_{\mu \nu \la}^{\dot b a} \, \pa S_a(w)  \ + \ \ldots
\label{5,97X2} \\
S^{\dot b}(z) \, \pa \psi_\mu(w) \ \ &\sim \ \ \frac{ \sib_\mu^{ \dot b a } \, S_{a}(w) }{2\sqrt{2} \, (z-w)^{3/2}} \ - \ \frac{ S_\mu^{\dot b}(w) }{2 \, (z-w)^{1/2}} \ + \ \frac{\sib_\mu^{\dot b a} \, \pa S_{a}(w) }{ \sqrt{2} \, (z-w)^{1/2}} \ + \ \ldots \ . \label{5,97X3}
\end{align}
With two R sector states involved, the OPEs are sensitive to their relative chirality:
\begin{align}
S_a(z) \, S_b(w) \ \ &\sim \ \ \frac{\vep_{ab}}{(z-w)^{1/2}} \ - \  \frac{1}{4} \; (z-w)^{1/2} \, (\si^{\mu \nu} \vep )_{a b} \,  \psi_\mu \, \psi_\nu(w) \ + \ \ldots \label{5,98X0} \\
S_a(z) \, S^{\dot b}(w) \ \ &\sim \ \ \frac{(\si^\mu \vep)_{a}{}^{ \dot b} \, \psi_\mu(w)}{\sqrt{2}} \ + \  (z-w) \,  \frac{(\si^\mu \vep )_{a}{}^{\dot b} \, \pa \psi_\mu(w)}{2 \sqrt{2}} \notag \\
& \hskip2.5cm - \ (z-w) \, \frac{ (\si^{\mu \nu \la} \vep)_{a}{}^{\dot b} \, \psi_\mu \, \psi_\nu \, \psi_\la(w)}{12 \sqrt{2}}   \ + \ \ldots \label{5,98X1} \\
S_a(z) \, S_\mu^{\dot b}(w) \ \ &\sim \ \ \frac{ (\si^\nu \vep)_a{}^{\dot b} \, \psi_\mu\, \psi_\nu(w)  }{\sqrt{2} \, (z-w)^{1/2}}    \ + \ \ldots \label{5,98X2a} \\
S_a(z) \, S_b^\mu(w) \ \ &\sim \ \ \frac{  \vep_{ab} \, \psi^\mu(w)  }{ (z-w)} \  - \ \frac{\vep_{ab} \, \pa \psi^\mu(w) }{2 } \ -  \ \frac{(\si_{\nu \la} \vep)_{ab} \, \psi^\mu \, \psi^\nu \, \psi^\la(w)}{4}  \ + \ \ldots \label{5,98X2} \\
S_a(z) \, \pa S_{b}(w) \ \ &\sim \ \ \frac{ \vep_{ab}}{ 2 \, (z-w)^{3/2} } \  + \ \frac{(\si^{\mu \nu } \vep)_{a b} \, \psi_\mu \, \psi_\nu(w)}{8 \, (z-w)^{1/2}} \ + \ \ldots \label{5,98X3a} \\
S_a(z) \, \pa S^{\dot b}(w) \ \ &\sim \ \ \frac{ (\si^\mu \vep )_{a}{}^{\dot b} \, \pa \psi_\mu(w)}{2 \sqrt{2} } \  + \ \frac{(\si^{\mu \nu \la} \vep)_{a}{}^{\dot b} \, \psi_\mu \, \psi_\nu \, \psi_\la(w)}{12\sqrt{2}} \ + \ \ldots \label{5,98X3}
\end{align}

\subsection{Internal CFT for ${\cal N}=4$ SUSY}
\label{appN=4SUSY}

The internal components of the ten-dimensional NS fermion are denoted by $\Psi_m$ with vector index $m$ for the $SO(6)$ R-symmetry. Accordingly, the associated $h=\frac{3}{8}$ spin fields $\Si^I,\bar \Si_{\bar J}$ have $SO(6)$ spinor indices $I,\bar J=1,2,3,4$. Their mutual OPEs can be covariantly expressed in terms of $SO(6)$ gamma matrices:
\begin{align}
\Psi_m(z) \, \Si^{I}(w) \ \ &\sim \ \ \frac{\ga_m^{I \bar J} \, \bar \Si_{\bar J}(w)}{\sqrt{2} \, (z-w)^{1/2}}  \ + \ (z-w)^{1/2} \, \left[ \, \Si_m^I(w) \ + \ \frac{2\, \ga_m^{I \bar J} }{3\sqrt{2}} \; \pa \bar \Si_{\bar J}(w) \, \right] \ + \
 \ldots \label{so6,1} \\
 \Psi^m(z) \, \bar  \Si^n_{\bar J}(w) \ \ &\sim \ \ \frac{\de^{mn}_{(6)} \, \bar \Si_{\bar J}(w)}{(z-w)^{3/2}}  \ + \ \frac{ \bar \ga^m_{\bar J I} \, \Si^{n,I}(w) }{\sqrt{2}\, (z-w)^{1/2}} \ - \ \frac{2\, \de^{mn}_{(6)} \, \pa \bar \Si_{\bar J}(w)}{3 \, (z-w)^{1/2}}
\ + \
  \ldots \label{so6,2} \\
 \Psi_m(z) \, \pa \Si^{I}(w) \ \ &\sim \ \ \frac{\ga_m^{I \bar J} \, \bar \Si_{\bar J}(w)}{2\sqrt{2} \, (z-w)^{3/2}}  \ - \ \frac{ \Si^I_m(w) }{2\, (z-w)^{1/2}}
 \ + \  \frac{2 \, \ga_m^{I \bar J} \, \pa \bar \Si_{\bar J}(w)}{3\sqrt{2} \, (z-w)^{1/2}} \ + \
 \ldots \label{so6,3}
\end{align}
We need the following OPEs for computing SUSY transformations of bosons:
\begin{align}
\bar \Si_{\bar J}(z) \, \Psi^m(w)  \ \ &\sim \ \ \frac{ \bar \ga^m_{\bar JI} \, \Si^I(w) }{\sqrt{2} \, (z-w)^{1/2}} \ + \ (z-w)^{1/2} \, \left[ \, \bar \Si^m_{\bar J}(w)  \ + \  \frac{ \bar \ga^m_{\bar JI} }{3\sqrt{2}} \; \pa \Si^I(w) \, \right] \ + \ \ldots
\label{so6,4a} \\
\bar \Si_{\bar J}(z) \, \Psi^m \, \Psi^n (w) \ \ &\sim \ \ \frac{ - \, (\gab^{mn})_{\bar J}{}^{\bar I} \, \bar \Si_{\bar I}(w)}{2 \, (z-w)} \ + \ \sqrt{2} \, \gab^{[n| \bar J I} \, \Si^{m]}_I(w) \ + \ \frac{ 1}{6} \; (\gab^{mn})_{\bar J}{}^{\bar I} \, \pa \bar \Si_{\bar I} (w) \ + \ \ldots \\
\bar \Si_{\bar J}(z) \, \Psi^m \, \Psi^n \, \Psi^p(w) \ \ &\sim \ \ \frac{- \, 1 }{2 \sqrt{2} \, (z-w)^{3/2}} \; \bar \ga^{mnp}_{\bar JI} \, \Si^I(w) \ - \ \frac{3}{2 \, (z-w)^{1/2}} \; (\bar \ga^{[mn})_{\bar J}{}^{\bar I} \, \bar \Si^{p]}_{\bar I}(w)  \notag \\
& \hskip2.5cm+ \  \frac{ 1}{2\sqrt{2} \,  (z-w)^{1/2} } \; \bar \ga^{mnp}_{\bar JI} \, \pa \Si^{I}(w)  \ + \ \ldots
\label{so6,5a} \\
\bar \Si_{\bar J}(z) \, \pa \Psi^m(w) \ \ &\sim \ \ \frac{ \bar \ga^m_{ \bar J I } \, \Si^I(w) }{2\sqrt{2} \, (z-w)^{3/2}} \ - \ \frac{ \bar \Si^m_I(w) }{2 \, (z-w)^{1/2}} \ + \ \frac{5\, \bar \ga^k_{\bar J I} \, \pa \Si^I(w) }{6 \sqrt{2} \, (z-w)^{1/2}} \ + \ \ldots \ . \label{so6,6a}
\end{align}
Again, OPEs between R sector states depend on the relative chirality:
 \begin{align}
\Si^I(z) \, \bar \Si_{\bar J}(w) \ \ &\sim \ \ \frac{ C^I {}_{\bar J} }{ (z-w)^{3/4} } \ - \ \frac{1}{4} \, (z-w)^{1/4} \, (\ga_{mn} \, C)^I{}_{\bar J} \, \Psi^m \, \Psi^n(w) \ + \ \ldots \label{so6,4} \\
\Si^{ I}(z) \, \Si^{ J}(w) \ \ &\sim \ \ \frac{ (\ga_m \, C)^{IJ} \, \Psi^m(w) }{\sqrt{2} \, (z-w)^{1/4} } \ + \ (z-w)^{3/4} \, \frac{ (\ga_m \, C)^{IJ} \, \pa \Psi^m(w) }{2\sqrt{2} } \notag \\
& \hskip2.5cm - \ (z-w)^{3/4} \, \frac{ (\ga_{mnp} \, C)^{IJ} \, \Psi^m \,\Psi^n \, \Psi^p(w) }{12 \sqrt{2}} \ + \
 \ldots \label{so6,5} \\
\Si^{ I}(z) \, \Si^{ J}_m(w) \ \ &\sim \ \ \frac{ (\ga^n \, C)^{IJ} \, \Psi_m \, \Psi_n(w) }{\sqrt{2} \, (z-w)^{3/4} } \ + \ \ldots
\label{so6,6}\\
\Si^{ I}(z) \, \bar \Si^m_{\bar J}(w) \ \ &\sim \ \ \frac{ C^I {}_{\bar J} \, \Psi^m(w) }{(z-w)^{5/4}} \ - \  \frac{ C^I {}_{\bar J} \, \pa \Psi^m(w) }{2 \, (z-w)^{1/4}} \ - \ \frac{ (\ga_{np} \, C)^I {}_{\bar J} \, \Psi^m\, \Psi^n \, \Psi^p(w) }{4 \, (z-w)^{1/4}} \ + \ \ldots
\label{so6,7} \\
\Si^I(z) \, \pa \bar \Si_{\bar J}(w) \ \ &\sim \ \ \frac{3 \, C^I {}_{\bar J} }{4\,  (z-w)^{7/4} } \ + \ \frac{ (\ga_{mn} \, C)^I{}_{\bar J}  \, \Psi^m \,\Psi^n(w)}{16\,(z-w)^{3/4}} \ + \ \ldots \label{so6,8} \\
\Si^{ I}(z) \, \pa \Si^{ J}(w) \ \ &\sim \ \ \frac{ (\ga_m \, C)^{IJ} \, \Psi^m(w)}{4\sqrt{2} \, (z-w)^{5/4} } \ + \ \frac{5 \, (\ga_m \, C)^{IJ} \, \pa \Psi^m(w)}{ 8\sqrt{2} \, (z-w)^{1/4} } \notag \\
&\hskip2.5cm + \ \frac{(\ga_{mnp} \, C)^{IJ} \, \Psi^m \, \Psi^n \, \Psi^p(w)}{16\sqrt{2}\, (z-w)^{1/4} } \ + \
 \ldots \label{so6,9}
\end{align}

\subsection{Internal CFT for ${\cal N}=1$ SUSY}
\label{appN=1SUSY}

Most of the OPEs relevant for the internal $c=9$ SCFT described in subsection \ref{sec:N=1cft} can be derived from the CFT of a free boson:
\begin{align}
i \pa H(z) \, \ee^{iqH}(w) \ \ &\sim \ \ \left[ \frac{q}{z-w} \ + \ i \pa H(w) \ + \ \ldots \right] \, \ee^{iq H}(w)
\label{1,6a} \\
\ee^{iqH}(z) \, i \pa H(w) \ \ &\sim \ \ \left[ \frac{q}{z-w} \ + \ (q^2\, - \, 1) \, i \pa H(w) \ + \ \ldots \right] \, \ee^{iq H}(w) \label{1,6aa} \\
\ee^{iq_1 H}(z) \, \ee^{iq_2 H}(w) \ \ &\sim \ \ (z-w)^{q_1 q_2} \, \Big[ 1 \ + \ q_1 \, (z-w) \, i\pa H \ + \ \ldots \Big] \, \ee^{i(q_1+q_2) H}(w)
\label{1,6b}
\end{align}
This allows to reproduce (\ref{1,3a}) and (\ref{1,3b}) from the bosonized representations (\ref{1,5}) of the operators ${\cal J}$, $\Si^{\pm}$ and ${\cal O}^{\pm}$. Moreover, we have
\begin{align}
\Si^{\pm}(z) \, {\cal J}(w) \ \ &\sim \ \ \frac{\pm \, \sqrt{3} \, \Si^{\pm}(w)}{2 \, (z-w)} \ \mp \ \frac{\pa \Si^{\pm}(w)}{2\sqrt{3}} \ + \ \ldots  \label{1,6p} \\
\Si^{\pm}(z) \, {\cal O}^{\mp}(w) \ \ &\sim \ \ (z-w)^{-3/2} \, \Si^{\mp}(w) \ - \ (z-w)^{-1/2} \, \pa \Si^{\mp}(w) \ + \ \ldots \label{1,6q}
\end{align}
The excited spin fields $\tilde \Si^{\pm} = g^{\mp} \ee^{\pm iH / \sqrt{12}}$ are canonically normalized
\begin{align}
\tilde \Si^{\pm}(z) \, \tilde \Si^{\mp}(w) \ \ &\sim \ \  \frac{1}{(z-w)^{11/4}} \ \pm \ \frac{ i \pa H (w) }{2\sqrt{3} \, (z-w)^{7/4}} \ + \ \ldots
\label{1,11a} \\
\tilde \Si^{\pm}(z) \, \tilde \Si^{\pm}(w) \ \ &\sim \ \ \frac{ g^{\mp } \, g^{\mp } \, \ee^{\pm \frac{iH}{\sqrt{3}}}(w) }{(z-w)^{1/4}} \ + \ \ldots
\label{1,11b}
\end{align}
such that the mutual singularities between standard and excited spin fields are given by
\begin{align}
\tilde \Si^{\pm}(z) \, \Si^{\pm}(w) \ \ &\sim \ \ (z-w)^{1/4} \, g^{\mp} \, \ee^{\pm \frac{2i}{\sqrt{3}} H}(w) \ + \ \ldots
\label{1,12a} \\
\tilde \Si^{\pm}(z) \, \Si^{\mp}(w) \ \ &\sim \ \ \sqrt{ \frac{2}{3} } \; \frac{G^{\mp}_{\te{int}}(w)}{ (z-w)^{1/4} } \ + \ \ldots
\label{1,12b}
\end{align}
Moreover, in presence of the internal supercurrents $G_{\te{int}}^{\pm} = \sqrt{ \frac{3}{2} } \ee^{\pm i H/\sqrt{3}}  g^{\pm}$,
\begin{align}
G^{\pm}_{\te{int}}(z) \, \tilde \Si^{\pm}(w) \ \ &\sim \ \ \sqrt{ \frac{3}{2}} \; \frac{ \Si^{\pm}(w) }{(z-w)^{5/2}} \ + \ \sqrt{ \frac{2}{3}} \; \frac{ \pa \Si^{\pm}(w)}{(z-w)^{3/2}} \ + \ \ldots \label{1,13a} \\
G^{\pm}_{\te{int}}(z) \, \tilde \Si^{\mp}(w) \ \ &\sim \ \ \sqrt{ \frac{3}{2}} \;  \frac{g^{\pm} \, g^{\pm} \, \ee^{\pm \frac{i}{2\sqrt{3}} H}(w) }{(z-w)^{1/2}} \ + \ \ldots
\label{1,13b} \\
\tilde \Si^{\pm}(z) \, {\cal J}(w) \ \ &\sim \ \ \pm \; \frac{ \tilde \Si^{\pm}(w)}{2\sqrt{3} \, (z-w)} \ + \ \ldots
\end{align}

\section{Spinor helicity methods for massive wavefunctions}
\label{app:hel}

Before we proceed to introduce the massive version of the spinor helicity
formalism, we will make a short review for the helicity formalism
of massless spinors. For massless spin-$\frac{1}{2}$ spinors, we use the following notations,
\begin{flalign}
|i\rangle & =|k_{i}\rangle=u_{+}(k_{i})=v_{-}(k_{i})=\binom{0}{k_{i}^{*\dot{a}}},\\
|i] & =|k_{i}]=u_{-}(k_{i})=v_{+}(k_{i})=\binom{k_{i,a}}{0},\\
{}[i| & =[k_{i}|=\bar{u}_{+}(k_{i})=\bar{v}_{-}(k_{i})=\left(k_{i}^{a},0\right),\\
\langle i| & =\langle k_{i}|=\bar{u}_{-}(k_{i})=\bar{v}_{+}(k_{i})=\left(0,k_{i,\dot{a}}^{*}\right).
\end{flalign}
Here the momentums with spinor indices denote two component commutative
spinors. They are defined by
\begin{gather}
P^{\dot{a}a}=p_{\mu}\bar{\sigma}^{\mu\dot{a}a}=-p^{*\dot{a}}p^{a},\\
P_{a\dot{a}}=p_{\mu}\sigma_{a\dot{a}}^{\mu}=-p_{a}p_{\dot{a}}^{*},
\end{gather}
where $p^{*\dot{a}}=(p^{a})^{*}$ and $p_{\dot{a}}^{*}=(p_{a})^{*}$.
Spinor indices could be raised (lowered) by $\varepsilon^{ab}$
($\varepsilon_{ab}$) or $a,b$ with dots,
\begin{equation}
p^{a}=\varepsilon^{ab}p_{b},\qquad p^{*\dot{a}}=\varepsilon^{\dot{a}\dot{b}}p_{\dot{b}}^{*}.
\end{equation}
Then we can define the notations for the spinor products,
\begin{gather}
\langle pq\rangle=\langle p|q\rangle=\bar{u}_{-}(p)u_{+}(q)=p_{\dot{a}}^{*}q^{*\dot{a}},\\
{}[pq]=[p|q]=\bar{u}_{+}(p)u_{-}(q)=p^{a}q_{a},
\end{gather}
so that simply we have
\begin{gather}
[pq]=-[qp],\qquad\langle pq\rangle=-\langle qp\rangle,\\
\langle pq\rangle^{*}=-[pq],\qquad\langle pp\rangle=[pp]=0,
\end{gather}
and
\begin{equation}
\langle pq\rangle[qp]=-2(p\cdot q).
\end{equation}

\subsection{Massive spin one boson}

A spin $J$ particle contains $2J+1$ spin degrees of freedom associated
to the eigenstates of $J_{z}$. The choice of the quantization axis
$z$ can be handled in an elegant way by decomposing the momentum
$k$ into two arbitrary light-like reference momenta $p$ and $q$:
\begin{equation}
k^{\mu}=p^{\mu}+q^{\mu},\qquad k^{2}=-m^{2}=2pq,\qquad p^{2}=q^{2}=0.
\end{equation}
Then the spin quantization axis is chosen to be the direction of $q$
in the rest frame. The $2J+1$ spin wavefunctions depend of $p$
and $q$, however this dependence drops out in the amplitudes summed
over all spin directions and in \textquotedblleft{}unpolarized\textquotedblright{}
cross sections.

\medskip
The massive spin one wavefunctions $\xi_{\mu}$ (transverse, i.e.,
$\xi_{\mu}k^{\mu}=0$) are given by the following polarization vectors
\cite{HWF1,HWF2}, up to a phase factor,
\begin{flalign}
\text{\ensuremath{\xi}}_{+}^{\mu}(k) & =\frac{1}{\sqrt{2}m}p_{\dot{a}}^{*}\bar{\sigma}^{\mu\dot{a}a}q_{a},\\
\text{\ensuremath{\xi}}_{0}^{\mu}(k) & =\frac{1}{2m}\bar{\sigma}^{\mu\dot{a}a}(p_{\dot{a}}^{*}p_{a}-q_{\dot{a}}^{*}q_{a}),\\
\text{\ensuremath{\xi}}_{-}^{\mu}(k) & =-\frac{1}{\sqrt{2}m}q_{\dot{a}}^{*}\bar{\sigma}^{\mu\dot{a}a}p_{a}.
\end{flalign}

\subsection{Massive spin two boson}

The massive spin two boson $\text{\ensuremath{\alpha}}^{\mu\nu}$
satisfies the following conditions,
\begin{gather}
\text{\ensuremath{\alpha}}^{\mu\nu}(k,\lambda)=\text{\ensuremath{\alpha}}^{\nu\mu}(k,\lambda),\\
k_{\mu}\text{\ensuremath{\alpha}}^{\mu\nu}(k,\lambda)=0,\\
g_{\mu\nu}\text{\ensuremath{\alpha}}^{\mu\nu}(k,\lambda)=0,
\end{gather}
where $\lambda$ expresses the helicity of $\text{\ensuremath{\alpha}}^{\mu\nu}$.
We do the same decomposition of the momentum, and the wavefunction
of a spin two boson can be written as \cite{HWF1},
\begin{flalign}
\text{\ensuremath{\alpha}}^{\mu\nu}(k,+2) & =\frac{1}{2m^{2}}\bar{\sigma}^{\mu\dot{a}a}\bar{\sigma}^{\nu\dot{b}b}p_{\dot{a}}^{*}q_{a}p_{\dot{b}}^{*}q_{b}\ ,\\
\text{\ensuremath{\alpha}}^{\mu\nu}(k,+1) & =\frac{1}{4m^{2}}\bar{\sigma}^{\mu\dot{a}a}\bar{\sigma}^{\nu\dot{b}b}\left[(p_{\dot{a}}^{*}p_{a}-q_{\dot{a}}^{*}q_{a})p_{\dot{b}}^{*}q_{b}+p_{\dot{a}}^{*}q_{a}(p_{\dot{b}}^{*}p_{b}-q_{\dot{b}}^{*}q_{b})\right],\\
\text{\ensuremath{\alpha}}^{\mu\nu}(k,~0~) & =\frac{1}{2\sqrt{6}m^{2}}\bar{\sigma}^{\mu\dot{a}a}\bar{\sigma}^{\nu\dot{b}b}\left[(p_{\dot{a}}^{*}p_{a}-q_{\dot{a}}^{*}q_{a})(p_{\dot{b}}^{*}p_{b}-q_{\dot{b}}^{*}q_{b})-p_{\dot{a}}^{*}q_{a}q_{\dot{b}}^{*}p_{b}-q_{\dot{a}}^{*}p_{a}p_{\dot{b}}^{*}q_{b}\right],\\
\text{\ensuremath{\alpha}}^{\mu\nu}(k,-1) & =\frac{1}{4m^{2}}\bar{\sigma}^{\mu\dot{a}a}\bar{\sigma}^{\nu\dot{b}b}\left[(q_{\dot{a}}^{*}q_{a}-p_{\dot{a}}^{*}p_{a})q_{\dot{b}}^{*}p_{b}+q_{\dot{a}}^{*}p_{a}(q_{\dot{a}}^{*}q_{a}-p_{\dot{b}}^{*}p_{b})\right],\\
\text{\ensuremath{\alpha}}^{\mu\nu}(k,-2) & =\frac{1}{2m^{2}}\bar{\sigma}^{\mu\dot{a}a}\bar{\sigma}^{\nu\dot{b}b}q_{\dot{a}}^{*}p_{a}q_{\dot{b}}^{*}p_{b}\ .
\end{flalign}

\subsection{Massive spin $1/2$ fermions}

Massive spin $\frac{1}{2}$ fermions satisfy the Dirac equation,
\begin{flalign}
(\slashed k+m)u(k) & =0,\\
(\slashed k-m)v(k) & =0,
\end{flalign}
where $u(k)$ and $v(k)$ are positive and negative energy solutions
with momentum $k^{\mu}$, which correspond to fermion and anti-fermion
wavefunctions respectively. Since we do not deal with the wavefunctions
of the negative energy solutions, we will only present $u(k)$ wavefunction here.
$u(k)$ satisfies the spin-sum relations,
orthogonal condition and the normalization condition,
\begin{gather}
\sum_{spin}u_{\pm}(k)\bar{u}_{\pm}(k)=-\slashed k+m,\\
\bar{u}_{\pm}(k)u_{\mp}(k)=0,\\
\bar{u}_{\pm}(k)u_{\pm}(k)=2m,
\end{gather}
Writing the four component spinor $u(k)$ as
\begin{equation}
u=\binom{\chi_{a}}{\bar{\eta}^{\dot{a}}}
\end{equation}
and plugging it into the Dirac equation, we get
\begin{equation}
k_{\mu}\left(\begin{array}{cc}
0 & \sigma_{a\dot{a}}^{\mu}\\
\bar{\sigma}^{\mu\dot{a}a} & 0
\end{array}\right)\binom{\chi_{a}}{\bar{\eta}^{\dot{a}}}=-m\binom{\chi_{a}}{\bar{\eta}^{\dot{a}}}.
\end{equation}
The Dirac equation is decomposed to,
\begin{flalign}
k_{\mu}\bar{\sigma}^{\mu\dot{a}a}\chi_{a} & =-m\bar{\eta}^{\dot{a}},\\
k_{\mu}\sigma_{a\dot{a}}^{\mu}\bar{\eta}^{\dot{a}} & =-m\chi_{a}.
\end{flalign}
Making the same decomposition of the momentum $k^{\mu}=p^{\mu}+q^{\mu}$, we can obtain the wavefunction of the massive spin $\frac{1}{2}$
fermion \cite{HWF2},
\begin{flalign}
u_{+}(k) & =\binom{\frac{\langle qp\rangle}{m}q_{a}}{p^{*\dot{a}}},\\
u_{-}(k) & =\binom{p_{a}}{\frac{[qp]}{m}q^{*\dot{a}}}.
\end{flalign}

\subsection{Massive spin $3/2$ fermions}

A massive spin $\frac{3}{2}$ fermion are described by a Rarita-Schwinger
spinor-vector $\Psi^{A,\mu}$ which satisfies equations,
\begin{flalign}
(i\slashed\partial-m)_{\phantom{A}B}^{A}\Psi^{B,\mu} & =0,\label{eq:32M-C1}\\
(\gamma_{\mu})_{\phantom{A}B}^{A}\Psi^{B,\mu} & =0,\label{eq:32M-C2}\\
\partial_{\mu}\Psi^{B,\mu} & =0,\label{eq:32M-C3}
\end{flalign}
where $A$ and $B$ are spinor indices. Again we only consider the
positive energy solution $U$, it satisfies,
\begin{flalign}
({\displaystyle {\not}k}+m)_{\phantom{A}B}^{A}U(k)^{B,\mu} & =0,\\
\bar{U}_{A,\mu}(k,\lambda)U^{A,\mu}(k,\lambda^{'}) & =2m\delta_{\lambda\lambda^{'}}.
\end{flalign}
The wavefunction of $U$ can be written as \cite{HWF2},
\begin{flalign}
U^{A,\mu}(+\frac{3}{2}) & =\frac{1}{\sqrt{2}m}\binom{\frac{\langle qp\rangle}{m}q_{a}}{p^{*\dot{a}}}(p_{\dot{b}}^{*}\bar{\sigma}^{\mu\dot{b}b}q_{b})\ ,\\
U^{A,\mu}(+\frac{1}{2}) & =\frac{\bar{\sigma}^{\mu\dot{b}b}}{\sqrt{6}m}\left[\binom{\frac{\langle qp\rangle}{m}q_{a}}{p^{*\dot{a}}}(p_{\dot{b}}^{*}p_{b}-q_{\dot{b}}^{*}q_{b})+\binom{\frac{\langle qp\rangle}{m}p_{a}}{-q^{*\dot{a}}}(p_{\dot{b}}^{*}q_{b})\right],\\
U^{A,\mu}(-\frac{1}{2}) & =\frac{\bar{\sigma}^{\mu\dot{b}b}}{\sqrt{6}m}\left[\binom{p_{a}}{\frac{[qp]}{m}q^{*\dot{a}}}(p_{\dot{b}}^{*}p_{b}-q_{\dot{b}}^{*}q_{b})+\binom{-q_{a}}{\frac{[qp]}{m}p^{*\dot{a}}}(q_{\dot{b}}^{*}p_{b})\right],\\
U^{A,\mu}(-\frac{3}{2}) & =\frac{1}{\sqrt{2}m}\binom{p_{a}}{\frac{[qp]}{m}q^{*\dot{a}}}(q_{\dot{b}}^{*}\bar{\sigma}^{\mu\dot{b}b}p_{b})\ .
\end{flalign}


\begin{thebibliography}{99}

\bibitem{Feng:2010yx}
  W.~-Z.~Feng, D.~L\"ust, O.~Schlotterer, S.~Stieberger, T.~R.~Taylor,
  ``Direct Production of Lightest Regge Resonances,''
  Nucl.\ Phys.\  {\bf B843 } (2011)  570-601,
  [arXiv:1007.5254 [hep-th]].

\bibitem{Amati:1988tn}
  D.~Amati, M.~Ciafaloni and G.~Veneziano,
  ``Can Spacetime Be Probed Below the String Size?''
  Phys.\ Lett.\ B {\bf 216} (1989) 41.

\bibitem{Lust:2008qc}
  D.~L\"ust, S.~Stieberger and T.~R.~Taylor,
  ``The LHC String Hunter's Companion,''
  Nucl.\ Phys.\ B {\bf 808} (2009) 1,
  [arXiv:0807.3333 [hep-th]].

\bibitem{Lust:2009pz}
  D.~L\"ust, O.~Schlotterer, S.~Stieberger and T.~R.~Taylor,
  ``The LHC String Hunter's Companion (II): Five-Particle Amplitudes and Universal Properties,''
  Nucl.\ Phys.\ B {\bf 828} (2010) 139,
  [arXiv:0908.0409 [hep-th]].




\bibitem{Koh:1987hm}
  I.~G.~Koh, W.~Troost, A.~Van Proeyen,
  ``Covariant Higher Spin Vertex Operators In The Ramond Sector,''
  Nucl.\ Phys.\  {\bf B292 } (1987)  201.

\bibitem{Berkovits:2002qx}
  N.~Berkovits and O.~Chandia,
  ``Massive Superstring Vertex Operator in D = 10 Superspace,''
  JHEP {\bf 0208} (2002) 040
  [arXiv:hep-th/0204121].

\bibitem{Sagnotti:2010at}
  A.~Sagnotti and M.~Taronna,
  ``String Lessons for Higher-Spin Interactions,''
  Nucl.\ Phys.\ B {\bf 842} (2011) 299
  [arXiv:1006.5242 [hep-th]].

\bibitem{Bianchi:2010es}
  M.~Bianchi, L.~Lopez, R.~Richter,
  ``On Stable Higher Spin States in Heterotic String Theories,''
  JHEP {\bf 1103 } (2011)  051,
  [arXiv:1010.1177 [hep-th]].

\bibitem{Park:2011if}
  I.~Y.~Park,
  ``Scattering of Massive Open Strings in Pure Spinor,''
  Nucl.\ Phys.\ B {\bf 852} (2011) 287,
  [arXiv:1101.1204 [hep-th]].

\bibitem{Feng:2011qc}
  W.~-Z.~Feng, T.~R.~Taylor,
  ``Higher Level String Resonances in Four Dimensions,''
  Nucl. Phys. B {\bf 856} (2012) 247-277,
  [arXiv:1110.1087 [hep-th]].



\bibitem{FMS}
  D.~Friedan, E.J.~Martinec and S.H.~Shenker,
  ``Conformal Invariance, Supersymmetry And String Theory,''
  Nucl.\ Phys.\  B {\bf 271} (1986) 93.

\bibitem{BD1}
T.~Banks, L.J.~Dixon, D.~Friedan and E.J.~Martinec,
``Phenomenology and Conformal Field Theory Or Can String Theory Predict the Weak Mixing Angle?,''
  Nucl.\ Phys.\  B {\bf 299} (1988) 613.

\bibitem{BD2}
  T.~Banks and L.J.~Dixon,
 ``Constraints on String Vacua with Spacetime Supersymmetry,''
  Nucl.\ Phys.\  B {\bf 307} (1988) 93.

\bibitem{BD3}
S.~Ferrara, D. L\"ust and S. Theisen,
``World-sheet Versus Spectrum Symmetries In Heterotic And Type II Superstrings.,''
  Nucl.\ Phys.\  B {\bf 325}, 501 (1989).



\bibitem{MHV} S.~Stieberger and T.R.~Taylor,
``Supersymmetry Relations and MHV Amplitudes in Superstring Theory,''
  Nucl.\ Phys.\  B {\bf 793}, 83 (2008),
  [arXiv:0708.0574 [hep-th]].


\bibitem{Cohn:1986bn}
  J.~Cohn, D.~Friedan, Z.~-a.~Qiu, S.~H.~Shenker,
  ``Covariant Quantization Of Supersymmetric String Theories: The Spinor Field Of The Ramond-neveu-schwarz Model,''
  Nucl.\ Phys.\  {\bf B278 } (1986)  577.

\bibitem{Kostelecky:1986xg}
  V.~A.~Kostelecky, O.~Lechtenfeld, W.~Lerche, S.~Samuel, S.~Watamura,
  ``Conformal Techniques, Bosonization and Tree Level String Amplitudes,''
  Nucl.\ Phys.\  {\bf B288 } (1987)  173.




\bibitem{Haertl:2009yf}
  D.~Haertl, O.~Schlotterer, S.~Stieberger,
  ``Higher Point Spin Field Correlators in D=4 Superstring Theory,''
  Nucl.\ Phys.\  {\bf B834 } (2010)  163-221.
  [arXiv:0911.5168 [hep-th]].

\bibitem{Hartl:2010ks}
  D.~Haertl, O.~Schlotterer,
  ``Higher Loop Spin Field Correlators in Various Dimensions,''
  Nucl.\ Phys.\  {\bf B849 } (2011)  364-409.
  [arXiv:1011.1249 [hep-th]]

\bibitem{Thesis}
O.~Schlotterer,
  ``Scattering Amplitudes in Open Superstring Theory,''
 (2011) PhD thesis.

\bibitem{progress}
  O.~Schlotterer, S.~Theisen
  ``The Covariant RNS Operator Algebra,'' work in progress.



\bibitem{HWF1}
D. Spehler and S. F. Novaes,
``Helicity Wave Functions for Massless and Massive Spin-2 Particles,''
Phys. Rev. D 44 (1991) 3990.

\bibitem{HWF2}
S. F. Novaes and D. Spehler,
``Weyl-van der Waerden Spinor Technique for Spin-$\frac{3}{2}$ Fermions,''
Nucl. Phys., B371 (1992) 618.

\bibitem{Boels:2011zz}
  R.~H.~Boels, C.~Schwinn,
  ``On-shell Supersymmetry for Massive Multiplets,''
  Phys.\ Rev.\  {\bf D84 } (2011)  065006,
  [arXiv:1104.2280 [hep-th]].



\bibitem{Nair}
V. P. Nair
``A Current Algebra for Some Gauge Theory Amplitudes,''
Phys. Lett. B 214, 215 (1988).

\bibitem{Coherent}
M. Bianchi, H. Elvang and D. Z. Freedman
``Generating Tree Amplitudes in $N=4$ SYM and $N=8$ SG,''
JHEP 0809 (2008) 063,
[arXiv:0805.0757 [hep-th]];
J. M. Drummond, J. Henn, G. P. Korchemsky and E. Sokatchev,
``Dual Superconformal Symmetry of Scattering Amplitudes in $N=4$ Super-Yang-Mills Theory,''
Nucl. Phys. B 828 (2010) 317-374,
[arXiv:0807.1095 [hep-th]];
J. M. Drummond, J. Henn, G. P. Korchemsky and E. Sokatchev,
``Generalized Unitarity for $N=4$ Super-Amplitudes,''
[arXiv:0808.0491 [hep-th]];
N. Arkani-Hamed, F. Cachazo and J. Kaplan
``What is the Simplest Quantum Field Theory?,''
[arXiv:0808.1446 [hep-th]];
H. Elvang, Y.-t. Huang and C. Peng
``On-shell Superamplitudes in $N<4$ SYM,''
[arXiv:1102.4843v1 [hep-th]].



\bibitem{Hanany:2010da}
  A.~Hanany, D.~Forcella, J.~Troost,
  ``The Covariant Perturbative String Spectrum,''
  Nucl.\ Phys.\  {\bf B846 } (2011)  212-225,
  [arXiv:1007.2622 [hep-th]].


\bibitem{Schlotterer:2010kk}
  O.~Schlotterer,
  ``Higher Spin Scattering in Superstring Theory,''
  [arXiv:1011.1235 [hep-th]].

\bibitem{Mafra:2011nv}
  C.~R.~Mafra, O.~Schlotterer and S.~Stieberger,
  ``Complete N-Point Superstring Disk Amplitude I. Pure Spinor Computation,''
  [arXiv:1106.2645 [hep-th]].

\bibitem{Mafra:2011nw}
  C.~R.~Mafra, O.~Schlotterer and S.~Stieberger,
  ``Complete N-Point Superstring Disk Amplitude II. Amplitude and Hypergeometric Function Structure,''
  [arXiv:1106.2646 [hep-th]].

\bibitem{Boels:2012ie}
  R.~H.~Boels and D.~O'Connell,
  ``Simple Superamplitudes in Higher Dimensions,''
  [arXiv:1201.2653 [hep-th]].

\bibitem{Boels:2012if}
  R.~H.~Boels,
  ``Three Particle Superstring Amplitudes with Massive Legs,''
  [arXiv:1201.2655 [hep-th]].



\bibitem{Berkovits:2000fe}
  N.~Berkovits,
  ``Super Poincare Covariant Quantization of the Superstring,''
  JHEP {\bf 0004} (2000) 018,
  [arXiv:hep-th/0001035].

\bibitem{Berkovits:1996bf}
  N.~Berkovits,
  ``A New Description of the Superstring,''
  In *Rio de Janeiro 1995, Particles and fields* 390-418,
  [arXiv:hep-th/9604123].

\bibitem{Berkovits:1997M1}
N.~Berkovits, M.~M.~Leite
``First Massive State of the Superstring in Superspace,''
Phys. Lett. B415 (1997) 144-148,
[arXiv:hep-th/9709148].

\bibitem{Berkovits:1998M2}
N.~Berkovits, M.M.~Leite
``Superspace Action for the First Massive States of the Superstring,''
Phys. Lett. B454 (1999) 38-42,
[arXiv:hep-th/9812153].




\bibitem{gamma1}
  P.~C.~West,
  ``Supergravity, Brane Dynamics and String Duality,''
  [arXiv:hep-th/9811101].

\bibitem{gamma2}
J.~Polchinski, ``Superstring Theory and beyond, Volume II,'' CUP 2005.

\bibitem{gamma3}
J.~A.~Strathdee,
  ``Extended Poincar\'e Supersymmetry,''
  Int.\ J.\ Mod.\ Phys.\  A {\bf 2} (1987) 273.

\end{thebibliography}
\end{document}